\documentclass[prodmode,acmtods]{acmsmall} 
\let\subcaption\relax
\pdfoutput=1

\usepackage{amsmath}
\usepackage{amssymb}
\usepackage{pifont}
\usepackage{graphicx}
\usepackage{listings}
\usepackage{paralist}
\usepackage{xspace}
\usepackage{tablefootnote}
\usepackage{caption, subcaption}
\usepackage{xifthen}
\usepackage{balance}
\usepackage{tikz}
\usetikzlibrary{shapes,arrows}
\usepackage{pgfplots}
\usepackage{capt-of}
\usepackage[T1]{fontenc}
\usepackage[scaled]{beramono}
\usepackage{booktabs}
\usepackage{epstopdf}
\usepackage[ruled]{algorithm2e}

\SetAlFnt{\small}
\SetAlCapFnt{\small}
\SetAlCapNameFnt{\small}
\SetAlCapHSkip{0pt}
\IncMargin{-\parindent}

\acmVolume{X}
\acmNumber{X}
\acmArticle{X}
\acmYear{2016}
\acmMonth{5}

\doi{0000001.0000001}

\issn{1234-56789}

\newcommand\MyFontSize{\fontsize{8}{9.5}\selectfont} 

\lstdefinelanguage{Scala}%
{morekeywords={abstract,%
  case,catch,char,class,%
  def,else,extends,final,finally,for,%
  if,import,implicit,%
  match,module,%
  new,null,%
  object,override,%
  package,private,protected,public,%
  for,public,return,super,%
  this,throw,trait,try,type,%
  val,var,%
  with,while,%
  yield%
  },%
  sensitive,%
  morecomment=[l]//,%
  morecomment=[s]{/*}{*/},%
  morestring=[b]",%
  morestring=[b]',%
  showstringspaces=false%
}[keywords,comments,strings]%

\lstdefinestyle{scala_numbered}{
  language=Scala,
  breaklines=true,
  showspaces=false,
  showtabs=false,
  showstringspaces=false,
  breakatwhitespace=true,
  numbers=left,
  numbersep=4pt,
  numberstyle=\scriptsize\ttfamily,
  basicstyle=\MyFontSize\ttfamily,
  keywordstyle=\MyFontSize\ttfamily\bfseries,%
  columns=fullflexible,
  escapeinside={(*@}{@*)}
}

\lstdefinestyle{scala_unnumbered}{
  language=Scala,
  breaklines=true,
  showspaces=false,
  showtabs=false,
  showstringspaces=false,
  breakatwhitespace=true,
  numbers=none,
  numberstyle=\scriptsize\ttfamily,
  basicstyle=\MyFontSize\ttfamily,
  keywordstyle=\MyFontSize\ttfamily\bfseries,%
  columns=fullflexible,
  escapeinside={(*@}{@*)}
}

\newcommand{\code}[1]{\lstinline[
  language=Scala,
  breaklines=true,
  showspaces=false,
  showtabs=false,
  showstringspaces=false,
  breakatwhitespace=true,
  numbers=none,
  numberstyle=\ttfamily,
  basicstyle=\ttfamily,
  keywordstyle=\ttfamily\bfseries,%
  columns=fullflexible,
  escapeinside={(*@}{@*)}    
]|#1|}

\lstdefinestyle{sqlstyle}{
  language=SQL,
  breaklines=true,
  showspaces=false,
  showtabs=false,
  showstringspaces=false,
  breakatwhitespace=true,
  numbers=none,
  numberstyle=\scriptsize\ttfamily,
  basicstyle=\MyFontSize\ttfamily,
  keywordstyle=\MyFontSize\ttfamily\bfseries,%
  columns=fullflexible,
  escapeinside={(*@}{@*)}
}

\lstdefinestyle{cstyle}{
  language=C,
  breaklines=true,
  showspaces=false,
  showtabs=false,
  showstringspaces=false,
  breakatwhitespace=true,
  numbers=none,
  numberstyle=\scriptsize\ttfamily,
  basicstyle=\MyFontSize\ttfamily,
  keywordstyle=\MyFontSize\ttfamily\bfseries,%
  columns=fullflexible,
  escapeinside={(*@}{@*)}
}

\newcommand{\commentout}[1]{}

\def\tpch{\mbox{TPC-H}\xspace}
\def\timesten{\mbox{DBX}\xspace}
\def\clang{\mbox{CLang}\xspace}
\def\glib{\mbox{GLib}\xspace}

\def\legobase{\mbox{LegoBase}\xspace}
\def\compiler{\mbox{SC}\xspace}

\newcommand{\systemname}{\legobase} 

\def\groupby{\mbox{group by}\xspace}
\def\datastructure{data-stru\-cture\xspace}
\def\datastructures{data-stru\-ctures\xspace}
\def\hyper{HyPer\xspace}
\def\gcc{GCC\xspace}

\def\lmsdefined{Lightweight Modular Staging (LMS)}
\def\lms{LMS}
\def\paper{article}

\def\prossign{{\color[rgb]{0,0.5,0}\checkmark}}

\graphicspath{{figures/}}

\pgfplotsset{compat=newest}

\begin{document}

\pagenumbering{gobble}
\title{Building Efficient Query Engines in a High-Level Language}

\author{
Amir Shaikhha
\affil{\'Ecole Polytechnique F\'ed\'erale de Lausanne}
Yannis Klonatos
\affil{\'Ecole Polytechnique F\'ed\'erale de Lausanne}
Christoph Koch
\affil{\'Ecole Polytechnique F\'ed\'erale de Lausanne}
}

\begin{abstract}
Abstraction without regret refers to the vision of using high-level 
programming languages for systems development without 
experiencing a negative impact on performance. A database system designed according to 
this vision offers both increased productivity and high performance, 
instead of sacrificing the former for the latter as is the case with 
existing, monolithic implementations that are hard to 
maintain and extend. 

In this \paper{}, we realize this vision in the domain of analytical 
query processing. We present {\systemname}, a query engine written in 
the \textit{high-level} programming language Scala. The key technique 
to regain efficiency is to apply \textit{generative} programming: 
\systemname{} performs source-to-source compilation and optimizes the
\textit{entire} query engine by converting the high-level Scala code to 
specialized, low-level C code. We show how generative programming allows to \textit{easily} 
implement a wide spectrum of optimizations, such as introducing data partitioning 
or switching from a row to a column data layout, which are difficult to achieve 
with existing low-level query compilers that handle \textit{only} queries. We 
demonstrate that sufficiently powerful abstractions are essential for dealing 
with the complexity of the optimization effort, shielding developers from 
compiler internals and decoupling individual optimizations from each other. 

We evaluate our approach with the \tpch benchmark 
and show that: (a) With all optimizations enabled, our architecture significantly outperforms a
commercial in-memory database as well as an existing query compiler. (b) Programmers need to provide just a few hundred lines of high-level code for implementing the optimizations, instead of complicated 
low-level code that is required by existing query compilation approaches. 
(c) These optimizations may potentially come at the cost of using more 
system memory for improved performance.
(d) The compilation overhead is low compared to the overall execution time, thus 
making our approach usable in practice for compiling query engines.
\end{abstract}

\begin{bottomstuff}
This work was supported by ERC grant 279804 and NCCR MARVEL of the Swiss National Science Foundation.
\end{bottomstuff}

\maketitle

\section{Introduction}
\label{sec:intro}
During the last decade, we have witnessed a shift towards the use of high-level programming languages for systems development. 
Examples include the Singularity Operating System~\cite{singularity}, the Spark~\cite{spark-zaharia} and DryadLINQ~\cite{dryadlunc} 
frameworks for efficient, distributed data processing, the FiST platform for specifying stackable file systems~\cite{zadok-incremental-fs} 
and GPUs programming~\cite{rust-gpu-prog}. All these approaches collide with the traditional wisdom which calls for using 
low-level languages like C for building high-performance systems.

This shift is necessary as the productivity of developers is severely diminished in the presence of complicated, monolithic,
\textit{low-level} code bases, making their debugging and maintenance very costly. High-level 
programming languages can remedy this situation in two ways. First, by offering advanced software features (modules, interfaces, 
object orientation, etc.), they allow the same functionality to be implemented with significantly less code (compared to low-level 
languages). Second, by providing powerful type systems and well-defined design patterns, they allow programmers not only to create 
abstractions and protect them from leaking but also to quickly define system modules that are 
\textit{reusable} (even in contexts 
very different from the one these were created for) and easily composable~\cite{odersky_scalable}. All these properties 
can reduce the number of software errors of the systems and facilitate their verification.

\vspace{0.4cm}
\noindent\textit{Yet, despite these benefits, database  systems are still written using low-level languages.}
\vspace{0.4cm}

The reason is that increased productivity comes at a cost: high-level languages increase indirection, which in turn has a 
pronounced negative impact on performance. For example, abstraction generally necessitates the need of containers, leading 
to costly object creation and destruction operations at runtime. Encapsulation is provided through object copying rather 
than object referencing, thus similarly introducing a number of expensive memory allocations on the critical path. 
Even primitive types such as integers are often converted to their object counterparts for use with general-purposes 
libraries. As a result of these overheads, the use of high-level languages for developing high-performance databases 
seems (deceptively) prohibited.

The \textit{abstraction without regret} vision~\cite{koch,kochmanifesto} argues that it is indeed possible to use high-level languages for building database systems that allow for \textit{both} productivity and high performance, instead of trading off the former for the 
latter. By programming databases in a high-level style and still being able to get good performance, the time saved can be 
spent implementing more data\-base features and optimizations. In addition, the language features of high-level languages can 
grant flexibility to developers so that they can easily experiment with various design choices.
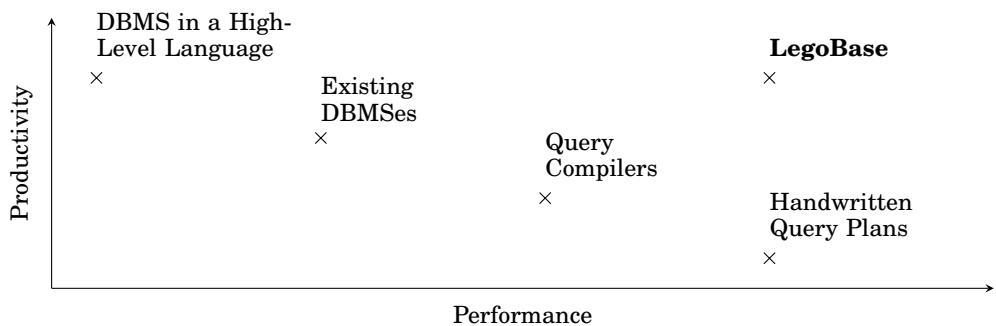
\begin{figure}[t!]
\centering
\begin{tikzpicture}

\pgfplotsset{
    every axis/.append style={
        scale only axis,
        width=0.9\columnwidth,
        height=3.6cm,
        xtick={0,0.05,0.1},
    },
    /tikz/every picture/.append style={
        trim axis left,
        trim axis right,
    }
}

 \begin{axis}[
      axis lines=left, xtick=\empty, ytick=\empty,
      xmin=0.8, ymin=0.5,
      xmax=5, ymax=5,
      nodes near coords,
      every node near coord/.append style={
          font=\small,         
          align=left,
          xshift=1.3cm,
          yshift=0.1cm,
          text width=2.59cm},
      x label style={at={(axis description cs:0.5,-0.03)},anchor=north},
      y label style={at={(axis description cs:-0.01,.5)},anchor=south},
      font=\small,
    xlabel=Performance,
    ylabel=Productivity]
  \addplot+[
      only marks, mark=x, mark size=3,
      point meta=explicit symbolic, color=black] 
   coordinates {
      (4, 4) [\textbf{\systemname}]
      (4, 1) [Handwritten\\Query Plans]
      (3, 2) [Query\\Compilers]
      (2, 3) [Existing\\DBMSes]
      (1, 4) [DBMS in a High-Level Language]
  };
  \end{axis}
\end{tikzpicture}
\caption{Comparison of the performance/productivity trade-off for all approaches
presented in this \paper{}.}
\label{designtradeoffs}
\end{figure}

In this \paper{}, we realize the abstraction without regret vision on the domain of \mbox{ad-hoc}, analytical query processing. We make the following contributions:

\begin{itemize}
\addtolength{\itemsep}{1ex}
\item We present {\systemname}, an in-memory query execution engine written in the high-level programming language, Scala, 
being the first step towards providing a full DBMS written in a high-level language. 
\vspace{1ex}

To avoid the overheads of a high-level language (e.g.\ complicated memory management) while maintaining well-defined 
abstractions, we opt for using \emph{generative} programming~\cite{taha00staging}, a technique that allows for programmatic 
removal of abstraction overhead through source-to-source compilation. This is a key benefit as, in contrast to 
traditional, general-purpose compilers -- which need to perform complicated and so\-metimes brittle analyses 
before \textit{maybe} optimizing programs -- generative programming in Scala takes advantage of the type system of the language 
to provide programmers with strong \textit{guarantees} about the structure of the generated code. For example, developers 
can specify optimizations that are applied during compilation in order to ensure that certain abstractions (e.g.\ generic 
data structures and function calls) are definitely optimized away during compilation. 
\vspace{1ex}

Generative programming can be used to optimize \textit{any} piece of Scala code. This allows {\systemname} to perform 
\textit{whole-system} specialization and compile \textit{all} components, data structures and auxiliary functions used 
inside the query engine to efficient C code. This design significantly contrasts our approach with existing \textit{query} compilation 
approaches (e.g.\ the one proposed in~\cite{neumann11}) for three reasons. First, a compiler that handles 
 \textit{only} queries 
cannot optimize and inline their code with the remaining code of the database system (which is typically \textit{precompiled}), 
thus missing a number of optimization opportunities. Second, in their purest form, query compilation approaches simply 
optimize or inline the code of \textit{individual} operators in the physical query plan, thus making cross-operator code 
optimization inside the query compiler impossible. Finally, existing approaches perform compilation using low-level code 
generation templates. These essentially come in stringified form, making their development and automatic type checking very 
difficult\footnote{For example, templates can be used to convert the code of individual query operators --  typically written today in C/C++ -- to optimized LLVM code. In that case, developers must handle a number of low-level concerns themselves, like register allocation.}.

\item The \systemname{} query engine uses a \textit{new} optimizing compiler called \compiler{}. When 
performing \textit{whole-system} compilation, an optimizing compiler effectively needs to specialize \textit{high-level} 
systems code which will naturally employ a hierarchy of components and libraries from relatively high to very low level 
of abstraction. To scale to such complex code bases, an optimizing compiler must guarantee two properties, not offered 
by existing compiler frameworks for applying generative programming.
\vspace{1ex}

First, to achieve maximum efficiency, developers must have tight control on the compiler's phases -- admitting 
custom optimization phases and phase orderings. This is necessary as code transformers with different optimization 
objectives may have to be combined in every possible ordering, depending on architectural, data, or query characteristics. 
However, existing generative programming frameworks do not offer much control over the compilation process\footnote{
For instance, \lmsdefined~\cite{lms} applies \textit{all} user-specified, domain-specific optimizations in a \textit{single} 
optimization step. It does so to avoid the well-known \textit{phase-ordering} problem in compilers, where applying two 
(or more) optimizations in an improper order can lead not only to suboptimal performance but also to programs that are semantically incorrect~\cite{tiark-phd-thesis}. We analyze how the design of the new optimizing compiler, \compiler, differs from 
that of \lms{} in Section~\ref{sec:system} of this \paper{}.}. 
This absence of control effectively forces developers to provision for \textit{all} possible optimization orderings. This 
pollutes the code base of individual optimizations, making some of them dependent on other, 
possibly semantically independent, optimizations. In general, the code complexity grows exponentially with the 
number of supported transformations\footnote{
As an example, consider the case of a compiler that is to support only two optimizations: 1) data-layout optimizations 
(i.e.\ converting a row layout to a column or PAX-like layout~\cite{Ailamaki:2001:WRC:645927.672367}) and 2) 
\datastructure{} specialization (i.e.\ adapting the definition of a data structure to the particular context in which 
it is used). This means that if the second optimization handles three different types of specialization, one has to provision 
for $2 \times 3 = 6$ cases to handle all possible combinations of these optimizations.
}.
\vspace{1ex}

Second, existing optimizing compilers expose a large number of low-level, compiler internals such as nodes of an 
intermediate representation (IR), dependency information encoded in IR nodes, and code generation templates to their 
users. This interaction with low-level semantics when coding optimizations, but also the 
introduction of the IR as an additional level of abstraction, both significantly increase the difficulty of debugging as 
developers cannot easily track the relationship between the source code, the optimization for it -- expressed using IR 
constructs 
-- and the final, generated code~\cite{yinyang,forge}.
\vspace{1ex}

Instead, the \compiler compiler was designed from the beginning so that it allows developers to have full control 
over the optimization process without exporting compiler internals such as code generation templates. It does so 
by delivering sufficiently powerful programming abstractions to developers like those afforded by modern high-level 
programming languages. The \compiler compiler along with all optimizations are both written in plain Scala, thus 
allowing developers to be highly productive when optimizing all components of the
query engine.

\item We demonstrate the ease of use of the new \compiler compiler for optimizing system components that differ 
significantly in structure and granularity of operations. We do so by providing (i) an 
in-depth presentation of the optimizations applied to the \systemname{} query engine and (b) a description of the 
\textit{high-level} compiler interfaces that database developers need to interact with when coding optimizations. 
\vspace{1ex}

We show that the design and interfaces of our optimizing compiler provide a number of nice properties for the \systemname{} optimizations.
These are expressed as library components, providing a \textit{clean} separation 
from the base code of 
{\systemname} (e.g. that of query operators), but also from each other. This is achieved, (as explained later in more detail
in Section~\ref{sec:system}) by applying them in multiple, \textit{distinct} optimization phases. Optimizations are (a) 
adjustable to the characteristics of workloads and architectures, (b) configurable, so that they can be turned on and off 
on demand and (c) composable, so that they can be easily chained but also so that higher-level optimizations can be built 
from lower-level ones. 
\vspace{1ex}

For each such optimization, we present: (a) the \textit{domain-specific} conditions that need to be satisfied in order to apply 
it (if any) and (b) possible trade-offs (e.g.\ improved execution time versus increased memory consumption). Finally, we examine 
which categories of database systems can benefit from applying each of our optimizations by providing a classification of the 
\systemname{} optimizations.

\item We perform an experimental evaluation in the domain of analytical query processing using the \tpch 
benchmark~\cite{tpch}. We show how our optimizations can lead to a system that has
performance competitive to that of a standard, commercial in-memory database called \timesten (that does not
employ compilation) and the code generated by the query compiler of the \hyper database~\cite{neumann11}. 
In addition, we illustrate that these performance improvements do not require significant programming effort 
as even complicated optimizations can be coded in \systemname{} with only a few hundred lines of code. 
We also provide insights on the performance characteristics and trade-offs of individual optimizations. We do so by 
comparing major architectural decisions as fairly as possible, using a shared codebase that only differs by the effect 
of a single optimization. Finally, we conclude our analysis by demonstrating that our \textit{whole-system} compilation 
approach incurs negligible overhead to query execution. 
\end{itemize}

\begin{figure}[t!]
\lstset{style=sqlstyle}
\centering

\pgfdeclarelayer{background}
\pgfsetlayers{background,main}

\begin{tikzpicture}[level/.style={sibling distance=2cm/#1, level distance=1.3cm}]


\node[anchor=center, left] (note1){
\begin{lstlisting}
SELECT * FROM 
  (SELECT S.D,
    SUM(1-S.B) AS E,
    SUM(S.A*(1-S.B)),
    SUM(S.A*(1-S.B)*(1+S.C))
  FROM S 
  GROUP BY S.D) T, R
WHERE T.E=R.Z AND R.Q=3
\end{lstlisting}
};

\node (join) at (2.8, 1.3) {\ \ $\bowtie_{E = Z}$}
   child {node (proj) {$\Gamma_{D,aggs}$}      
      child {node (s) {$S$} [solid, ultra thick]} 
      [solid, ultra thick]
   }
   child {node (sel) {$\sigma_{Q=3}$} 
      child {node (r) {$R$} [ultra thick] }
      [ultra thick]
   };

\end{tikzpicture}

\caption{Motivating example showing missed optimizations opportunities by
existing query compilers that use template expansion.}
\label{fig:motivating_example}
\end{figure}
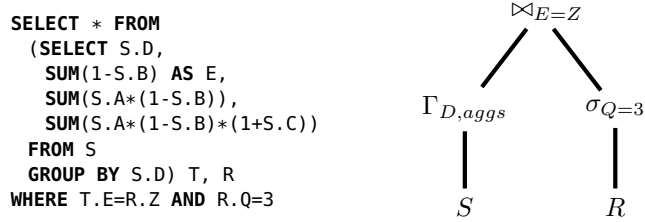

\noindent \textbf{Motivating Example.} To better understand the differences of
our work with previous approaches, consider the SQL query shown in
Figure~\ref{fig:motivating_example}.  This query first calculates some
aggregations from relation \textit{S} in the \groupby operator $\Gamma$.  Then,
it joins these aggregations with relation \textit{R}, the tuples of which are
filtered by the value of column \textit{Q}.  The results are then returned to
the user. Careful examination of the execution plan of this query, shown in the
same figure, reveals the following three basic optimization opportunities missed
by existing query compilers that use template expansion:
\begin{itemize}
\addtolength{\itemsep}{1ex}
\item First, the limited scope of existing approaches usually results in
performing the evaluation of aggregations in \textit{precompiled} DBMS code. Thus, each
aggregation is evaluated \textit{consecutively} and, as a result, common
sub-expression elimination cannot be performed in this case (e.g.\ in the
calculation of expressions \code{1-S.B} and \code{S.A*(1-S.B)}). This shows that,
if we include the evaluation of all aggregations in the \textit{compiled} final
code, we can get an additional performance improvement. This motivates us to
extend the scope of compilation in this work.

\item Second, template-based approaches may result in unnecessary computation. 
This is because operators are not aware of each other. 
In this example, the
generated code includes two materialization points: (a) at the \groupby and (b)
when materializing the left side of the join. However, there is no need to
materialize the tuples of the aggregation in two different data structures as
the aggregations can be immediately materialized in the data structure of the
join.  Such \textit{inter-operator} optimizations are hard to express using
\textit{template-based} compilers. By high-level programming, we can instead
easily pattern match on the operators, as we show in
Section~\ref{subsec:eliminate_materializations}.
 
\item Finally, the data structures have to be \textit{generic} enough for all
queries. As such, they incur significant abstraction overhead, especially when
these structures are accessed millions of times during query evaluation. Current
query compilers cannot optimize the data structures since these belong to the
precompiled part of the DBMS. Our approach eliminates these overheads as it
performs \textit{whole-program} optimization and compiles, along with the
operators, the data structures employed by a query. This
significantly contrasts our approach with previous work.  
\end{itemize}

The rest of this \paper{} is organized as follows. Section~\ref{sec:system}
presents the overall design of {\systemname}, along with a
detailed description of the APIs provided by the new \compiler{} optimizing compiler. 
Section~\ref{sec:optimizations} gives an in-depth presentation of all supported
compiler optimizations of our system in
multiple domains.
Section~\ref{sec:eval} presents our evaluation, where we
experimentally show that our approach using the \compiler{} optimizing compiler can lead to significant 
benefits compared to (i) a commercial DBMS that does not employ compilation and 
(ii) a database system that uses low-level, code-generation templates during query compilation.
We also give insights about 
the memory footprint, data loading time and programming effort required when working with the \systemname{} system. 
Section~\ref{sec:related} presents related work in the area of compilation and compares our
approach with existing query compilers and engines. Finally,
Section~\ref{sec:conclusions} concludes. 
\section{System Design}
\label{sec:system}
In this section, we present the design of the \legobase{} system. First, we describe 
the overall system architecture of our approach (Subsection~\ref{subsec:general}).
Then, we describe in detail the \compiler{} compiler that is the core of our proposal
(Subsection~\ref{subsec:compiler})
as well as how we efficiently convert the \textit{entire} high-level Scala code of the query engine (not just that of individual operators) to optimized C code for each incoming query (Subsection~\ref{subsec:ccodegen}). 
While doing so, we give concrete code examples of how (a) physical query operators, (b) physical query plans, and, (c) compiler 
interfaces look like in our system.

\subsection{Overall System Architecture}
\label{subsec:general}
\legobase implements the typical query plan operators found in traditional database systems, including equi, semi, anti, and 
outer joins, all on a high level. In addition, {\systemname} supports both a classical Volcano-style~\cite{Volcano} 
query engine as well as a push-style query interface~\cite{neumann11}\footnote{In a push engine, the meaning of child and parent operators is 
reversed compared to the usual query plan terminology: Data flows from the leaves (the ancestors, usually being scan operators) to 
the root (the final descendant, which computes the final query results that are returned to the user).}. 

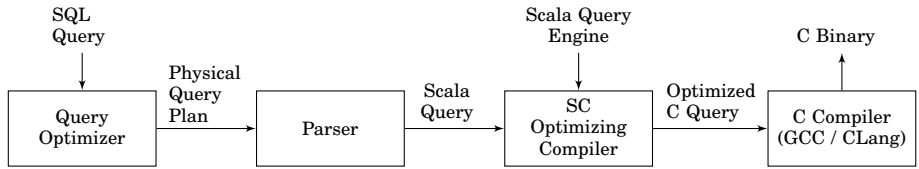
\begin{figure}
\scriptsize
\centering
\tikzstyle{block} =[rectangle, minimum width=1.8cm, minimum height=1cm, text centered, text width=1.8cm, draw=black]
\tikzstyle{line} = [draw, -latex']
\begin{tikzpicture}[node distance = 3.3cm, auto]
   \node[block](oracle_query_optimizer){Query \\ Optimizer};
   \node[block, right of=oracle_query_optimizer](lego_base){Parser};
   \node[block, right of=lego_base](compiler){\compiler{}\\Optimizing Compiler};
   \node[block, right of=compiler, xshift=0.2cm](c_compiler){C Compiler {(\gcc} / {\clang)}};
   \path[line](oracle_query_optimizer) -- (lego_base) node[midway, above, text width=1cm]{Physical Query Plan} (lego_base);
   \path[line](lego_base) -- (compiler) node[xshift=-0.1cm, midway, above, text width=1cm, align=center]{Scala Query};
   \path[line](compiler) -- (c_compiler.west) node[xshift=0.02cm,midway, above, text width=1.2cm]{Optimized C Query};
   \path[line]([yshift=0.5cm]oracle_query_optimizer.north) -- (oracle_query_optimizer.north) node[yshift=0.5cm,above, text width=0.8cm]{SQL Query};      
   \path[line]([yshift=0.5cm]compiler.north) -- (compiler.north) node[yshift=0.5cm,above, text width=1.5cm,align=center]{Scala Query\\Engine};
   \path[line](c_compiler.north) -- ([yshift=0.5cm]c_compiler.north) node[above, text width=1.2cm]{C Binary};
\end{tikzpicture}
\caption{Overall system architecture. The domain-specific optimizations of
{\systemname} are applied during the \compiler{} compiler optimization phase.}
\label{fig:system}
\end{figure}

The overall system architecture of \legobase{} is shown in Figure~\ref{fig:system}.  
First, for each incoming SQL query, we must get a query plan which describes the
physical query operators needed to process this query.  For this work, we
consider traditional query optimization (e.g.\ determining join ordering) as an
orthogonal problem and we instead focus more on experimenting with the different
optimizations that can be applied \textit{after} traditional query optimization.
Thus, to obtain a physical query plan, we pass the incoming query through \textit{any
existing} query optimizer. For example, for our evaluation, we choose the query
optimizer of a commercial, in-memory database system. 

Then, we pass the generated physical plan to {\systemname}. Our system, in turn,
parses this plan and instantiates the corresponding Scala implementation of the 
operators. Figure~\ref{fig:operatorexample} presents an example of how
query plans and operators are written in {\systemname}, respectively. That is, the Scala code example shown in
Figure~\ref{fig:operatorexample_qp} loads the data,
builds a functional tree from operator objects and then starts executing the query by calling \code{next} 
for the root operator. 

It is important to note that
operator implementations like the one presented in 
Figure~\ref{fig:operatorexample_agg} are exactly what one would write for a simple query
engine that does not involve compilation at all.  However, without further
optimizations, this engine cannot match the performance of existing databases:
it consists of generic data structures (e.g. the one declared in line~\ref{genericds} of Figure~\ref{fig:operatorexample_agg}) 
and involves expensive memory allocations on the critical path\footnote{Note that such memory 
allocations are not always explicit (i.e.~at object definition time through the \code{new} 
keyword in object-oriented languages like Java and Scala). For instance, in 
line~\ref{memalloc-expand} of Figure~\ref{fig:operatorexample_agg}, the HashMap data structure 
may have to expand (in terms of allocated memory footprint) and be reorganized by the Scala runtime in 
order to more efficiently store data for future lookup operations. We talk more about this issue and its consequences to 
performance later in this \paper{}.}, 
both properties that can significantly affect performance. 
\begin{figure}[t!]
\lstset{style=scala_numbered,xleftmargin=0.2cm}
\begin{tabular}{cc}
\begin{subfigure}[b]{.46\textwidth}
\begin{lstlisting}
def Q6() {
  val lineitemTable = loadLineitem()
  val scanOp = new ScanOp(lineitemTable)
  val startDate = parseDate("1996-01-01")
  val endDate = parseDate("1997-01-01")
  val selectOp = new SelectOp(scanOp)
    (x =>
      x.L_SHIPDATE >= startDate && 
      x.L_SHIPDATE < endDate && 
      x.L_DISCOUNT >= 0.08 && 
      x.L_DISCOUNT <= 0.1 && 
      x.L_QUANTITY < 24
    )
  val aggOp = new AggOp(selectOp)
    (x => "Total")
    ((t, agg) => { agg +
      (t.L_EXTENDEDPRICE * t.L_DISCOUNT)
    })
  val printOp = new PrintOp(aggOp)(
    kv => printf("%.4f\n", kv.aggs(0))
  )
  printOp.open
  printOp.next
}
\end{lstlisting}
\caption{}
\label{fig:operatorexample_qp}
\end{subfigure}
&
\begin{subfigure}[b]{.53\textwidth}
\begin{lstlisting}
class AggOp[B](child:Operator, grp:Record=>B, 
               aggFuncs:(Record,Double)=>Double*) 
  extends Operator { 
    val hm = HashMap[B, Array[Double]]()(*@\label{genericds}@*) 
    def open() { parent.open }
    def process(aggs:Array[Double], t:Record){
      var i = 0
      aggFuncs.foreach { aggFun => (*@\label{startloop}@*)
        aggs(i) = aggFun(tuple, aggs(i))
        i += 1
      } (*@\label{endloop}@*) 
    }
    def consume(tuple:Record) {
      val key = grp(tuple)
      val aggs = hm.getOrElseUpdate(key,(*@\label{memalloc-expand}@*)  
          new Array[Double](aggFuncs.size))
      process(aggs, tuple)
    }
    def next() : Record = {
      hm.foreach { pair => child.consume(
        new AGGRecord(pair._1, pair._2)
      ) }
    }
}
\end{lstlisting}
\caption{}
\label{fig:operatorexample_agg}
\end{subfigure}
\\
\end{tabular}
\vspace{-0.1cm}
\caption{Example of a query plan and an operator implementation in {\systemname}. The SQL query used as an input 
here is actually Query 6 of the \tpch workload. The operator implementation presented here uses
the Push-style interface~\protect\cite{neumann11}.}
\label{fig:operatorexample}
\end{figure}

However, in our system, the \compiler{} optimizing compiler specializes the code of the \textit{ 
entire} query engine on the fly (including  the code of individual operators, all data structures used as well as any required 
auxiliary functions), and progressively optimizes the code using our domain-specific 
optimizations (described in detail in Section~\ref{sec:optimizations}). For example, it
optimizes away the \code{HashMap} abstraction and transforms it to efficient low-level constructs (Section~\ref{subsec:dsopt}). In addition, \compiler{} utilizes the available \textit{query-specific} information during compilation. For instance, it will inline the code of all individual 
operators and, for the example of Figure~\ref{fig:operatorexample_agg}, it automatically unrolls the 
loop of lines \ref{startloop}-\ref{endloop}, since the number of aggregations can be
statically determined based on how many aggregations the input SQL query has. Such
fine-grained optimizations have a significant effect on performance, as they
improve branch prediction.  Finally, our system generates the optimized C code\footnote{
In this work, we choose C as our code-generation language 
as this is the language traditionally used for building high-performance
database systems. However, \compiler is not particularly
aware of C and can be used to generate programs in other languages as well (e.g.\ optimized Scala).},
which is compiled using any existing C compiler (e.g.\ we use the
CLang\footnote{http://clang.llvm.org/} frontend
of LLVM~\cite{llvm} for compiling the generated C code in our evaluation). We then 
return the query results to the user.

\commentout{
We also use \compiler to recompile the components of our query engine that are affected
by a runtime change. We motivate this decision based on the observation that, for
database systems, many configuration variables are set \textit{only} at start time 
but are unnecessarily checked multiple times at runtime. In {\systemname}, the
compiled final code does not include any \code{if} condition that checks
configuration parameters. Instead, at startup time we generate only the proper
branch according to the current value of the parameter.  Then, we re-compile components
\textit{only if} any of their corresponding parameters changes at runtime. For example, 
if the user disables logging, we recompile the code to remove all logging statements. 
This \textit{continuous} cycle of optimization and execution is only made
possible by on-the-fly compilation and, to our knowledge, is not provided by
existing query compilers. 
}

\subsection{The \compiler Compiler Framework}
\label{subsec:compiler}
{\systemname} makes key use of the \compiler framework, which provides
runtime compilation and code generation facilities for the Scala programming
language, as follows.
\begin{figure}[t!]
\lstset{style=scala_unnumbered,xleftmargin=-0.15cm}
\begin{tabular}{cc}
\begin{subfigure}[b]{.47\textwidth}
\begin{lstlisting}
analysis += statement { 
  case sym -> code"new MultiMap[_, $v]" 
    if isRecord(v) => allMaps += sym
}
analysis += rule {
  case loop @ code"while($cond) $body" => 
    currentWhileLoop = loop
}

rewrite += statement {
  case sym -> (code"new MultiMap[_, _]") 
    if allMaps.contains(sym) => 
      createPartitionedArray(sym)
}
rewrite += remove {
  case code"($map: MultiMap[Any, Any])
    .addBinding($elem, $value)" 
      if allMaps.contains(map) =>
}
rewrite += rule {
  case code"($map: MultiMap[Any, Any])
    .addBinding($elem, $value)" 
      if allMaps.contains(map) =>
      /* Code for processing add Binding */
}  
\end{lstlisting}
\caption{}
\label{fig:transformation-api}
\end{subfigure}
&
\begin{subfigure}[b]{.52\textwidth}
  \centering
  \begin{lstlisting}
pipeline += OperatorInlining
pipeline += SingletonHashMapToValue
pipeline += ConstantSizeArrayToValue
pipeline += ParamPromDCEAndPartiallyEvaluate
if (settings.partitioning) {
  pipeline += PartitioningAndDateIndices
  pipeline += ParamPromDCEAndPartiallyEvaluate
}
if (settings.hashMapLowering) 
  pipeline += HashMapLowering
if (settings.stringDictionary) 
  pipeline += StringDictionary
if (settings.columnStore) {
  pipeline += ColumnStore
  pipeline += ParamPromDCEAndPartiallyEvaluate
}
if (settings.dsCodeMotion) {
  pipeline += HashMapHoisting
  pipeline += MallocHoisting
  pipeline += ParamPromDCEAndPartiallyEvaluate
}
if (settings.targetIsC) 
  pipeline += ScalaToCLowering
// else: handle other languages, e.g. Scala
pipeline += ParamPromDCEAndPartiallyEvaluate
\end{lstlisting}
\caption{}
\label{fig:legobase-pipeline}
\end{subfigure}
\\
\end{tabular}
\vspace{-0.3cm}
\caption{(a) The analysis and transformation APIs provided by \compiler{}. 
(b) The \compiler{} transformation pipeline used by \legobase. Details for the optimizations listed in this pipeline are presented in Section~\ref{sec:optimizations}.}
\end{figure}

To begin with, in contrast to low-level compilation frameworks like LLVM -- which express 
optimizations using a low-level, compiler-internal intermediate representation (IR) 
that operates on the level of registers and basic blocks -- programmers in 
\compiler{} specify the result of a program transformation as a high-level, 
\textit{compiler-agnostic} Scala program.
\compiler{} offers
two \textit{high-level} programming primitives named \code{analyze} and \code{rewrite} for this purpose, which 
are illustrated in Figure~\ref{fig:transformation-api} and which analyze and manipulate 
statements and expressions of the input program, respectively.
For example, our \datastructure specialization (Section~\ref{subsubsec:maps-to-native-arrs}) 
replaces 
operations on hash maps with 
operations on native arrays.  
By expressing optimizations at a high level, our approach enables a user-friendly way 
to describe these domain-specific optimizations that humans can easily identify, without 
imposing the need to interact with compiler internals\footnote{Of course, every compiler needs 
to represent code through an intermediate representation. The difference between
\compiler and other optimizing compilers is that the IR of our compiler is completely hidden
from developers: both the input source code and all of its optimizations are written
in plain Scala code, which is then translated to an internal IR through Yin-Yang~\cite{yinyang}.}. We use 
this optimization interface to provide database-specific optimizations as a library and 
to aggressively optimize our query engine.

Then, to allow for maximum efficiency when specializing all components of the query engine, 
developers must be able to easily experiment with different optimizations and optimization 
orderings (depending on the characteristics of the input query or the properties of 
the underlying architecture). 
In \compiler{}, developers do so by explicitly specifying a \textit{transformation pipeline}. This 
is a straightforward task as \compiler transformers act as black boxes, which can be plugged in at 
any stage in the pipeline. For instance, for the transformation pipeline of \legobase, shown in 
Figure~\ref{fig:legobase-pipeline}, Parameter Promotion, Dead Code Elimination and Partial 
Evaluation are all applied at the end of each of the custom, domain-specific optimizations. 
Through this transformation pipeline, developers can easily turn optimizations on and off at demand 
(e.g.~by making their application dependant on simple runtime or configuration conditions) as well 
as specifying which optimizations should be applied \textit{only} for specific hardware platforms. 

Even though it has been advocated in previous work~\cite{rompf13popl} that having multiple 
transformers can cause phase-ordering problems, our experience is that system 
developers are empowered by the control they have when coding optimizations with \compiler and 
rise to the challenge of specifying a suitable order of transformations as they 
design their system and its compiler optimizations.
As we show in Section~\ref{sec:eval}, with a relatively small number of transformations
we can get a significant performance improvement in \legobase{}.

\compiler already provides many generic compiler optimizations like function inlining,
common subexpression and dead code elimination, constant propagation,
scalar replacement, partial evaluation, 
and code motion. In this work, we extend this set to
include DBMS-specific optimizations (e.g.\ using the popular columnar layout for
data processing).
We describe these optimizations in more detail 
in Section~\ref{sec:optimizations}.

\subsection{Efficiently Compiling High-Level Query Engines}
\label{subsec:ccodegen}
Database systems comprise many components of significantly different nature 
and functionality, thus typically resulting in very big code bases. 
To efficiently optimize those, developers must be able to express new 
optimizations without the having to modify neither (i) the base code of the 
system nor (ii) previously developed optimizations. As discussed previously,
compilation techniques based on template expansion do not scale to the task, 
as their single-pass approach forces developers to deal with a number of 
low-level concerns, making their debugging and development costly. 

\begin{figure}[t]
\centering
\hspace{-1cm}
\includegraphics[scale=0.6]{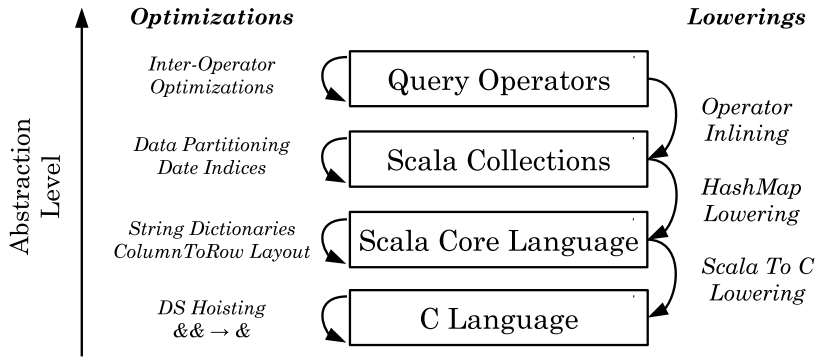}
\caption{Source-to-source compilation expressed through the progressive lowering approach -- there different optimizations are applied in different optimization stages, thus guaranteeing the notion of separation of concerns.}
\label{fig:dsl-stack}
\end{figure}

To this end, the \compiler compiler framework is built around the principle that, instead of using template 
expansion to directly generate low-level code from a high-level program in a \textit{single} macro 
expansion step, an optimizing compiler should instead \textbf{progressively lower the level of 
abstraction} until we reach the lowest possible level of representation, and only then generating 
the final, low-level code. This design is illustrated in Figure~\ref{fig:dsl-stack}. 

Each level of abstraction and all associated optimizations operating in it can be seen as 
independent modules, enforcing the principle of \textit{separation of concerns}. 
Higher levels are generally more declarative, thus allowing for increased productivity,
while lower levels are closer to the underlying architecture, thus making it possible to
more easily perform low-level performance tuning. For example, optimizations such as join 
reordering are only feasible in higher abstraction levels (where the operator objects are 
still present in the code), while register allocation decisions can only be expressed in very low 
abstraction levels. This design provides the 
nice property that generation of the final code basically becomes a trivial and naive 
stringification of the lowest level representation.
Table~\ref{table:ip-vs-dp} provides a brief summary of the benefits of
imperative and declarative languages in general.

More precisely, in order to reach the abstraction level of C code in \legobase{} (the lowest level 
representation for the purposes of this \paper{}),  transformations in \compiler also include multiple 
\emph{lowering} steps that \textit{progressively} map Scala constructs to (a set) of C constructs. 
Most Scala abstractions (e.g.\ objects, classes, inheritance) are optimized away in one of
these intermediate stages (for example, hash maps are converted to arrays through the domain-specific optimizations described in more detail in 
Section~\ref{sec:optimizations}), and for the remaining constructs (e.g.\ loops, variables, arrays) 
there exists a one-to-one correspondence between Scala and C. \compiler{} already offers such lowering 
transformers for an important subset of the Scala programming language.
For example, classes are converted to structs, strings to arrays of bytes, etc. In
general, composite types are handled in a recursive way, by first lowering their
fields and then wrapping the result in a C struct. The final result is a struct
of only primitive C constructs. 

This way of lowering does not require any modifications to the database code or effort from database 
developers other than just specifying in \compiler{} how and after which abstraction level custom 
data types and abstractions should be lowered.
More importantly, such a design allows developers to create new abstractions in one of their 
optimizations, which can be in turn optimized away in subsequent optimization passes.
After all lowering steps have been performed, developers can now apply low-level, 
architecture-dependent optimizations, as the code is now close to the semantics offered by low-level 
programming languages (e.g.~includes pointers for explicitly referencing memory). Then, a final 
iteration 
emits the actual C code. 
\begin{table}[t!]
\begin{center}
  \begin{tabular}{ l p{0.5\columnwidth} }
    \toprule
    Paradigm & Advantages \\
    \midrule
    Declarative &    
            \begin{minipage}[t]{\textwidth}
            \begin{itemize}
            \renewcommand\labelitemi{\prossign}
              \item Concise programs
              \item Simple to analyze and verify
              \item Simple to parallelize
            \end{itemize}
            \end{minipage}
            \\ 
            \midrule
    Imperative &
            \begin{minipage}[t]{\textwidth}
            \begin{itemize}
            \renewcommand\labelitemi{\prossign}
              \item Efficient data structures
              \item Precise control of execution flow
              \item More predictable performance
            \end{itemize}
            \end{minipage}
			\\
  \bottomrule
  \end{tabular}
  \end{center}
  \vspace{-0.1cm}
  \caption{Comparison of declarative and imperative language characteristics. We use both paradigms for different steps of our progressive lowering compilation approach.}
  \label{table:ip-vs-dp}
  \vspace{-0.5cm}
\end{table}
\begin{figure*}[t!]
\def\captiondistdef{-0.3cm}
\def\figfirstheight{1.6cm}
\def\figcomplheight{8.1cm}
\def\figsepheight{0.5cm}
\begin{minipage}[c][\figfirstheight][t]{.45\textwidth}
\lstset{style=sqlstyle}
\begin{lstlisting}
SELECT COUNT(*) 
FROM R, S 
WHERE R.name == "R1" 
  AND R.id == S.id
\end{lstlisting}
\vspace{\captiondistdef}
\subcaption{The example query in SQL.}
\label{fig:dsls-example-query-sql}
\end{minipage}
\begin{minipage}[c][\figfirstheight][t]{.5\textwidth}
\lstset{style=scala_unnumbered}
\begin{lstlisting}
AggOp(HashJoinOp(
  SelectOp(ScanOp(R), r => r.name == "R1"), 
  ScanOp(S), (r,s) => r.id == s.id
), (rec, count) => count + 1)
\end{lstlisting}
\vspace{\captiondistdef}
\subcaption{The physical plan of the example query.}
\label{fig:dsls-example-query-qopl}
\end{minipage}
\begin{minipage}[c]{\textwidth}
\vspace{\figsepheight}
\begin{minipage}[c][\figcomplheight][t]{.32\textwidth}
\lstset{style=scala_unnumbered}
\begin{lstlisting}
val hm = new MultiMap[Int,R]

for(r <- R) {
  if(r.name == "R1") {
    

    hm.addBinding(r.id, r)



  }
}
var count = 0
for(s <- S) {
   hm.get(s.id) match {
      case Some(rList) =>
         for(r <- rList) {
            if(r.id == s.id)
               count += 1
         }
      case None        => ()
   }
}
return count
\end{lstlisting}
\vspace{\captiondistdef}
\subcaption{}
\label{fig:dsls-example-query-mchl}
\end{minipage}
\hspace{0.02\textwidth}
\begin{minipage}[c][\figcomplheight][t]{.32\textwidth}
\lstset{style=scala_unnumbered}
\begin{lstlisting}
val MR: Array[Seq[R]] = 
 new Array[Seq[R]](BUCKETSZ)
for(r <- R) {
  if(r.name == "R1") {


    MR(r.id) += r



  }
}
var count = 0
for(s <- S) {
  
  val rList = MR(s.id)
  for(r <- rList) {
    if(r.id == s.id)
      count += 1

  }
}

return count
\end{lstlisting}
\vspace{\captiondistdef}
\subcaption{}
\label{fig:dsls-example-query-mcl}
\end{minipage}
\vspace{\figsepheight}
\hspace{0.02\textwidth}
\begin{minipage}[c][\figcomplheight][t]{.29\textwidth}
\lstset{style=scala_unnumbered}
\begin{lstlisting}
val MR: Array[R] = 
  new Array[R](BUCKETSZ)
for(r <- R) {
  if(r.name == "R1") {
    if(MR(r.id) == null) {
      MR(r.id) = r
    } else {
      r.next = MR(r.id)
      MR(r.id) = r
    }
  }
}
var count = 0
for(s <- S) {
  
  var r: R = MR(s.id)
  while(r != null) {
    if(r.id == s.id)
      count += 1
    r = r.next
  }
}

return count
\end{lstlisting}
\vspace{\captiondistdef}
\subcaption{}
\label{fig:dsls-example-query-scalaql}
\end{minipage}
\begin{minipage}[c][\figcomplheight][t]{.48\textwidth}
\lstset{style=scala_unnumbered}
\begin{lstlisting}
val MR: Array[Pointer[R]] = 
  malloc[Pointer[R]](BUCKETSZ)
for(r <- R) {

  if(r->name == "R1") {
    if(MR(r->id) == null) MR(r->id) = r
    else {
      r->next = MR(r->id)
      MR(r->id) = r
    }
  }
}
var count = 0
for(s <- S) {
  
  var r: Pointer[R] = MR(s->id)
  while(r != null) {
    if(r->id == s->id)
      count += 1
    r = r->next
  }
}
return count
\end{lstlisting}
\vspace{\captiondistdef}
\subcaption{}
\label{fig:dsls-example-query-cscalaql}
\end{minipage}
\begin{minipage}[c][\figcomplheight][t]{.5\textwidth}
\lstset{style=cstyle}
\begin{lstlisting}
R** MR = (R**)
  malloc(BUCKETSZ * sizeof(R*))
for(int i=0; i < R_REL_SIZE; i++) { 
  R* r = R[i];
  if(strcmp(r->name, "R1") == 0) {
    if(MR[r->id] == NULL) MR[r->id] = r;
    else {
      r->next = MR[r->id];
      MR[r->id] = r;
    }
  }
}
int count = 0
for(int i=0; i < S_REL_SIZE; i++) {
  S* s = S[i];
  R* r = MR[s->id]
  while(r != NULL) {
    if(r->id == s->id)
      count += 1;
    r = r->next;
  }
}
return count;
\end{lstlisting}
\vspace{\captiondistdef}
\subcaption{}
\label{fig:dsls-example-query-cql}
\end{minipage}
\end{minipage}
\caption{Progressively lowering an example query to C code with \compiler.}
\label{fig:dsls-example-query}
\end{figure*}

Finally, there are two additional implementation details of our source-to-source compilation from Scala to C that require special mentioning. 

First, the final code produced by {\systemname}, with all optimizations enabled, does not require 
library function calls. For example, all collection data structures like hash maps are converted in 
\systemname{} to primitive arrays (Section~\ref{subsec:dsopt}). Thus, lowering such 
library calls to C is not a big issue. However, we view {\systemname} as a platform for easy 
experimentation of database optimizations. As a result, our architecture must also be able to 
support traditional collections as a library and convert, whenever necessary, Scala collections to 
corresponding ones in C. We have found \glib~\cite{GLib} to be efficient enough for this purpose. 

Second, and more importantly, the two languages handle memory management in a
totally different way: Scala is garbage collected, while C has explicit memory
management. Thus, when performing source-to-source compilation from Scala
to C, we must take special care to free the memory that would normally be
garbage collected in Scala in order to avoid memory overflow. This is a hard
problem to solve automatically, as garbage collection may have to occur for objects
allocated outside the DBMS code, e.g.\ for objects allocated inside the Scala libraries.  
For the scope of this work, we follow a conservative approach and make, whenever needed,
allocations and deallocations explicit in the Scala code. We also free the allocated memory after 
each query execution.

\subsection{Putting it all together -- A compilation example}
To better illustrate the various steps of our progressive lowering, we analyze 
how \systemname{} converts the example SQL query shown in Figure~\ref{fig:dsls-example-query-sql} 
to efficient C code. 

To begin with, the query plan, shown in Figure~\ref{fig:dsls-example-query-qopl}, is parsed 
and converted to the program shown in Figure~\ref{fig:dsls-example-query-mchl}. This step inlines 
the code of all relational operators present in the query plan and implements the equijoin using a hash table. This is the natural way database 
developers would typically implement a join operator using high-level collections programming.
Then, this hash-table data structure is lowered to an array of linked lists 
(Figure~\ref{fig:dsls-example-query-mcl}). However, these lists are not really required, as we 
can chain the records together using their \textit{next} pointer. This optimization, which 
is presented in more detail in Section~\ref{subsec:dsopt}, takes place in the next step 
(Figure~\ref{fig:dsls-example-query-scalaql}). Finally, the code is converted to an 
embedded~\cite{hudak-dsl} version of the C language in Scala 
(Figure~\ref{fig:dsls-example-query-cscalaql}) and, only then, \compiler generates the final C 
program out of this embedding (Figure~\ref{fig:dsls-example-query-cql}). 

This example clearly illustrates that our optimizing compiler applies
different optimizations in distinct transformation phases, thus 
guaranteeing the separation of concerns among different optimizations. For 
example, operator inlining is applied in the very first, high-level 
representation, which only describes operator objects. Performance 
concerns for data structures are then handled in subsequent optimization 
steps. Finally, low-level, code generation concerns are addressed only in the 
last, low-level representation. Next, we give more details about our 
individual optimizations. 
\section{Compiler Optimizations}
\label{sec:optimizations}
\begin{figure}[t!]
\lstset{style=scala_unnumbered}
\begin{lstlisting}
def Q12() {
  val ordersScan     = new ScanOp(loadOrders())
  val lineitemScan   = new ScanOp(loadLineitem())
  val lineitemSelect = new SelectOp(lineitemScan)(record =>
    record.L_RECEIPTDATE >= parseDate("1994-01-01") && 
    record.L_RECEIPTDATE <  parseDate("1995-01-01") && 
   (record.L_SHIPMODE == "MAIL" || record.L_SHIPMODE == "SHIP") &&
    record.L_SHIPDATE < record.L_COMMITDATE && record.L_COMMITDATE < record.L_RECEIPTDATE
  )
  val jo = new HashJoinOp(ordersScan, lineitemSelect) // Join Predicate and Hash Functions
    ((ordersRec,lineitemRec) => ordersRec.O_ORDERKEY == lineitemRec.L_ORDERKEY)
    (ordersRec => ordersRec.O_ORDERKEY)(lineitemRec => lineitemRec.L_ORDERKEY)
  val aggOp = new AggOp(jo)(t => t.L_SHIPMODE) // L-SHIPMODE is the Aggregation Key
    ((t, agg) => {
      if (t.O_ORDERPRIORITY == "1-URGENT" || t.O_ORDERPRIORITY == "2-HIGH") agg + 1 else agg 
    },
    (t, agg) => {
      if (t.O_ORDERPRIORITY != "1-URGENT" && t.O_ORDERPRIORITY != "2-HIGH") agg + 1 else agg 
    })
  val sortOp = new SortOp(aggOp)((x, y) => x.key - y.key)
  val po = new PrintOp(sortOp)(kv => { 
    printf("%s|%.0f|%.0f\n", kv.key, kv.aggs(0), kv.aggs(1))
  })
  po.open
  po.next
}
\end{lstlisting}
\caption{Example of an input query plan (\tpch Q12). We use this query to explain various characteristics of
our domain-specific optimizations in Section~\ref{sec:optimizations}.}
\label{fig:query_example2}
\end{figure}
In this section, we present examples of compiler optimizations in six domains:
(a) inter-operator optimizations for query plans, (b) transparent
\datastructure modifications, (c) changing the data layout, (d) using string dictionaries for efficient processing
of string operations, (e) domain-specific code motion, and, 
finally, (f) traditional compiler optimizations like dead code elimination. The purpose of
this section is to demonstrate the ease-of-use of our methodology: that by
programming at the high-level, such optimizations are easily expressible 
without requiring changes to the base code of the query engine or interaction with compiler 
internals. 
Throughout this section we
use, unless otherwise stated, Q12 of \tpch, shown in Figure~\ref{fig:query_example2}, as a guiding example in order to better 
illustrate various important characteristics of our optimizations. 

\subsection{Inter-Operator Optimizations -- Eliminating Redundant Materializations}
\label{subsec:eliminate_materializations}
Consider again the motivating example of our introduction. We observed that
existing query compilers use template-based generation and, thus, in such
schemes operators are not aware of each other. This can cause redundant
computation: in this example there are two materialization points (in the
\groupby and in the left side of the hash join) where there could be only a
single one. 

By expressing optimizations at a higher level, \systemname{} can optimize code 
across operator interfaces. For this example, we can treat operators as
objects in Scala, and then match specific optimizations to certain chains of
operators.  Here, we can completely remove the aggregate operator and merge it
with the join, thus eliminating the need of maintaining two distinct
data structures. The code of this optimization is shown in Figure~\ref{operatorremoval}.
\begin{figure}[t!]
\lstset{style=scala_numbered,xleftmargin=0.4cm}
\begin{lstlisting}
def optimize(op: Operator): Operator = op match {
  case joinOperator@HashJoinOp(aggOp:AggOp, rightChild, joinPred, leftHash, rightHash) =>
    new HashJoinOp(aggOp.leftChild, rightChild, joinPred, leftHash, rightHash) {
      override def open() {
        // leftChild is now the child of aggOp (relation S)
        leftChild foreach { t => 
          // leftHash hashes according to the attributes referenced in the join condition
          val key = leftHash(aggOp.grp(t))
          // Get aggregations from the hash map of HashJoin
          val aggs = hm.getOrElseUpdate(key, new Array[Double](aggOp.aggFuncs.size))
          // Process all aggregations using the original code of Aggregate Operator
          aggOp.process(aggs,t)
        }
      }
    }
  case op: Operator =>
    op.leftChild  = optimize(op.leftChild)
    op.rightChild = optimize(op.rightChild)
  // Operators with only one child have leftChild set, but rightChild null.
  case null => null	
}
\end{lstlisting}
\caption{Removing redundant materializations by high-level programming
(here between a \groupby and a join). The semantics (child-parent relationships)
of this code segment are adapted to a Volcano-style engine. However, 
the same optimization logic can be similarly applied to a push engine. The code
of the Aggregate Operator is given in Figure~\ref{fig:operatorexample_agg}. The
\code{next} function of the HashJoinOp operator remains the same.}
\label{operatorremoval}
\end{figure}

This optimization operates as follows. First, we call the optimize function,
passing it the top-level operator as an argument. The function then traverses
the tree of Scala operator objects, until it encounters a proper chain of
operators to which the optimization can be applied to. In the case of the example
this chain is (as shown in line 2 of Figure~\ref{operatorremoval}) a hash-join
operator connected to an aggregate operator.  When this pattern is detected, a
new HashJoinOp operator object is created, that is \textit{not} connected to the
aggregate operator, but instead to the child of the latter (first function argument in
line 3 of Figure~\ref{operatorremoval}). As a result, the materialization point of the
aggregate operator is completely removed.  However, we must still find a place
to (a) store the aggregate values and (b) perform the aggregation.  For this
purpose we use the hash map of the hash join operator (line 10), and we just call
the corresponding function of the Aggregate operator (line 12), respectively. The
processing of the tuples of the right-side relation (relation \code{R} in 
Figure~\ref{fig:motivating_example}), alongside with checking the join condition and
the rest of join-related processing, still takes place during the call of \code{next} 
function of the HashJoinOp operator, similarly to the original query operator code.

We observe that this optimization is programmed in the same level of
abstraction as the rest of the query engine: as normal Scala code. This allows to completely
avoid code duplication during development, but more importantly it demonstrates that when coding optimizations at a high level of 
abstraction (e.g.~to optimize the operators' interfaces), developers no longer have to worry about low-level concerns such as 
code generation (as is the case with existing approaches) -- these concerns are simply addressed by later stages in the 
transformation pipeline. Both these properties raise the productivity provided by our system, showing the merit of developing 
database systems using high-level programming languages.

\subsection{Data-Structure Specialization}
\label{subsec:dsopt}

Data-structure optimizations contribute significantly to the complexity of
database systems today, as they tend to be heavily specialized to be workload,
architecture and (even) query-specific. Our experience with the PostgreSQL\footnote{http://www.postgresql.org}
database management system reveals that there are many
distinct implementations of the memory page abstraction and B-trees. These versions
are \textit{slightly} divergent from each other, suggesting that the optimization scope
is limited. However, this situation significantly contributes to a
\textit{maintenance nightmare} as in order to apply any code update, many
different pieces of code have to be modified.

In addition, even though \datastructure specialization is important when
targeting high-performance systems, it is not provided, to the best of our knowledge, by any 
existing query compilation engine. Since our approach can be used to optimize the
\textit{entire} Scala code, and not only the operator
interfaces, it allows for various degrees of specialization in data structures,
as has been previous shown in~\cite{rompf13popl}.

In this article, we demonstrate such possibilities by explaining how our
compiler can be used to:
\begin{inparaenum}
\item Optimize the data structures used to hold in memory the data of the input relations,
\item Optimize Hash Maps which are typically used in intermediate computations like aggregations, and, finally,
\item Automatically infer and construct indices for SQL attributes of \code{date} type.
\end{inparaenum}
We do so in the next three sections.

\subsubsection{Data Partitioning}
\label{subsec:indexingAndPartitioning}
Optimizing the structures that hold the data of the input relations is an
important form of \datastructure specialization, as such optimizations generally enable
more efficient join processing throughout query execution. We have observed that this is true 
even for multi-way, join-intensive queries. In \legobase{}, we perform data partitioning when loading 
the input data. We analyze this optimization, the code of which can be found in Appendix~\ref{code:partitioning},
next. 

To begin with, in \legobase{} developers can annotate the primary and foreign keys of their 
input relations, at schema definition time. Using this information, our system then creates 
optimized data structures for those relations, as follows.

First, for each input relation, \legobase creates a data structure which is accessed 
through the primary key specified for that relation. There are two possibilities:
\begin{itemize}
\item For single-attri\-bute primary keys, the value of this attribute in each tuple is used to place 
the tuple in a continuous 1D-array. For example, for the relations of the \tpch workload this is a 
straightforward task as the  primary keys are typically integer values in the range of $[1...\#num\_tuples]$. 
However, even when the primary key is not in a continuous value range, \legobase currently
aggressively trades-off system memory for performance, and stores the input data into a sparse array.

\item For composite primary keys (e.g.\ those of the LINEITEM table of \tpch), creating an 1D array does not suffice, as
there may be multiple tuples with the same value for \textit{any one} of the attributes of the primary key (thus
causing conflicts when accessing the array). One possible solution would be to hash \textit{all} attributes of the primary
key and guarantee that we get a unique index value to access the 1D-array. However, deriving such a function in full generality 
requires knowledge of the whole dataset in advance (in order to know all possible combinations of the primary key). 
More importantly, it introduces additional computation on the critical path in order to perform the hash, a fact that, 
according to our observations, can lead to significant, negative impact on performance.
For this reason, \legobase does not create an 1D array and, instead, handles such primary keys similarly to the handling of
foreign key, as we discuss shortly. 
\end{itemize}

\noindent For the example given in Figure~\ref{fig:query_example2}, \legobase creates a 1D array for the ORDERS 
table, indexed through the O\_ORDERKEY attribute, but does not create a data structure
accessed through the primary key for LINEITEM (as this relation has a composite primary key
of the L\_ORDERKEY, L\_LINENUMBER attri\-butes).

Second, \legobase replicates and repartitions the data of the input relations based
on each specified foreign key. This basically leads to the creation of a
two-dimensional array, indexed by the foreign key, where each bucket holds
all tuples having a particular value for that foreign key. We also apply the same 
partitioning technique for relations that have composite primary keys, as we 
mentioned above. We resolve the case where the foreign key is not in a contiguous value 
range by trading-off system memory, in a similar way to how we handled primary keys. 

For the example of Q12, \legobase creates four partitioned tables: one for the foreign key 
of the ORDERS table (O\_CUSTKEY), one for the composite primary key of the LINEITEM 
table (as described above), and, finally, two more for the foreign keys of the LINEITEM table 
(on L\_ORDERKEY and L\_PARTKEY/L\_SUPPKEY respectively).

Observe that for relations that have multiple foreign keys, not all corresponding partitioned 
input relations need to be kept in memory at the same time, as an incoming SQL query may not need to use 
all of them.  
To decide which partitioned
tables to load, \legobase{} depends mainly on the derived physical query execution plan
(attributes referenced as well as select and join conditions of the input query), but also on
simple to estimate statistics, like cardinality estimation of the input relations.
For example, for Q12, out of the two partitioned, foreign-key data structures of LINEITEM, 
our optimized generated code for Q12 uses only the partitioned table on L\_ORDERKEY, as there 
is no reference to attributes L\_PARTKEY or L\_SUPPKEY in the query.
\begin{figure}[t]
\lstset{style=scala_numbered,xleftmargin=0.3cm}
\begin{lstlisting}
// Sequential accessing for the ORDERS table (since it has smaller size)
for (int idx = 0 ; idx < ORDERS_TABLE_SIZE ; idx += 1) {
  int O_ORDERKEY = orders_table[idx].O_ORDERKEY;
  struct LINEITEMTuple* bucket = lineitem_table[O_ORDERKEY];
  for (int i = 0; i < counts[bucket]; i+=1) {
    // process bucket[i] -- a tuple of the LINEITEM table
  }
}
\end{lstlisting}
\caption{Using primary and foreign keys in order to generate code for high-performance join processing. The underlying storage layout is that of a row-store for simplicity. The \code{counts} array holds the number of elements that exist in each bucket.}
\label{fig:join-processing-code}
\end{figure}

These data structures can be used to significantly improve join processing, as 
they allow to quickly extract matching tuples on a join between two relations on
attributes that have a primary-foreign key relationship. This is best illustrated through 
our running example of Q12 and the join between the LINEITEM and ORDERS tables. 
For this query, \legobase (a) infers that the ORDERKEY attribute actually represents a primary-foreign key relationship 
and (b) uses statistics to derive that ORDERS is the smaller of the two tables. 
By utilizing this information, \systemname{} can generate the code shown in Figure~\ref{fig:join-processing-code}
in order to directly get the corresponding bucket of the array of LINEITEM (by using the value of the ORDERKEY attribute), 
thus avoiding the processing of a possibly significant number of LINEITEM tuples. 

\legobase uses this approach for multi-way joins as well, to completely eliminate the overhead of intermediate data 
structures for most \tpch queries. This results in significant performance improvement as the corresponding tuple copying 
between these intermediate data structures is completely avoided, thus reducing memory pressure and improving cache locality. 
In addition, a number of expensive system calls responsible for the tuple copying is also avoided by applying this optimization.

After the aforementioned optimization has been performed, \legobase has removed the overhead of using generic data structures for join processing, but there are still some hash maps remaining in the generated code. These are primarily hash maps which correspond to aggregations, as in this case there is no primary/foreign key information that can be used to optimize these data structures away, but also hash maps which process joins on attributes that are not represented by a primary/foreign key relationship. In these cases, \legobase lowers these maps to two-dimensional arrays as we discuss in our hash map lowering optimization in the next section. 

\subsubsection{Optimizing Hash Maps}
\label{subsubsec:maps-to-native-arrs}
Next, we show how hash maps, which are the most commonly used
\datastructures along with Trees in DBMSes, can be specialized for
significant performance improvement by using schema and query knowledge.

By default, \legobase uses GLib~\cite{GLib} hash tables for generating C code out of the HashMap 
constructs of the Scala language. Close examination of these generic hash maps in the baseline
implementation of our operators (e.g.\ in the Aggregation of Figure~\ref{fig:operatorexample_agg}) reveals 
the following three main abstraction overheads.

First, for every \textit{insert} operation, a generic hash map must allocate a container
holding the key, the corresponding value as well as a pointer to the
next element in the hash bucket. This introduces a significant number of
expensive memory allocations on the critical path. Second, hashing and
comparison functions are called for every \textit{lookup} 
in order to acquire the correct bucket and element in the hash list. 
These function calls are
usually virtual, causing significant overhead on the critical path. Finally, the
data structures may have to be resized \textit{during runtime} in order to
efficiently accommodate more data. 
These operations typically correspond to (a) allocating a
bigger memory space, (b) copying the old data over to the new memory space and,
finally, (c) freeing the old space. 
These resizing operations are a significant bottleneck,
especially for long-running, computationally expensive queries. 
\begin{figure}[t!]
\lstset{style=scala_numbered}
\begin{lstlisting}[xleftmargin=.08\textwidth]
class HashMapToArray extends RuleBasedTransformer {
  rewrite += rule {
    case code"new HashMap[K, V]($size, $hashFunc, $equalFunc)" => {
      // Create new array for storing only the values
      val arr = code"new Array[V]($size)"
      // Keep hash and equal functions in the metadata of the new object
      arr.attributes += "hash" -> hashFunc
      arr.attributes += "equals" -> equalFunc
      arr // Return new object for future reference
    }
  }
  rewrite += rule {
    case code"($hm: HashMap[K, V]).getOrElseUpdate($key, $value)" => {
      val arr = transformed(hm) // Get the array representing the original hash map
      // Extract functions
      val hashFunc  = arr.attributes("hash")
      val equalFunc = arr.attributes("equals")
      code"""
        // Get bucket
        val h = $hashFunc($value)     // Inlines hash function
        var elem = $arr(h)
        // Search for element & inline equals function
        while (elem != null && !$equalFunc(elem, $key))
          elem = elem.next
        // Not found: create new elem / update pointers
        if (elem == null) {
          elem = $value
          elem.next = $arr(h)
          $arr(h) = elem
        }
        elem
      """
    }
  } 
  // Lowering of remaining operations is done similarly
}
\end{lstlisting}
\caption{Specializing HashMaps by converting them to native arrays. The corresponding 
operations are mapped to a set of primitive C constructs.}
\label{dsspel}
\vspace{-0.15cm}
\end{figure} 

Next, we resolve all these issues with our compiler, without cha\-nging a single
line of the base code of the operators that use these data structures, or the code of other optimizations.  
This
property shows that our approach, which is based on using a high-level compiler API, is practical for specializing DBMS components. 
The transformation, shown in Figure~\ref{dsspel}, is applied during the \textit{lowering} phase
of the compiler (Section~\ref{subsec:ccodegen}), where high-level Scala 
constructs are
mapped to low-level C constructs. The optimization lowers Scala HashMaps to native
C arrays and inlines the corresponding operations, by making use of the 
following three observations:
\begin{enumerate}
\item[1.] For our workloads, the information stored on the key is \textit{usually} a subset of the
attributes of the value. Thus, generic hash maps store redundant data. To avoid this, whenever
a functional dependency between key and value is detected, we convert the hash map to a native
array that stores only the values, and not the associated key (lines 2-11).
Then, since the inserted elements are anyway chained together in a hash
list, we provision for the next pointer when these are first
allocated (e.g.\ at data loading, \textit{outside the critical path}\footnote{The transformer shown in Figure~\ref{dsspel} is applied only
for the code segment that handles basic query processing. There is another transformer which handles the provision of the next pointer during data loading.}). 
Thus, we no longer need the key-value-next container and we manage
to reduce the amount of memory allocations significantly.

\item[2.] Second, the \compiler{} optimizing compiler offers function inlining for any Scala function out-of-the-box. 
Thus, our system can automatically inline the body of 
the hash and equal functions wherever they are
called (lines 20 and 23 of Figure~\ref{dsspel}). This significantly reduces 
the number of function calls (to almost zero), considerably improving query
execution performance.

\item[3.] Finally, to avoid costly maintenance operations on the critical path,
we preallocate in advance all the necessary memory space that \textit{may} be
required for the hash map during execution. This is done by specifying a size
parameter when allocating the data structure (line 3). Currently, we obtain
this size by performing worst-case analysis on a given query, which means that
we possibly allocate much more memory space that what is actually needed. However, 
database statistics can make this estimation very accurate, as we show in our
experiments section where we evaluate the overall memory consumption of our 
system in more detail.
\end{enumerate}

For our running example, the aggregation array, created in step 1 above, is accessed using the integer value 
obtained from hashing the \code{L_SHIPMODE} string. Then, the values located into the corresponding bucket of
the array are checked one by one, in order to see if the value of \code{L_SHIPMODE} exists and if a match is 
found, the aggregation entries are updated accordingly, or a new entry is initialized otherwise.

In addition to the above optimizations, the \compiler optimizing compiler also detects hash table data structures 
that receive only a single, \textit{statically-known} key and converts each such structure to a single value,
thus completely eliminating the unnecessary abstraction overhead of these tables. In this case, this optimization maps 
all related HashMap operations to operations in the single value. For example, we convert a \code{foreach} to a 
single value lookup. An example of such a lowering is in aggregations which calculate one single global aggregate 
(in this case \code{key = ''TOTAL''}). This happens for example in Q6 of the \tpch workload. 

Finally, we note that \datastructure specialization is an example of
intra-operator optimization and, thus, each operator can specialize its own
\datastructures by using similar optimizations as the one shown in Figure~\ref{dsspel}. 

\subsubsection{Automatically Inferring Indices on Date Attributes}
\label{subsec:date-indices}
Assume that an SQL query needs to \textit{fully} scan an input relation in order to extract tuples 
belonging to a particular year. 
A naive implementation would simply execute an \code{if} condition for each tuple of the relation and
propagate that tuple down the query plan if the check was satisfied. 
However, it is our observation that such conditions, as simple as they may be, can have a pronounced
negative impact on performance, as they can significantly increase the total number of CPU instructions 
executed in a query. 

Thus, for such cases, \legobase uses the aforementioned partitioning mechanism in order to automatically 
create indices, at data loading time, for all attributes of \code{date} type. 
It does so by grouping the tuples of a date attribute based on the year, thus forming a two-dimensional array
where each bucket holds all tuples of a particular year. This design allows to immediately skip, at query execution time,
all years for which this predicate is incorrect. That is, as shown in Figure~\ref{fig:dateindices}, the \code{if} condition now 
just checks whether the first tuple of a bucket is of a particular year and if not the whole bucket is skipped, as 
\textit{all} of its tuples have the same year and, thus, they \textit{all} fail to satisfy the predicate condition. 

These indices are particularly important for queries that process large input relations, 
whose date values are uniformly distributed across years. This is the case,
for example, for the LINEITEM and ORDERS tables of \tpch, whose date attributes are always populated
with values ranging from 1992-01-01 to 1998-12-31~\cite{tpch}. 

\begin{figure}[t!]
\lstset{style=scala_unnumbered,xleftmargin=-0.15cm}
\begin{tabular}{cc}
\begin{subfigure}[b]{.5\textwidth}
\begin{lstlisting}
// Sequential scan through table
for (int idx=0 ; idx<TABLE_SIZE ; idx+=1) {

  if (table[idx].date >= "01-01-1994" &&
       table[idx].date <= "31-12-1994")
      // Propagate tuple down the query plan
}
\end{lstlisting}
\caption{Original, naive code}
\end{subfigure}
&
\begin{subfigure}[b]{.5\textwidth}
  \centering
  \begin{lstlisting}
// Sequential scan through table
for (int idx=0 ; idx<NUM_BUCKETS ; idx+=1) {
  // Check only the first entry
  if (table[idx][0].date >= "01-01-1994" &&
       table[idx][0].date <= "31-12-1994") 
       // Propage all tuples of table[idx]
}
\end{lstlisting}
\caption{Optimized code}
\end{subfigure}
\\
\end{tabular}
\vspace{-0.3cm}
\caption{Using \code{date} indices to speed up selection predicates on large relations.}
\label{fig:dateindices}
\end{figure}

\subsection{Changing Data Layout}
\label{subsec:layout}
A long-running debate in data\-base literature is the one between row and
column stores~\cite{rowcolumn2,rowcolumn3,rowcolumn1}. Even though there are
many significant differences between the two approaches in all levels of the
database stack, the central contrasting point is the \textit{data-layout}, i.e.\
the way data is organized and grouped together. By default {\systemname}
uses the row layout, since this intuitive data organization facilitated fast
development of the relational operators. However, we quickly noted the
benefits of using a column layout for efficient data processing.  One solution
would be to go back and redesign the whole query engine; however this 
misses the point of our compiler framework.  In this section, we show how the
transition from row to column layout can be expressed as an
optimization\footnote{We must note that changing the data layout does not
mean that {\systemname} becomes a column store. There are other important aspects
which we do not yet handle, and which we plan to investigate in future work.}.

The optimization of Figure~\ref{aos2soa} performs a conversion from
an array of records (row layout) to a record of arrays (column layout), where
each array in the column layout stores the values for \textit{one} attribute.
The optimization basically overwrites the default lowering for arrays, thus providing the new behavior. 
As with all our optimizations, \textit{type information} determines the
applicability of an optimization: here it is performed only if the 
array elements are of record type (lines 3,13,26). Otherwise, this transformation is a NOOP and the
original code is generated (e.g.\ an array of Integers remains unchanged).
\begin{figure}[t!]
\lstset{style=scala_numbered,xleftmargin=0.35cm}
\begin{lstlisting}
class ColumnarLayoutTransformer extends RuleBasedTransformer {
  rewrite += rule {
    case code"new Array[T]($size)" if typeRep[T].isRecord => typeRep[T] match {
      case RecordType(recordName, fields) => {
        val arrays = 
          for((name, tp: TypeRep[Tp]) <- fields) yield 
            name -> code"new Array[Tp]($size)"
        record(recordName, arrays)
      }
    }
  }
  rewrite += rule {  
    case code"(arr:Array[T]).update($idx,$v)" if typeRep[T].isRecord => typeRep[T] match {
      case RecordType(recordName, fields) => {
        val columnarArr = transformed(arr) // Get the record of arrays
          for((name, tp: TypeRep[Tp]) <- fields) {
            code """
              val fieldArr: Array[Tp] = record_field($columnarArr, $name)
              fieldArr($idx) = record_field($v, $name)
            """
          }
      }
    }
  }
  rewrite += rule {  
    case code"(arr:Array[T]).apply($index)" if typeRep[T].isRecord => typeRep[T] match {
      case RecordType(recordName, fields) => {
        val columnarArr = transformed(arr) // Get the record of arrays
        val elems = for((name, tp: TypeRep[Tp]) <- fields) yield {
          name -> code """
            val fieldArr: Array[Tp] = record_field($columnarArr, $name)
            fieldArr($index)
          """
        }
        record(recordName, elems)
      }
    }
  }
  // Fill remaining operations accordingly
}
\end{lstlisting} 
\caption{Changing the data layout (from row to column) expressed as an
optimization. Scala's \code{typeRep} carries type information, which is used 
to differentiate between Array[Rec] and other non-record arrays (e.g.\ an array of integers).}
\label{aos2soa}
\end{figure}
\begin{figure}[t]
\centering
\lstset{style=scala_unnumbered,numbersep=0pt}
\begin{subfigure}[c]{0.20\columnwidth}
\begin{lstlisting}
val a1 = a.L1
val a2 = a.L2
val e1 = a1(i)
val e2 = a2(i)
val r =
  record(L1->e1,
           L2->e2)
r.L1
\end{lstlisting}
\end{subfigure}$\mapsto\hspace{0.55cm}$%
\begin{subfigure}[c]{0.20\columnwidth}
\begin{lstlisting}
val a1 = a.L1
val a2 = a.L2
val e1 = a1(i)
val e2 = a2(i)
val r =
  record(L1->e1,
           L2->e2)
e1
\end{lstlisting}
\end{subfigure}$\mapsto\hspace{0.55cm}$%
\begin{subfigure}[c]{0.20\columnwidth}
\begin{lstlisting}
val a1 = a.L1
val a2 = a.L2
val e1 = a1(i)
val e2 = a2(i)
e1
\end{lstlisting}
\end{subfigure}$\mapsto\hspace{0.55cm}$%
\begin{subfigure}[c]{0.20\columnwidth}
\begin{lstlisting}
val a1 = a.L1
val e1 = a1(i)
e1
\end{lstlisting}
\end{subfigure}
\caption{Dead code elimination (DCE) can remove intermediate
materializations, e.g.\ row reconstructions when using a column layout.
Here \textit{a}
is a record of arrays (column-layout) and \textit{i} is an integer. The records have only two attributes \textit{L1} 
and \textit{L2}. The notation \code{L1->v} associates the label (attribute name) \text{L1} with value \textit{v}.}
\label{dce}
\end{figure}

Each optimized operation is basically a straightforward rewriting to a set of
operations on the underlying record of arrays.  Consider, for example, an update
to an array of records (\verb|arr(n) = v|), where \textbf{v} is a record. We
know that the \textit{new} representation of \verb|arr| will be a record of arrays (column layout),
and that \verb|v| has the same attributes as the elements of \verb|arr|.  So, for each of
those attributes we extract the corresponding array from \verb|arr| (line 18) and field
from \verb|v| (line 19); then we can perform the update operation on the extracted array
(line 19) using the same index.

This optimization also reveals another benefit of using an optimizing compiler:
developers can create \textit{new} abstractions in their optimizations, which
will be in turn optimized away in \textit{subsequent} optimization passes. For
example, array\_apply results in \textit{record
reconstruction} by extracting the individual record fields from the record of
arrays (lines 29-34) and then building a new record to hold the result (line 35). This intermediate
record can be \textit{automatically} removed using dead code elimination (DCE),
as shown in Figure~\ref{dce}. Similarly, if \compiler can statically determine that
some attribute is never used (e.g.\ by having all queries given in advance), then
this attribute will just be an unused field
in a record, which the optimizing compiler will be able to optimize away (e.g.\
attribute L2 in Figure~\ref{dce}).

We notice that, as was the case with previously presented 
optimizations, the transformation described in this section does not have 
any dependency on other optimizations or the code of the query 
engine. This is because it is applied in the distinct optimization phase 
that handles \textit{only} the optimization of arrays. This separation of 
concerns leads, as discussed previously, to a significant increase in 
productivity as, for example, developers that tackle the optimization of 
individual query operators do not have to worry about optimizations 
handling the data layout (described in this section). 

\subsection{String Dictionaries}
\label{subsec:stringdict}
Operations on non-primitive data types, such as strings, incur a very high performance overhead.
This is true for two reasons. First, there is typically a function call required. Second,
most of these operations typically need to execute loops to process the encapsulated data. For 
example, \textit{strcmp} needs to iterate over the underlying array of characters, comparing 
one character from each of the two input strings on each iteration. Thus, such operations 
significantly affect branch prediction and cache locality. 

\legobase{} uses String Dictionaries to remove the abstraction overhead of Strings. Our system
maintains one dictionary for every attribute of String type, which generally operates as follows.
First, at data loading time, each string value of an attribute is mapped to an integer value. This value 
corresponds to the index of that string in a single linked-list holding the \textit{distinct} string values
of that attribute. The list basically constitutes the dictionary itself. In other words, each 
time a string appears for the first time during data loading, a unique integer is assigned to it; 
if the same string value reappears in a later tuple, the dictionary maps this string to the previously assigned
integer. Then, at query execution time, string operations are mapped to their integer counterparts,
as shown in Table~\ref{table:string_compression}. This mapping allows to significantly improve the 
query execution performance, as it typically eliminates underlying loops and, thus, significantly
reduces the number of CPU instructions executed. For our running example, \legobase compresses the 
attributes L\_SHIPMODE and O\_ORDERPRIORITY by converting the six string equality checks into corresponding
integer comparisons.
\begin{table}[t!]
\centering
\begin{tabular}{l l l l }
\toprule
String & & Integer & Dictionary \\
Operation & C code & Operation & Type \\
\midrule
equals & strcmp(x, y) == 0 & x == y & Normal \\ 
notEquals & strcmp(x, y) != 0 & x != y & Normal \\
startsWith & strncmp(x, y, strlen(y)) == 0 & x>=start \&\& x<=end & Ordered \\
indexOfSlice & strstr(x, y) != NULL & N/A & Word-Token \\
\bottomrule
\end{tabular}
\caption{Mapping of string operations to integer operations through the corresponding type of 
string dictionaries. Variables $x$ and $y$ are strings arguments which are mapped to integers. The rest of
string operations are mapped in a similar way.} 
\label{table:string_compression}
\end{table}

Special care is needed for string operations that require ordering. For example, Q2 and Q14
of \tpch need to perform the \textit{endsWith} and \textit{startsWith} string operations with a constant
string, respectively. This requires that we utilize a dictionary that maintains the data in order;
that is, if $string_x$ $<$ $string_y$ lexicographically, then $Int_x$ < $Int_y$ as well. To do so, 
we take advantage of the fact that in \legobase all input data is already materialized, and thus we can first
compute the list of distinct values, as mentioned above, then sort this list lexicographically, and afterwords
make a second pass over the data to assign integer values to the string attribute.
By doing so, the constant string is then converted to a $[start,end]$ range, by iterating over the list of distinct values and 
finding the first and last strings which start or end with that particular string. This range is then used when 
lowering the operation, as shown in Table~\ref{table:string_compression}. This \textit{two-phase} string dictionary
allows to map all operations that require some notion of ordering in string operations. 

In addition, there is one additional special case where the string attributes need to be tokenized on
a word granularity. This happens for example in Q13 of \tpch. This is because queries like that one need
to perform the \textit{indexOfSlice} string operation, where the slice represents a word. \legobase provides a \textit{word-tokenizing} string 
dictionary that contains all words in the strings instead of the string attributes themselves to handle such 
cases. Then, searching for a word slice is equal to looking through all the integer-typed words in that string for a match 
during query execution. This is the only case where the integer counterparts of strings operations contain a 
loop. It is however our experience, that even with this loop through the integer vales, the obtained performance 
significantly outperforms that of the \code{strstr} function call of the C library.
This may be because such loops can be more easily vectorized by an underlying C compiler like \clang, compared to 
the corresponding loops using the string types.

Finally, it is important to note that string dictionaries, even though they significantly improve query execution 
performance, they have an even more pronounced negative impact on the performance of data loading. This is particularly 
true for the word-tokenizing string dictionaries, as the impact of tokenizing a string is significant. In addition, 
string dictionaries can actually degrade performance when they are used for primary keys or for attributes that contain 
many distinct values (as in this case the string dictionary significantly increases memory consumption). In such cases,
\legobase can be configured so that it does not use string dictionaries for those attributes, through proper usage of the optimization pipeline described in Section~\ref{sec:system}.

\subsection{Domain-Specific Code Motion}
\label{subsec:dscodemotion}
Domain-Specific code motion includes optimizations that remove 
code segments that have a negative impact on query execution performance from the critical path and instead 
executes the logic of those code segments during data loading. Thus, the optimizations in 
this category trade-off increased loading time for improved query execution performance. 
There are two main optimizations in this category, described next.
\subsubsection{Hoisting Memory Allocations}
Memory allocations can cause significant performance degradation in query 
execution. Our experience shows that, by taking advantage of type information 
available in each SQL query, we can completely eliminate such allocations from 
the critical path. The \systemname{} system provides the following optimization 
for this purpose.

At query compilation time, information is gathered regarding the
data types used throughout an incoming SQL query. This is done through an analysis phase,
where the compiler collects all \code{malloc} nodes in the program, once the latter has 
been lowered to the abstraction level of C code. This is necessary to be done at this level,
as high-level programming languages like Scala provide implicit memory management, which the
\compiler optimizing compiler cannot currently optimize. The obtained types correspond either 
to the initial database relations (e.g.\ the LINEITEM table of \tpch) or to types
required for intermediate results, such as aggregations. Based on this information, 
\compiler initializes memory pools during data loading, one for each type.

Then, at query execution time, the corresponding \code{malloc} statements are replaced
with references to those memory pools. We have observed that this
optimization significantly reduces the number of CPU instructions executed during the query evaluation,
and significantly contributes to improving cache locality. This is because the memory space allocated for 
each pool is contiguous and, thus, each cache miss brings useful records to the cache (this is not the case for
the fragmented memory space returned by the \code{malloc} system calls). 
We also note that it is necessary to resolve 
dependencies between data types. This is particularly true for composite types,
which need to reference the pools of the native types (e.g. the
pool for Strings). We resolve such dependencies by first applying topological sorting
on the obtained type information and only then generating the pools in the proper
order. 

Finally, we must mention that the size of the
memory pools is estimated by performing worst-case analysis on a given query. This means that 
\legobase may allocate much more space than needed. However, we have confirmed that our estimated statistics are accurate enough so 
that the pools do not unnecessarily create memory pressure, thus negatively affecting query performance. 

\subsubsection{Hoisting Data-Structure Initialization}
The proper initialization and maintenance of any data structure needed during query execution 
generally require specific code to be executed in the critical path. This is typically true for 
data structures representing some form of \textit{key-value} stores, as we describe next. 

Consider the case of \tpch Q12, for which a data structure is needed to store the results of the 
aggregate operator. Then, 
when evaluating the aggregation during query execution, we must check whether the
corresponding key of the aggregation has been previously inserted in the aggregation data structure. 
In this case, the code must check whether a specific value of O\_ORDER\-PRIORITY
is already present in the data structure. If so, it would return the existing aggregation.
Otherwise, it would insert a new aggregation into the data structure. This means that 
\textit{at least one} \code{if} condition must be evaluated for \textit{every} tuple that 
is processed by the aggregate operator. We have observed that such \code{if} conditions, which 
exist purely for the purpose of \datastructure initialization, significantly affect branch 
prediction and overall query execution performance.

\legobase provides an optimization to remove such \datastructure initialization
from the critical path by utilizing domain-specific knowledge. For
example, \legobase takes advantage of the fact 
that aggregations can usually be statically initialized with the value zero, for each 
corresponding key. To infer all these possible key values (i.e. infer the 
\textit{domain} of that attribute), \legobase{} utilizes the statistics collected  
during data loading for the input relations. Then, at query execution time, the corresponding \code{if}
condition mentioned above no longer needs to be evaluated, as the aggregation
already exists and can be accessed directly. We have observed that by removing
code segments that perform \textit{only} \datastructure initialization, branch 
prediction is improved and the total number of CPU instructions executed is
significantly reduced as well.  

Observe that this optimization is not possible in its full generality, as it directly
depends on the ability to predict the possible key values in advance, during data loading. 
However, we note three things. First, once our partitioning optimization 
(Section~\ref{subsec:indexingAndPartitioning}) has been applied, \legobase requires intermediate data structures
mostly for aggregate operators, whose initialization code segment we remove,
as described above. Second, particularly for \tpch, there is no key that is the result of an intermediate 
join operator deeply nested in the query plan. Instead, \tpch uses attributes of the original
relations to access most data structures, attributes whose value range can be accurately estimated 
during data loading through statistics, as we discussed previously. Finally, for \tpch queries the 
key value range is very small, typically ranging
up to a couple of thousand sequential key values\footnote{A notable exception is \tpch Q18 which 
uses O\_ORDERKEY as a key, which has a sparse distribution of key values. 
\legobase generates a specialized data structure for this case.
}.
These three properties allow to completely remove 
initialization overheads and the associated unnecessary computation for all \tpch queries. 

\subsection{Traditional Compiler Optimizations}
In this section, we present a number of traditional compiler optimizations that originate mostly from work in the PL community.
Most of them are generic in nature, and, thus, they are offered out-of-the-box by the \compiler optimizing compiler.

\subsubsection{Removal of Unused Relational Attributes}
In Section~\ref{subsec:layout} we mentioned that \legobase provides an optimization for removing 
relational attributes that are not accessed by a particular SQL query, assuming that this query is known \textit{in advance}. 
For example, the Q12 running example references eight relational attributes. However, the relations LINEITEM and ORDERS 
contain 25 attributes in total. \legobase avoids loading these unnecessary attributes into memory at data loading time. 
It does so by analyzing the input SQL query and removing the set of unused fields from the record definitions. This 
reduces memory pressure and improves cache locality. 

\subsubsection{Removing Unnecessary Let-Bindings}
\label{subsubsec:letbind}
The \compiler compiler uses the Administrative Normal Form (ANF) when generating 
code. This simplifies code generation for the compiler. However, it has the negative effect of 
introducing many unnecessary intermediate variables. We have observed that this form of code 
generation not only affects code compactness but also significantly increases register pressure. 
To improve upon this situation, \compiler uses a technique first introduced by the programming 
language community~\cite{anftoast}, which removes any intermediate variable that satisfies 
the following three conditions: the variable (a) is set only once, (b) has no side effects, and, finally, 
(c) is initialized with a single value (and thus its initialization does not correspond to
executing possibly expensive computation). \compiler then replaces any appearance of this variable later
in the code with its initialization value. We have noticed that this optimization makes the
generated code much more compact and reduces register pressure, resulting in improved performance. Moreover, 
we have observed that since the variable initialization time may 
take place significantly earlier in the code of the program than its actual use, this does not allow 
for this optimization opportunity to be detected by low-level compilers like LLVM. 

Finally, our compiler applies a technique known as \textit{parameter promotion}\footnote{This technique 
is also known as Scalar Replacement in the PL community.}. This 
optimization removes \textit{structs} whose fields can be flattened to local variables. This optimization
has the effect of removing a memory access from the critical path as the field of a struct can be 
referenced immediately without referencing the variable holding the struct itself. We have 
observed that this optimization significantly reduces the number of memory accesses occurring during query execution. 

\subsubsection{Fine-grained Compiler Optimizations}
\label{subsec:finegrain}
Finally, there is a category of fine-grained compiler optimizations that are
applied last in the compilation pipeline. These optimizations target optimizing
very small code segments (even individual statements) under particular
conditions. We briefly present three such optimizations next. 

First, \compiler can transform arrays to a set of local variables. This 
lowering is possible only when (a) the array size is statically known at
compile time, (b) the array is relatively small (to avoid increasing register pressure) and, finally, (c) the index 
of every array access can be inferred at compile time (otherwise, the compiler is not
able to know to which local variable an array access should be mapped to). 

Second, the compiler provides an optimization to change the boolean condition \code{x && y} to \code{x & y}
where \code{x} and \code{y} both evaluate to boolean and the second operand does not have any side effect.
According to our observations, this optimization can significantly improve branch prediction, when the aforementioned
conditions are satisfied.

Finally, the compiler can be instructed to apply tiling to \code{for} loops whose range 
are known at compile time (or can be accurately estimated). 

It is our observation
that all these fine-grained optimizations (as described above), which can be typically written in less than a 
hundred lines of code, can help to improve the performance of certain queries. More importantly,
since they have very fine-grained granularity, their application does not introduce 
additional performance overheads.


\subsection{Discussion}
\label{sec:classifying}
\begin{figure}[t]
\includegraphics[width=\columnwidth]{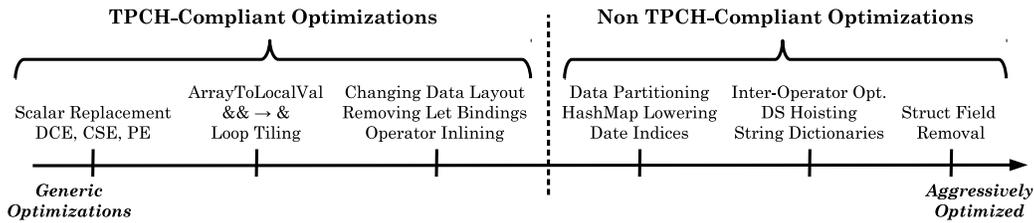}
\caption{Classification of \legobase optimizations.}
\label{fig:classification}
\end{figure}

In this section, we classify the \legobase optimizations according to (a) their generality and
(b) whether they follow the rules of the \tpch benchmark, which we use in
our evaluation. These two metrics are important for a more thorough understanding of which
categories of database systems can benefit from these optimizations. We detect six groups of optimizations, 
illustrated in Figure~\ref{fig:classification}, described next in the order they appear from left to right in the figure.
\\

\noindent \textbf{Generic Compiler Optimizations:} In this category, we include optimizations which
are also applied by traditional compilers, such as LLVM. These include Dead Code Elimination (DCE),
Common Subexpression Elimination (CSE), Partial Evaluation (PE) and the Scalar Replacement optimization presented
in Section~\ref{subsubsec:letbind}. These optimizations are \tpch compliant and do 
not require any particular domain-specific knowledge; thus they can be applied for optimizing any input query as well as the code of the query engine.
\\

\noindent \textbf{Fine-grained Optimizations:} In this \tpch compliant category we include, as described in 
Section~\ref{subsec:finegrain}, fine-grained optimizations that aim to transform and improve
the performance of individual statements (or a small number of contiguous statements). We do not
list this category alongside the generic compiler optimizations, as whether they improve
the performance or not depends on the characteristics of the input query. Thus, \compiler needs to analyze the 
program before detecting whether the application of one of the optimizations in this group is beneficial. 
\\

\noindent \textbf{Optimizing Data Accesses:} The two optimizations presented in Sections~\ref{subsec:layout}
and~\ref{subsubsec:letbind}, alongside the generic \textit{operator inlining} optimization, aim to 
improve performance by minimizing the number of function 
calls and optimizing data accesses and code compactness. Even though they are coarse-grained in nature, affecting 
large code segments, they are still \tpch compliant, as they are neither query specific nor depend on type information. 
\\

\noindent \textbf{Partitioning and Indexing Optimizations:} This class of optimizations, presented in detail in Section~\ref{subsec:dsopt}, 
aims to improve query execute performance. However, even though they provide significant performance improvement (as we show in our evaluation), they 
are not \tpch compliant, as this workload does not allow logical replication of data. Similarly, our HashMap lowering optimization requires
knowledge of the domain of the aggregation keys in advance. 
Still, there is a class of database systems that can greatly benefit from such indexing and partitioning transformations. These include systems that have all their data known in advance (e.g.\ OLAP style processing) or systems where we can introduce pre-computed indexing views, as in the case of Incremental View Maintenance (IVM).
\\

\noindent \textbf{Inter-Operator, String Dictionaries, and Domain-Specific Hoisting Optimizations:} The three optimizations in this category, presented in Sections~\ref{subsec:eliminate_materializations},~\ref{subsec:stringdict} and~\ref{subsec:dscodemotion} respectively, aim to remove unnecessary materialization points and computation from the critical path. However, they are \textit{query specific}, as they can only be applied if a query is known in advance. This is the primary characteristic that differentiates this category of optimizations from the previous one. They also depend on
type information and introduce auxiliary data structures. Thus, they are not \tpch compliant. 
\\

\noindent \textbf{Struct Field Removal Optimization:} The most aggressive optimization that \legobase applies removes unnecessary relational attributes from C structs. This optimization is query specific and is highly dependant on type information. It also requires specializing the data structures during data loading (to remove the unnecessary fields). Thus, it is not \tpch compliant. 

\section{Evaluation}
\label{sec:eval}
In this section, we evaluate the realization of the abstraction without regret vision in 
the domain of analytical query processing. After briefly presenting 
our experimental platform, we address the following topics and open questions related to \systemname{}:

\begin{enumerate}

\item How well can general-purpose compilers, such as LLVM or GCC, optimize query 
engines? We show that these compilers fail to detect many high-level 
optimization opportunities and, thus, they perform poorly compared to our system
(Section~\ref{lego-eval-one}).

\item Is the code generated by \legobase{} competitive, performance-wise, 
to (a) traditional database systems and (b) query compilers based on template 
expansion? We show that by utilizing query-specific knowledge and by extending 
the scope of compilation to optimize the \textit{entire} query engine, we can 
obtain a system that significantly outperforms both alternative approaches (Section~\ref{lego-eval-two}).

\item We experimentally validate that the source-to-source compilation from Scala to 
efficient, low-level C binaries is necessary as even highly optimized Scala programs 
exhibit a considerably worse performance than C (Section~\ref{lego-eval-three}).

\item What insights can we gain by analyzing the performance improvement of 
individual optimizations? Our analysis reveals that important optimization 
opportunities have been so far neglected by compilation approaches that optimize 
\textit{only} queries. To demonstrate this, we compare architectural decisions as 
fairly as possible, using a shared codebase that only differs by the effect of a 
single optimization (Section~\ref{lego-eval-four}).

\item How much are the overall memory consumption and data loading speed of our system? These two metrics are of importance to main-memory databases, as a query engine must perform well in both directions to be usable in practice (Section~\ref{lego-eval-five}). 

\item We analyze the amount of effort required when coding query engines in 
{\systemname} and show that, by programming in the abstract, we can derive a fully 
functional system in a relatively short amount of time and coding effort (Section~\ref{lego-eval-six}). 

\item We evaluate the compilation overheads of our approach. We show that 
\compiler{} can efficiently compile query engines even for the complicated, 
multi-way join queries typically found in analytical query processing (Section~\ref{lego-eval-seven}).
\end{enumerate}

\subsection{Experimental Setup}
Our experimental platform consists of a server-type x86 machine equipped with
two Intel Xeon E5-2620 v2 CPUs running at 2GHz each, 256GB of DDR3 RAM at 1600Mhz
and two commodity hard disks of 2TB storing the experimental datasets. The
operating system is Red Hat Enterprise 6.7. 
For all experiments, we have disabled huge pages in the kernel, since this
provided better results for all tested systems (described in more detail in Table~\ref{systems}).
For compiling the generated programs
throughout the evaluation section, we use version 2.11.4 of the Scala compiler and
version 3.4.2 of the \clang front-end for LLVM~\cite{llvm}, with the default
optimization flags set for both compilers. For the Scala programs, we configure the 
Java Virtual Machine (JVM) to run with 192GB of heap space, while 
we use the \glib library (version 2.38.2)~\cite{GLib} whenever we need to generate generic data structures in C.
\begin{table}[t!]
\centering
\begin{tabular}{p{1.83cm} p{3.5cm} p{2.8cm} p{1.7cm} p{2cm}}
\toprule
System & Description & Compiler \mbox{optimizations} & \tpch{} compliant & Uses query-specific info \\
\midrule
\timesten{} & Commercial, \mbox{in-memory} DBMS & No \mbox{compilation} & Yes & No \\\rule{0pt}{2.5ex}
Compiler \mbox{of \hyper} & Query compiler of the HyPer DBMS 
& Operator inlining, push engine & Yes & No \\\rule{0pt}{2.5ex}
\systemname{} (Naive) & A naive engine with the minimal number of optimizations
& Operator inlining, push engine & Yes & No \\\rule{0pt}{2.5ex}
\systemname{} (\tpch/C) & \tpch{} compliant engine & Operator inlining, push engine, data partitioning & Yes\tablefootnote{
We note that according to the \tpch specification rules, a database system can employ data partitioning (as 
described in Section~\ref{subsec:indexingAndPartitioning}) and still be \tpch compliant. This is the case when 
all input relations are partitioned on \textit{one and only one} primary or foreign key attribute across all queries. 
The \systemname{}(\tpch/C) configuration of our system follows exactly this partitioning approach, which is also used 
by the \hyper{} system (but in contrast to \compiler{}, partitioning in \hyper is not expressed as a compiler optimization).
} & No \\\rule{0pt}{2.5ex}
\systemname{} (StrDict/C) & Non \tpch{} compliant engine with some optimizations applied & Like above, plus String Dictionaries & No & No \\\rule{0pt}{2.5ex}
\systemname{} (Opt/C) & Optimized push-style engine & All optimizations of this \paper{} & No & Yes \\\rule{0pt}{2.5ex}
\systemname{} (Opt/Scala) & Optimized push-style engine & All optimizations of this \paper{} & No & Yes \\
\bottomrule
\end{tabular}
\caption{Description of all systems evaluated in this section. Unless otherwise stated, all generated C programs of \systemname{} are compiled to a final C binary using \clang{}. All listed \systemname{} engines and optimizations are written with \textit{only} high-level Scala code, which is then optimized and compiled to C or Scala code with \compiler{}.}
\label{systems}
\end{table}

For our evaluation, we use the \tpch benchmark~\cite{tpch}. \tpch is a
data warehousing and decision support benchmark that issues business analytics
queries to a database with sales information. This benchmark suite includes
22 queries with a high degree of complexity that express most SQL features. 
We use all 22 queries to evaluate various design choices of our methodology. 
We execute each query five times and report the average performance of these
runs. Unless otherwise stated, the scaling factor of \tpch is set to 8 for all experiments.
It is important to note that the final generated optimized code of \legobase (configurations
\systemname{}(Opt/C) and \systemname{}(Opt/Scala) in Table~\ref{systems}) employs materialization 
(e.g. for the date indices) and, thus, this version of the code does \textit{comply} with 
the \tpch implementation rules. However, we also present a \tpch compliant configuration, 
\systemname{}(\tpch{}/C), for comparison purposes.

As a reference point for most results presented in this section, we use a	
commercial, \text{in-memory}, row-store database system called \timesten, which
does not employ compilation. We assign 192GB of DRAM as memory space in
\timesten, and we use the \timesten-specific data types instead of generic SQL
types. As described in Section~\ref{sec:system}, {\systemname} uses query plans 
from the \timesten data\-base. We also use the query compiler of the \hyper 
system~\cite{neumann11} (v0.4-452) as a point of comparison with existing query 
compilation approaches. Since parallel execution is 
still under development at the time of writing for \systemname{}, all systems have 
been forced to single-threaded execution, either by using the execution parameters 
some of them provide or by manually disabling the usage of CPU cores in the 
kernel configuration.
\begin{figure*}[t!]
\centering
\includegraphics[width=0.7\textwidth]{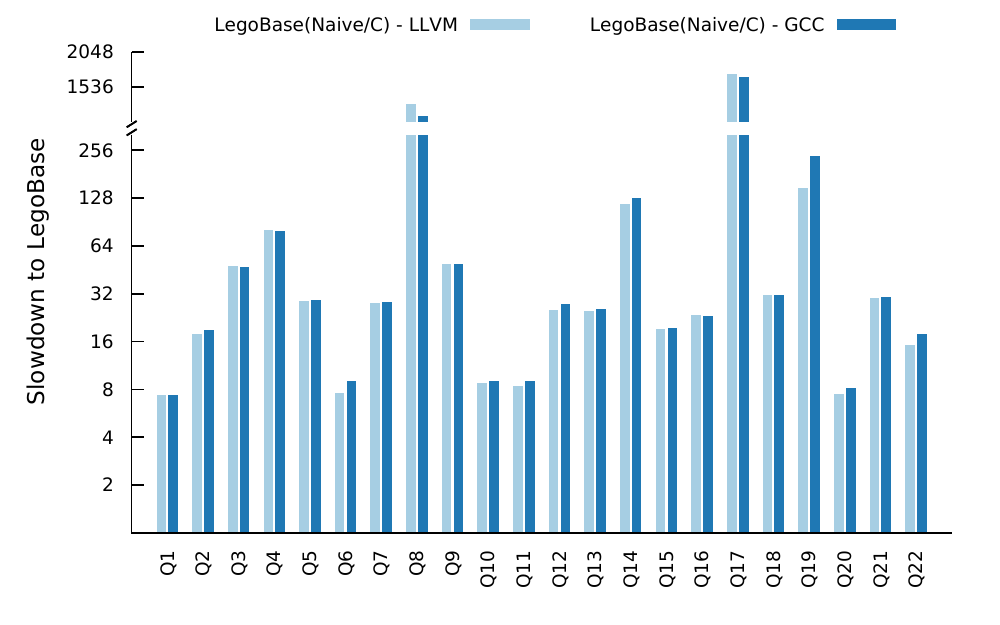}
\caption{Performance of a push-style engine compiled with LLVM and GCC. These engines are generated using 
\textit{only} operator inlining. The baseline is the performance of the optimal generated code, 
\systemname{}(Opt/C), with all optimizations enabled.}
\label{volcano}
\end{figure*}

\subsection{Optimizing Query Engines Using General-Purpose Compilers}
\label{lego-eval-one}
First, we show that low-level, general-purpose compilation frameworks, 
such as LLVM, are not adequate for efficiently optimizing query engines. To do 
so, we use \systemname{} to generate an \textit{unoptimized} push-style engine with 
only operator inlining applied, which is then compiled to a final C binary using 
LLVM. We choose this engine as a starting point since it allows the underlying C 
compiler to be more effective when optimizing the whole C program (as the presence of 
procedures may otherwise force the compiler to make conservative decisions or miss 
optimization potential during compilation).

As shown in Figure~\ref{volcano}, the achieved performance is very
poor: the unoptimized query engine, \systemname{}(Naive/C)--LLVM, is significantly slower for all
\tpch queries, requiring more than 16$\times$ the execution time of the optimal \systemname{}
configuration in most cases. This is because frameworks 
like LLVM cannot automatically detect all optimization opportunities that we support in
{\systemname} (as described thus far in this \paper{}). This is because either 
(a) the scope of an optimization is too coarse-grained to be detected by a low-level 
compiler or (b) the optimization relies on domain-specific 
knowledge that general-purpose optimizing compilers such as LLVM are not aware of. 

In addition, as shown in the same figure, compiling with LLVM does not
\textit{always} yield better results compared to using another traditional compiler
like \gcc\footnote{For this experiment, we use version 4.4.7 of the \gcc compiler.}.
We see that LLVM outperforms \gcc for \textit{only} 15 out of 22 queries (by
1.09$\times$ on average) while, for the remaining ones, the binary generated by 
\gcc performs better (by 1.03$\times$ on average). In general, the performance difference 
between the two compilers can be significant (e.g.\ for Q19, there is a 
1.58$\times$ difference). We also experimented with manually specifying 
optimizations flags to the two compilers, but this turns out to be a very delicate 
and complicated task as developers can specify flags which actually make 
performance worse. We argue
that it is instead more beneficial for developers to invest their
effort in developing high-level optimizations, like those presented in
this \paper{}.

\subsection{Optimizing Query Engines Using Template Expansion}
\label{lego-eval-two}
\begin{figure*}[t!]
\hspace{-0.2cm}
\includegraphics[width=1.04\textwidth]{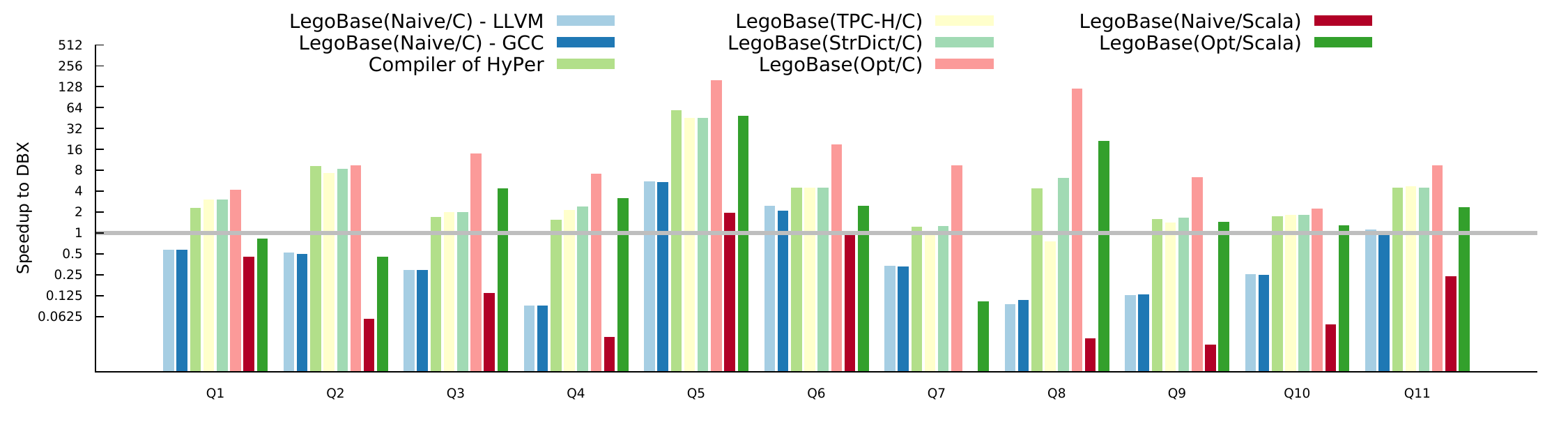}
\includegraphics[width=1.04\textwidth]{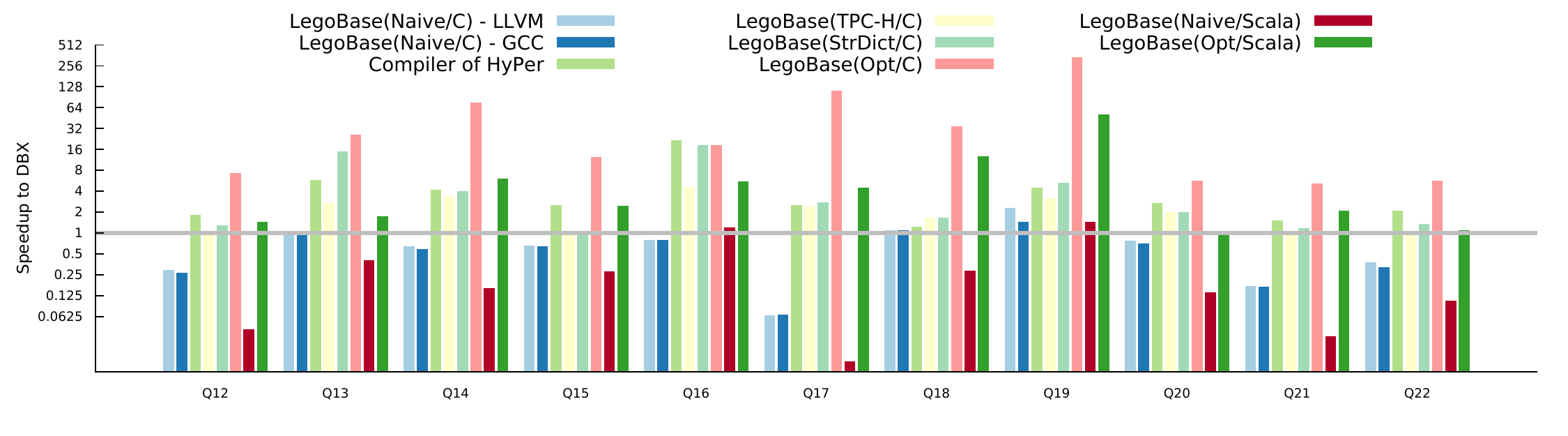}
\caption{Performance comparison of various {\systemname} configurations (C and Scala programs) 
with the code generated
by the query compiler of~\protect\cite{neumann11}. The baseline for all systems is the performance of the \timesten{} commercial database system. The absolute execution times for this figure
can be found in Appendix~\protect\ref{appendix:absolute-exec-times}. This graph also includes the performance of the
naive push-engines of Figure~\ref{volcano} for reference.}
\label{neumann}
\end{figure*}

Next, we compare our approach -- which compiles the \textit{entire} query engine and utilizes \textit{query-specific} 
information -- with the compiler of the \hyper database~\cite{neumann11}. \hyper performs template expansion through 
LLVM in order to inline the relational operators of a query executed on a push engine.
The results are presented in Figure~\ref{neumann}.  

We perform this analysis in two steps. First, we generate a query engine that (a) does not utilize any query-specific information 
and (b) adheres to the implementation rules of the \tpch workload. Such an engine represents a system where data are loaded
\textit{only once}, and all optimizations are applied before any query arrives (as happens with \hyper and any other traditional
DBMS). We show that this \systemname{} configuration, titled \systemname{}(\tpch/C), has performance competitive
to that of the \hyper database system, and that efficient handling of string operations is essential in order to have the performance 
of our system match that of \hyper. Second, we show that by utilizing query-specific knowledge and performing aggressive materialization
and repartition of input relations based on multiple attributes, we can generate a query engine, titled \systemname{}(Opt/C), which significantly
outperforms existing approaches. Such an engine corresponds to systems that, as discussed previously in Section~\ref{sec:classifying}, have all queries or data known in advance.

Figure~\ref{neumann} shows that by using the query compiler of \hyper, performance is improved by 
6.4$\times$ on average compared to \timesten. To achieve this performance improvement, \hyper uses a push engine, 
operator inlining, and data partitioning. In contrast, the \tpch-compliant configuration of our system, 
\legobase{}(\tpch{}/C), which uses the same optimizations, has an average execution time of only 4.4x the one of 
the \timesten{} system, across all \tpch queries. The main reason behind this significantly slower performance is, as 
we mentioned above, the inefficient handling of string operations in \legobase{}(\tpch{}/C). In this version, 
\systemname{} uses the \code{strcmp} function (and its variants). In contrast, \hyper uses the SIMD instructions 
found in modern instructions sets (such as SSE4.2) for efficient string handling~\cite{Boncz2014}, a design choice 
that can lead to significant performance improvement compared to \systemname{}(\tpch{}/C). To validate this analysis, 
we use a configuration of our system, called \legobase{}(StrDict/C), which additionally applies the string dictionary 
optimization. This configuration is no longer \tpch-compliant (as it introduces an auxiliary data structure), but is 
still does not require query-specific information. We notice that the introduction of this optimization is 
enough to make \legobase{}(StrDict/C) match the performance of \hyper: the two systems have \textit{only} a 
1.06$\times$ difference in performance.

\begin{figure}[t]
\centering
\includegraphics[width=\textwidth]{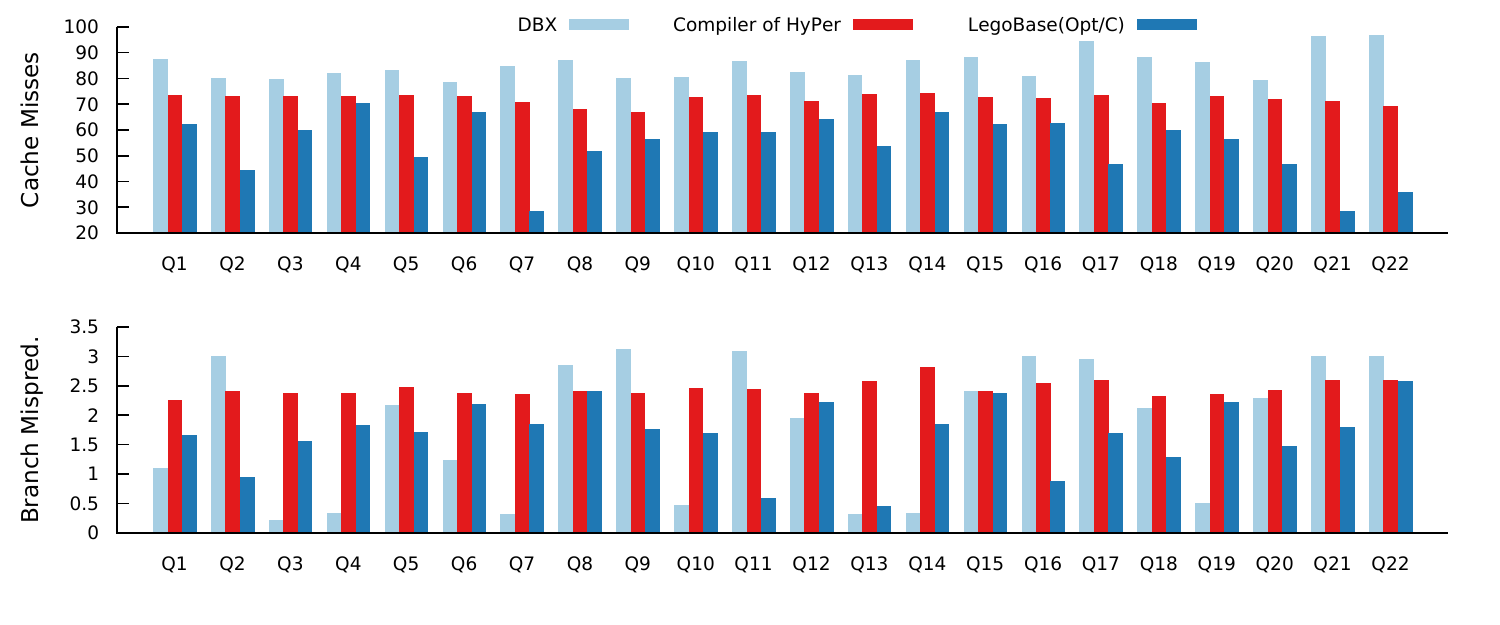}
\caption{Percentage of cache misses and branch mispredictions for \timesten,
\hyper and {\systemname}(Opt/C) for all 22 \tpch queries.}
\label{fig:branch-cache}
\end{figure}

Second, Figure~\ref{neumann} also shows that by using all other optimizations of {\systemname} 
(as they were presented in Section~\ref{sec:optimizations}), which are not performed by 
the query compiler of \hyper, we can get a total 45.4$\times$ performance improvement 
compared to \timesten with all optimizations enabled. This is because, for example,
\systemname{}(Opt/C) uses query-specific information to remove unused relational attributes or hoist out expensive 
computation (thus reducing memory pressure and decreasing the number of CPU instructions executed) and aggressively 
repartitions input data on multiple attributes (thus allowing for more efficient join processing). Such 
optimizations result in improved cache locality and branch prediction, as shown in Figure~\ref{fig:branch-cache}. 
More specifically, there is an improvement of 1.68$\times$ and 1.31$\times$ on average for the two metrics, 
respectively, between \timesten and {\systemname}.  In addition, the maximum, average and minimum difference in the 
number of CPU instructions executed in \hyper is 3.76$\times$, 1.61$\times$, and 1.08$\times$ that executed 
in \legobase. These results prove that the optimized code of \legobase{}(Opt/C) is competitive, performance wise, 
to both traditional database systems and query compilers based on template expansion.

Finally, we note that we plan to investigate even more aggressive and query-specific \datastructure optimizations in 
future work. Such optimizations are definitely feasible, given the easy extensibility of the \compiler{} compiler. 

\subsection{Source-to-Source Compilation from Scala to C}
\label{lego-eval-three}
Next, we show that source-to-source compilation from Scala to C is necessary
in order to achieve optimal performance in \systemname{}.
To do so, Figure~\ref{neumann} also presents
performance results for both a naive and an optimized Scala query engine, named \systemname{}(Naive/Scala) 
and \systemname{}(Opt/Scala), respectively. First, we notice that the optimized generated Scala code is significantly
faster than the naive counterpart, by 40.3$\times$ (excluding Q8 whose performance is prohibitively slow in the unoptimized Scala version). This shows that extensive optimization of the Scala code is essential
in order to achieve high performance. However,
we also observe that the performance of the optimized Scala program cannot compete with that of the optimized C code, 
and is on average 10$\times$ slower. 
Profiling information gathered with the \textit{perf}\footnote{https://perf.wiki.kernel.org/index.php/Main\_Page.} 
profiling tool of Linux reveals the following three reasons for this behavior: 
\begin{inparaenum}
\item[(a)] Scala causes an increase to branch mispredictions, by 1.8$\times$ compared to the branch mispredictions in C,
\item[(b)] The percentage of LLC misses is 1.3$\times$ to 2.4$\times$ those in Scala, and
more importantly,
\item[(c)] The number of CPU instructions executed in Scala is 6.2$\times$ the one executed by the C binary.
\end{inparaenum}
Of course, these inefficiencies are to a great part due to the
Java Virtual Machine and not specific to Scala.
Note that the optimized Scala program is competitive to \timesten
(especially for non-join-intensive queries, e.g.\ queries that have less than two joins):
for 19 out of 22 queries, Scala outperforms the commercial
\timesten system. This is because we remove all abstractions that incur
significant overhead for Scala. For example, the performance of Q18, which
builds a large hash map, is improved by 40$\times$ when applying the
\datastructure specialization provided by \compiler.

\subsection{Impact of Individual Compiler Optimizations}
\label{lego-eval-four}
\begin{figure*}
\hspace{-0.5cm}
\includegraphics[width=1.05\textwidth]{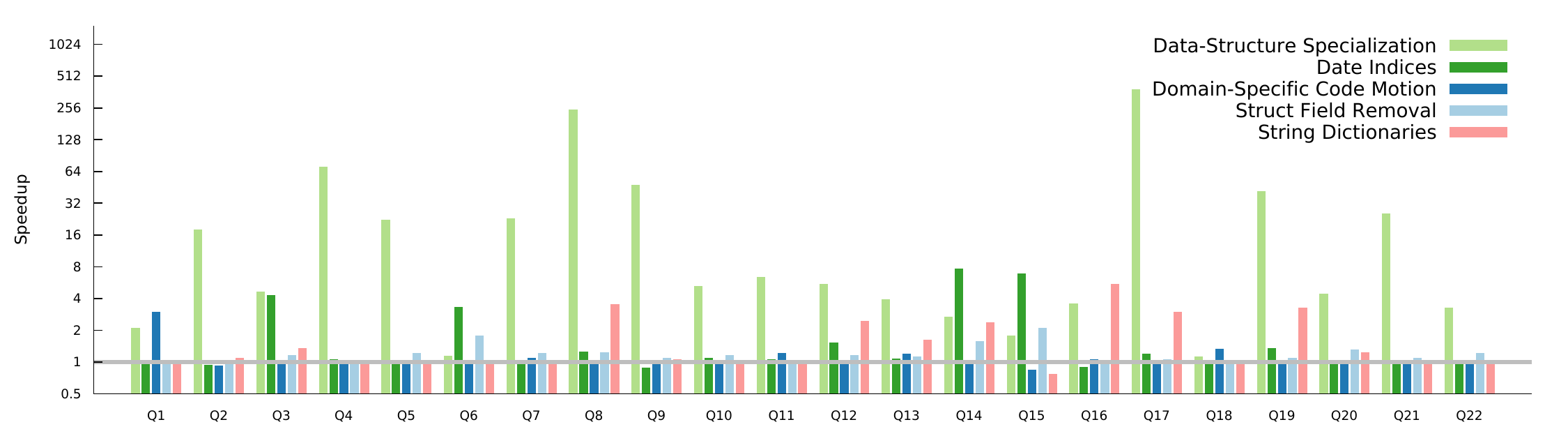}
\caption{Impact of different optimizations on query execution time. The baseline
is an engine that does not apply this optimization.} 
\label{fig:optimizations}
\end{figure*}

In this section, we provide additional information about the performance 
improvement expected when applying one of the compiler optimizations of \systemname{}. 
These results, illustrated in Figure~\ref{fig:optimizations}, aim to demonstrate that 
significant optimization opportunities have been ignored by existing compilation 
techniques that handle only queries.

To begin with, we can see in this figure that the most important 
transformation in \systemname{} is the \datastructure specialization (presented in 
Sections~\ref{subsec:indexingAndPartitioning} and \ref{subsubsec:maps-to-native-arrs}). 
This form of optimization is not provided by existing approaches, 
as data structures are typically precompiled in existing database systems. We see that, in 
general, when \datastructure specialization is applied, the generated code has an 
average performance improvement of $30\times$ (excluding queries Q8 and Q17
where the partitioning optimization facilitates skipping the processing of the majority 
of the tuples of the input relations).  
Moreover, we note that the performance improvement is not directly dependent 
on the number of join operators or input relations in the query plan. 
For example, join-intensive queries such as Q5, Q7, Q8, Q9, Q21 obtain a speedup of at least 22$\times$ when applying 
this optimization. However, the single-join queries Q4 and Q19 also
receive similar performance benefit to that of multi-way join queries.
This is because query plans may filter input data early
on, thus reducing the need for efficient join data structures. Thus, selectivity information and analysis of
the whole query plan are essential for analyzing the potential performance benefit of this optimization. Note that, 
for similar reasons, date indices (Section~\ref{subsec:date-indices}) allow to avoid unnecessary tuple 
processing and thus lead to increased performance for a number of queries.

For the domain-specific code motion and the removal of unused relational attributes optimizations,
we observe that they both improve performance, by $1.12\times$ and $1.21\times$, respectively
on average for all \tpch queries. This improvement is not be as pronounced as that
of other optimizations of \systemname{} (like the one presented above). However, it is important
to note that they both significantly reduce memory pressure, thus allowing the 
freed memory space to be used for other optimizations, such as the partitioning specialization,
which in turn provide significant performance improvement. Nevertheless, these two optimizations
-- which are not provided by previous approaches (since they depend on query-specific knowledge) --
can provide considerable performance improvement by themselves for some queries. For example, for
\tpch Q1, performing domain-specific code motion leads to a speedup of $2.96\times$, 
while the removal of unused attributes gives a speedup of $2.11\times$ for Q15. 

Moreover, the same figure evaluates the speedup we gain when using string dictionaries. 
We observe that for the \tpch queries that perform a number of expensive string operations, 
using string dictionaries always leads to improved query execution 
performance: this speedup ranges from 1.06$\times$ to 5.5$\times$, with an average speedup of 
2.41$\times$\footnote{The 
rest of the \tpch queries (Q1, Q4, Q5, Q6, Q7, Q10, Q11, Q15, Q18, Q21, Q22) either did not have any string operation or the number of these operations on 
those queries was negligible.}. We also note that the speedup this optimization provides depends on the characteristics 
of the query. More specifically, if the query contains string operations on scan operators, 
as is the case with Q8, Q12, Q13, Q16, Q17, and Q19, then this optimization provides a greater performance 
improvement than when string operations occur in operators appearing later in the query plan. 
This is because, \tpch queries typically filter out more tuples as 
more operators are applied in the query plan. Stated otherwise, operators being executed in 
the last stages of the query plan do not process as many tuples as scan operators. Thus, the 
impact of string operations is more pronounced when such operations take place 
in scan operators.

It is important to note that using string dictionaries comes at a price. First, this optimization increases the 
loading time of the query. Second, this optimization requires keeping a dictionary between strings and integer 
values, a design choice which requires additional memory. This may, in turn, increase memory pressure, possibly causing
a drop in performance. However, it is our observation that, based on the individual use case and data characteristics 
(e.g. number of distinct values of a string attribute), developers can easily detect whether it makes sense
performance-wise to use this optimization or not. We also present a more detailed analysis of the memory consumption required by the overall \legobase system later in this section.

Then, the benefit of applying operator inlining (not shown) varies significantly between different \tpch
queries and ranges from a speedup of $1.07\times$ up to $19.5\times$, with an average performance
improvement of $3.96\times$. The speedup gained from applying this optimization depends on the 
complexity of the execution path of a query.  This is a hard metric to visualize, 
as the improvement depends not only on \textit{how many} operators are used but also 
on their type, their position in the overall query plan and how much each of them affects 
branch prediction and cache locality. For instance, queries Q5, Q7 and Q9
have the same number of operators, but the performance improvement gained varies 
significantly, by 10.4$\times$, 1.4$\times$ and 7.5$\times$, respectively. 
In addition, it is our observation that the benefit of inlining depends on which operators 
are being inlined. This is an
important observation, as for very large queries, the compiler may have to choose
which operators to inline (e.g.\ to avoid the code not fitting in the
instruction cache). In general, when such cases appear, we believe that the compiler
framework should merit inlining joins instead of simpler operators (e.g.\ scans
or aggregations).

Finally, for the column layout optimization (not shown), the performance improvement is
generally proportional to the percentage of attributes in the input relations that 
are actually used.  This is expected as the benefits of the column layout
are evident when this layout can ``skip'' loading into memory a number of unused 
attributes, thus significantly reducing cache misses. Synthetic queries on 
\mbox{TPC-H} data referencing 100\% of the attributes show that, in this case, the 
column layout actually yields no benefit, and it is slightly worse than the row 
layout. Actual \tpch queries reference 24\% - 68\% of the attributes and, for this range, 
the optimization gives a 2.5$\times$ to 1.05$\times$ improvement, which degrades as more
attributes are referenced. 
\begin{figure}
\centering 
\includegraphics[scale=0.9]{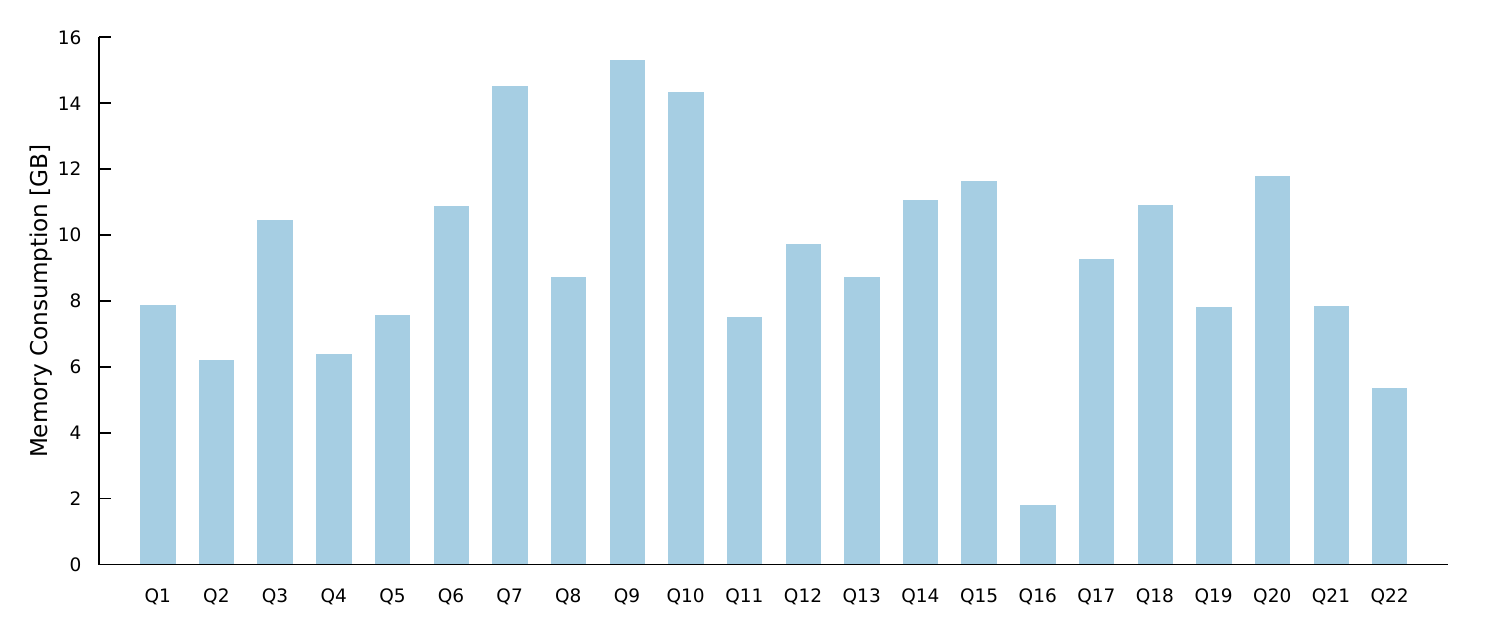}
\caption{Memory consumption of \systemname{}(Opt/C) for the \tpch{} queries.}
\label{fig:mem}
\end{figure}
\subsection{Memory Consumption and Overhead on Input Data Loading}
\label{lego-eval-five}
Figure~\ref{fig:mem} shows the memory consumption for all \tpch queries. We use Valgrind 
for memory profiling as well as a custom memory profiler, while the JVM is always first 
warmed up. We make the following observations. 
First, the allocated memory is at most twice the size of the input data for all \tpch 
queries (16GB of memory for 8GB of input data for all relations), while the \textit{average} 
memory consumption is only 1.16$\times$ the total size of the input relations. 
These results suggest that our approach is usable in practice, as even for complicated, 
multi-way join queries the memory used remains relatively small. The additional memory 
requirements come as a result of the fact that \legobase{} aggressively repartitions input 
data in many different ways (as was described in Section~\ref{subsec:dsopt}) in order to achieve 
optimal performance. Second, when all optimizations are enabled, \legobase{} consumes 
less memory than the total size of the input data, for a number of queries. For instance, Q16
consumes merely 2GB for all necessary data structures. This behavior is a result of removing unused 
attributes from relational tables when executing a query. In general, it is our observation that memory 
consumption grows linearly with the scaling factor of the \tpch workload. 

\begin{figure}
\centering
\includegraphics[scale=0.75]{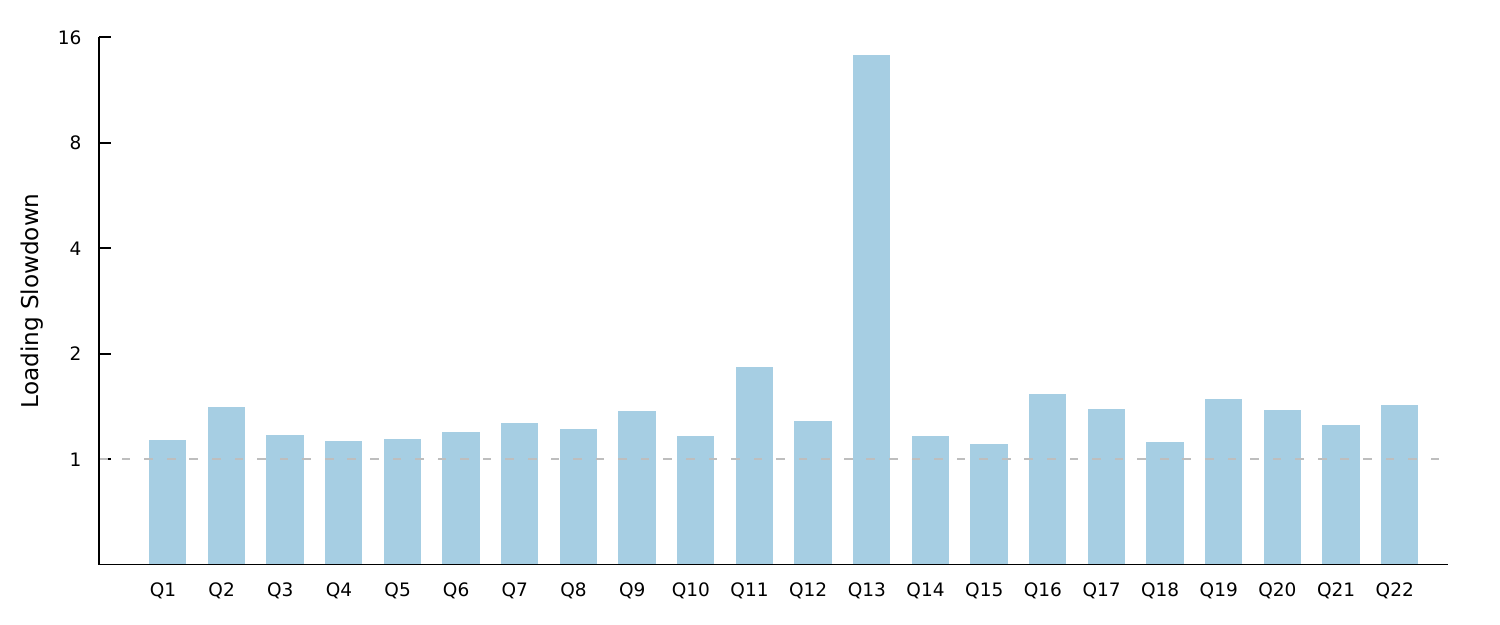}
\caption{Slowdown of input data loading occurring from applying all \systemname{} optimizations to the C programs of the TPC-H workload (scaling factor 8).}
\label{fig:loading}
\end{figure}

In addition, we have mentioned before that applying our compiler optimizations can lead to an increase in the loading time of the input data. Figure~\ref{fig:loading} presents the total slowdown on input data loading when applying all \legobase{} optimizations to the generated C programs (\systemname{}(Opt/C)) compared to the loading time of the unoptimized C programs (\systemname{}(Naive/C)). We observe that the total time spent on data loading, across all queries and with all optimizations applied, is \textit{not} (excluding Q13 which applies the word-tokenizing string dictionary) more than $1.5\times$ that of the unoptimized, push-style generated C code. This means that while our optimizations lead to significant performance improvement, they do not affect the loading time of input data significantly (there is an average slowdown of 1.88$\times$ including Q13). Based on these observations, as well as the absolute loading times presented in Appendix~\ref{appendix:absolute-exec-times}, we can see that the additional overhead of our optimizations is not prohibitive: it takes in average less than a minute for \systemname{} to load the 8GB \tpch{} dataset, repartition the data, and build all necessary auxiliary data structures for efficient query processing.  

\subsection{Productivity Evaluation}
\label{lego-eval-six}
An important point of this \paper{} is that the performance of query engines can be improved without much programming effort.
Next, we present the productivity/performance evaluation of our system, 
which is summarized in Table~\ref{table:loctrans}.

We observe three things. First, by programming at the high level, we can provide a
fully functional system with a small amount of effort.
Less development time was spent on debugging the system, thus allowing us to focus
on developing new useful optimizations. Development of the \systemname{} 
query engine alongside the domain-specific optimizations required, including 
debugging time, eight months for only two programmers. However, the 
majority of this effort was invested in building the new optimizing compiler 
\compiler (27K LOC) rather than developing the basic, unoptimized, query engine 
itself (1K LOC).

Second, each optimization requires only a few hundred lines of
high-level code to provide significant performance improvement. 
More specifically, for $\approx$3000 LOC in total, {\systemname} is improved by 45.4$\times$ compared
to the performance of \timesten, as we described previously. 
Source-to-source compilation is critical to achieving
this behavior, as the combined size of the operators and optimizations of {\systemname}
is around 40 times less than the generated code size for all 22 \tpch queries written in C.

Finally, from Table~\ref{table:loctrans} it becomes clear that new transformations 
can be introduced in \compiler with relative small programming effort. This becomes
evident when one considers complicated transformations like those of Automatic Index 
Inference and Horizontal Fusion\footnote{To perform a decent loop fusion, the
short-cut deforestation is not sufficient. Such techniques only provide 
\textit{vertical} loop fusion, in which one loop uses the result produced by another loop. 
However, in order to perform further optimizations one requires to perform 
\textit{horizontal} loop fusion, in which different loops iterating over the same 
range are fused into one loop~\cite{alg-opt-oop-ql,Goldberg:1984:SP:800055.802021}. 
A decent loop fusion is still an open topic in the PL 
community~\cite{Svenningsson:2002:SFA:581478.581491,Coutts07streamfusion,foldr-fusion-1}.} 
which can both be coded
for merely $\approx$500 lines of code. 
We also observe that around half of the code-base required to be introduced in \compiler
concerns converting Scala code to C.
However, this is a na\"ive task to be performed by \compiler developers, as it usually results 
in a one-to-one translation between Scala and C constructs. More importantly, this is a task that is required to be performed only \textit{once} for each language construct, and it needs to 
be extended \textit{only} as new constructs are introduced in \compiler (e.g. those required for custom 
data types and operations on those types).
\begin{table}[t!]
\centering
\begin{tabular}{l r}
\toprule
Data-Structure Partitioning & 505 \\
Automatic Inference of Date Indices & 318 \\ 
Memory Allocation Hoisting & 186 \\ 
Column Store Transformer & 184 \\ 
Constant-Size Array to Local Vars & 125 \\ 
Flattening Nested Structs & 118 \\ 
Horizontal Fusion & 152 \\
\midrule
Scala Constructs to C Transformer & 793 \\
Scala Collections to GLib Transformer & 411 \\ 
Scala Scanner Class to mmap Transformer & 90 \\ 
\midrule
Other miscellaneous optimizations & $\approx200$\\
\midrule
Total & 3082 \\
\bottomrule
\end{tabular}
\caption{Lines of code of several transformations in \systemname with the \compiler compiler.}
\label{table:loctrans}
\end{table}

\subsection{Compilation Overheads}
\label{lego-eval-seven}
\begin{figure}[t]
\centering
\includegraphics[scale=0.75]{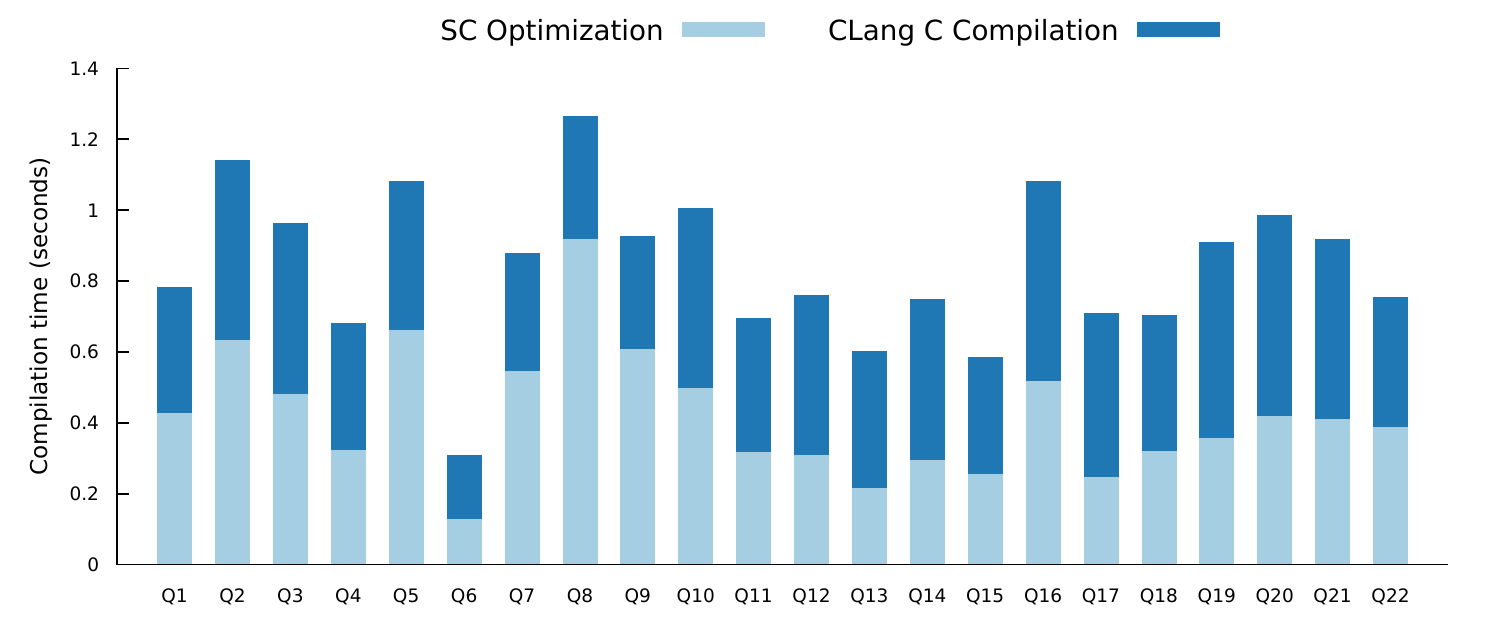}
\caption{Compilation time (in seconds) of all \systemname{}(Opt/C) programs.}
\label{compilation-overheads}
\end{figure}

Finally, we analyze the compilation time for the optimized C programs of \systemname{}(Opt/C) for
all \tpch queries. Our results are presented in Figure~\ref{compilation-overheads}, 
where the y-axis corresponds to the time to (a) optimize a query and generate the C code with \compiler{}, and, (b) the time 
\clang requires before producing the final C executable.

We see that, in general, all \tpch queries require less than 1.2 seconds of
compilation time. We argue 
that this is an acceptable compilation overhead, especially
for analytical queries like those in \tpch that are typically known in advance
and which process huge amounts of data. In this case, a compilation overhead of
some seconds is negligible compared to the total execution time.  This result proves
that our approach is usable in practice for quickly compiling \textit{entire} query 
engines written using high-level programming languages.
To achieve these results, special effort was made so that the \compiler{} compiler can quickly optimize input
programs. More specifically, our progressive lowering approach allows for quick application of optimizations,
as most of our optimizations operate on a relatively small number of language constructs, thus making it easy 
to quickly detect which parts of the input program should be modified at each transformation step, while the 
rest of them can be quickly skipped. In addition, we observe that the CLang C compilation time can be 
significant. This is because, by applying all the domain-specific optimizations of \systemname{} to an input 
query, we get an increase in the total program size that CLang receives from \compiler{}.

Finally, we note that if we generate Scala code instead of C, then the time required for 
compiling the final optimized Scala programs is 7.2$\times$ that of compiling the C programs 
with LLVM. To some extent this is expected as calling the Scala compiler is a heavyweight
process: for every query compiled there is significant startup overhead for loading the 
necessary Scala and Java libraries. By just optimizing a Scala program using optimizations 
written in the same level of abstraction, our architecture allows us to avoid these overheads, 
providing a much more lightweight compilation process.
\section{Related Work} \label{sec:related}
We outline related work in five areas:
\begin{inparaenum}
\item[(a)] Previous compilation approaches,
\item[(b)] Frameworks for applying intra-opera\-tor optimizations, 
\item[(c)] Orthogonal techniques to speed up query processing,
\item[(d)] a brief summary of work on Domain Specific Compilation in the Programming, and, finally,
\item[(e)] a comparison with a previous realization of the abstraction without regret vision
Languages (PL) community, a field of study that closely relates to ours.
\end{inparaenum}
We briefly discuss these areas below.
\\

\noindent \textbf{Previous Compilation Approaches.}
Historically, System R~\cite{systemR} first proposed code generation for query
optimization. However, the Volcano iterator model eventually dominated over
compilation, since code generation was very expensive to maintain. The
Daytona system~\cite{daytona} revisited compilation in the late nineties;
however, it heavily relied on the operating system for functionality that is
traditionally provided by the DBMS itself, like buffering.

The shift towards pure \textit{in-memory} computation in databa\-ses, evident in
the space of data analytics and transaction processing has lead developers to revisit compilation. The reason is that, as more and more
data is put in memory, query performance is increasingly determined by the
effective thr\-oughput of the CPU.  In this context, compilation strategies aim
to remove unnecessary CPU overhead. Examples of industrial systems in the area since the mid-2000s include SAP HANA~\cite{SAPHANA},
VoltDB~\cite{VoltDB,hstore-compilation} and Oracle's TimesTen~\cite{TimesTen}. In the academic 
context, interest in query compilation has also been renewed since 2009 and continues to grow
\cite{mohan-compilation,sql-generation-zhao,DBLP:journals/pvldb/AhmadK09,ferry-2,krikellas,neumann11,koch,DBLP:journals/corr/CrottyGDKCZ14,Nagel:2014:CGE:2732977.2732984,DBLP:journals/debu/ViglasBN14,spark-sql,crotty-udf-centric,goel-scalable-analytics}. We briefly discuss some of these systems next. 

Rao et al. propose to remove the overhead of virtual functions in the Volcano
iterator model by using a compiled execution engine built on top of the Java
Virtual Machine (JVM)~\cite{mohan-compilation}. The HIQUE system takes a step
further and completely eliminates the Volcano iterator model in the generated 
code~\cite{krikellas}. It does so by translating the algebraic representation to
C++ code using templates. In addition, Zane et al. have
shown how compilation can also be used to additionally improve operator
internals~\cite{sql-generation-zhao}.

The query compiler of the \hyper database system also uses query compilation, as 
described in~\cite{neumann11}. This work targets minimizing the CPU overhead of the
Volcano operator model while maintaining low compilation times. The authors use
a mixed LLVM/C++ execution engine where the algebraic representation of the
operators is first translated to low-level LLVM code, while the complex part of
the database (e.g.\  management of data structures and memory allocation) is
still \textit{precompiled} C++ code called periodically from the LLVM code
whenever needed. Two basic optimizations are presented: operator inlining and
reversing the data flow (to a push engine). 

All these works aim to improve database systems by removing unnecessary
abstraction overheads. However, these \textit{temp\-late-based} approaches
require writing low-level code which is hard to maintain and extend.  This fact
significantly limits their applicability. 
In contrast, our approach  
advocates a new methodology for programming query engines where the query engine 
and its optimizations are written in a \textit{high-level} language. This provides 
a programmer-friendly way to express optimizations and allows extending the scope
of optimization to cover the whole query engine. Finally, in contrast to previous work, we separate the optimization
and code generation phases. Even though~\cite{neumann11} argues that
optimizations should happen completely before code generation (e.g.\ in the
algebraic representation), there exist many optimization opportunities that
occur only \textit{after} one considers the complete generated code, e.g.\ after
operator inlining. Our compiler can detect such optimizations, thus providing
additional performance improvement over existing techniques. 
\\\\
\noindent \textbf{Frameworks for applying intra-opera\-tor optimizations.}
There has recently been extensive work on how to specialize the code of query
operators in a systematic way by using an approach called
Micro-Specialization~\cite{microspecialization2,microspecialization1,microspecialization3}.
In this line of work, the authors propose a framework to encode DBMS-specific
intra-operator optimizations, like unrolling loops and removing if conditions,
as precompiled templates in an extensible way. All these optimizations are
performed by default by the \compiler compiler in {\systemname}. However, in contrast
to our work, there are two main limitations in Micro-Specialization. First, the
low-level nature of the approach makes the development process very
time-consuming: it can take days to code a single intra-operator
optimization~\cite{microspecialization2}.  Such optimizations are very
fine-grained, and it should be possible to implement them quickly: for the same
amount of time we are able to provide much more coarse-grained optimizations in
{\systemname}. Second, the optimizations are limited to those that can be
statically determined by examining the DBMS code and cannot be changed at
runtime.  Our architecture maintains all the benefits of Micro-Specialization,
while it is not affected by these two limitations.
\\

\noindent \textbf{Techniques to speed up query processing.}
There are many works that aim to speed up query processing in general,
by focusing 
on improving the way data is processed rather than individual
operators. Examples of such works include block-wise processing~\cite{blockprocessing}, 
vectorized execution~\cite{vectorizevscompile}, compression techniques to provide
constant-time query processing~\cite{constanttimequeryprocessing} or a combination
of the above along with a column-oriented data layout~\cite{monetdb-mammals}.
These approaches are orthogonal to this work as our framework
provides a high-level framework for encoding \textit{all} such
optimizations in a user-friendly way (e.g.\ we present the transition from row
to column data layout in Section~\ref{subsec:layout}).
\\

\noindent \textbf{Domain-specific compilation}, which admits domain-specific optimizations, is
a topic of great current interest in multiple research communities. 
Once one limits the domain or language, program analysis can
be more successful. More powerful and global transformations then become
possible, yielding speedups that cannot be expected from classical compilers
for general purpose languages. To this end, multiple frameworks and research prototypes
\cite{hudak-dsl,khepera,annotated-dsl,telescoping,lms,jet2012,delite,yinyang,truffle-dsl}, 
have been proposed to easily introduce and perform domain-specific compilation and optimization
for systems. Of interest is the observation that domain-specificity has already benefited query 
optimization tremendously: Relational algebra is a domain-specific language, and yields
readily available associativity properties that are the foundation
of query optimization. Optimizing compilers can combine the performance benefits
of classical interpretation-based query engines with the benefits of
abstraction and indirection elimination by compilers. Finally, OCAS~\cite{ocas} has been developed within the context of domain-specific synthesis and attempts to automatically generate optimized out-of-core algorithms for a particular target memory hierarchy.
\\

\noindent \textbf{Previous realization of the abstraction without regret vision.} We have previously realized this vision for query engines 
in~\cite{legobase}. In this \paper, we provide a from scratch implementation of the vision using a \textit{new} optimizing compiler, 
called \compiler{}, developed specifically to handle the optimization needs of large-scale software systems. We also present a detailed analysis
of the compiler interfaces of \compiler{} as well as a significantly more thorough list of the optimizations supported by the \systemname{} 
system in order to demonstrate the ease-of-use of our compiler framework for optimizing database components that differ significantly in 
granularity and scope of operations. Finally, we provide a more extensive evaluation where, along with a renewed analysis of the previous 
results, we also evaluate three additional query engine configurations. We do so in order to compare as fairly as possible the performance of our 
system with that of previous work. 
\section{Conclusions}
\label{sec:conclusions}
{\systemname} is a new analytical database system currently under development.
In this \paper{}, we presented the current prototype of the query execution subsystem of {\systemname}.
Our approach suggests using high-level programming languages for DBMS development without having to pay 
the associated abstraction penalty. 
This vision has been previously called \textit{abstraction without regret}. The 
key technique to admit this productivity/efficiency combination is to apply generative programming and source-to-source 
compile the high-level Scala code to efficient low-level C code. 
We demonstrate how state-of-the-art compiler technology
allows developers to express database-specific optimizations naturally at a high level as a library and use it to optimize the \textit{entire} query engine. 
In \systemname{}, programmers need to develop just a few hundred lines of \textit{high-level} code
to implement techniques and optimizations that result in significant performance improvement. All these properties are very hard to achieve with existing compilers that handle \textit{only} queries and which are based on template expansion. 
Our experiments show that {\systemname} significantly
outperforms both a commercial in-memory database system as well as an existing query compiler.

\appendix
\section*{APPENDIX}
\setcounter{section}{0}
\section{Absolute Execution Times}
\label{appendix:absolute-exec-times}
For completeness, the following tables present the  {\em absolute performance results}\/ of all evaluated systems and metrics in this \paper{}.
\begin{table}[h!]
\begin{small}
\setlength{\tabcolsep}{2pt}
\begin{tabular}[b]{p{2.8cm} r@{~~~}r@{~~}r@{~~}r@{~~}r@{~~}r@{~}r@{~~}r@{~}r@{~~}r@{~~}r@{~~}r}
\toprule
System & Q1 & Q2 & Q3 & Q4 & Q5 & Q6 & Q7 & Q8 & Q9 & Q10 & Q11\\
\midrule
DBX&1790&396&1528&960&19879&882&969&2172&3346&985&461\\
Compiler of HyPer&779&43&892&622&338&198&798&493&2139&565&102\\
\systemname{} \mbox{(Naive/C) -- LLVM} &3140&755&5232&10742&3627&357&2901&23161&26203&3836&409\\
\systemname{} \mbox{(Naive/C) -- GCC}&3140&801&5204&10624&3652&423&2949&19961&25884&3966&445\\
\systemname{} \mbox{(Naive/Scala)}&3972&6910&11118&30103&10307&874&114677&72587&137369&20353&1958\\
\systemname{}\mbox{(TPC-H/C)}&593&55&767&445&440&199&975&2871&2387&546&98\\
\systemname{}(StrDict/C)&592&47&759&402&439&197&781&346&2027&544&103\\
\systemname{}(Opt/C)&426&42&110&134&126&47&104&18&530&439&49\\
\systemname{}(Opt/Scala)&2174&871&352&306&413&356&9496&104&2296&775&197\\
\bottomrule
\end{tabular}
\\\hspace{1cm}\\
\begin{tabular}[b]{p{2.8cm} r@{~~}r@{~~}r@{~~}r@{~~~}r@{~~}r@{~~~}r@{~~}r@{~~}r@{~~~}r@{~~}r@{~~}r}
\toprule
System & Q12 & Q13 & Q14 & Q15 & Q16 & Q17 & Q18 & Q19 & Q20 & Q21 & Q22\\
\midrule
DBX&881&13593&823&578&12793&1224&4535&6432&744&1977&447\\
Compiler of HyPer & 485&2333&197&229&590&490&3682&1421&277&1321&212\\
\systemname{} \mbox{(Naive/C) -- LLVM} &3037&12794&1289&889&16362&18893&4135&2810&974&11648&1187\\
\systemname{} \mbox{(Naive/C) -- GCC}&3286&13149&1398&899&16159&18410&4174&4460&1055&11848&1396\\
\systemname{} \mbox{(Naive/Scala)}&21735&33403&5163&2093&10568&86953&15798&4470&5301&61712&4207\\
\systemname{}\mbox{(TPC-H/C)}&891&5106&244&550&2774&513&2725&2020&370&1992&453\\
\systemname{}(StrDict/C)&688&910&204&535&702&445&2735&1222&370&1706&333\\
\systemname{}(Opt/C)&120&516&11&46&695&11&133&19&130&388&79\\
\systemname{}(Opt/Scala)&604&7743&136&234&2341&274&355&125&700&955&406\\
\bottomrule
\end{tabular}
\end{small}
\caption{Execution times (in milliseconds) of Figure~\ref{volcano} and Figure~\ref{neumann}. The various configurations of \systemname{} are explained in more detail in Table~\ref{systems} of this article.}
\label{table1}
\end{table}

\begin{table}[h!]
\begin{small}
\setlength{\tabcolsep}{5pt}
\begin{tabular}[b]{p{2.6cm} r@{~~~~~~}r@{~~}r@{~~}r@{~~}r@{~~~~~~}r@{~~}r@{~~}r@{~~}r@{~~~~}r@{~~~}r@{~~}r}
\toprule
& Q1 & Q2 & Q3 & Q4 & Q5 & Q6 & Q7 & Q8 & Q9 & Q10 & Q11\\
\midrule
\systemname{} \mbox{(Naive/C) -- LLVM}&3140&755&5232&10742&3627&357&2901&23161&26203&3836&409\\
+Struct Field Removal&3104&734&4480&10346&2983&202&2394&18707&24125&3323&403\\
+Domain-Specific Code Motion&1047&794&4283&10435&2902&196&2203&18507&23854&3177&332\\
+Data-Structure Specialization&497&44&918&148&130&172&96&75&498&610&52\\
+Date Indices&497&47&213&140&131&52&96&60&568&553&49\\
+String Dictionaries&497&43&158&140&130&51&94&17&533&552&47\\
\systemname{}(Opt/C)&426&42&110&134&126&47&104&18&530&439&49\\
\bottomrule
\end{tabular}
\end{small}
\end{table}

\begin{table}[h!]
\begin{small}
\setlength{\tabcolsep}{5pt}
\begin{tabular}[b]{p{2.6cm} r@{~~}r@{~~}r@{~~~~~}r@{~~}r@{~~}r@{~~}r@{~~~~}r@{~~~~}r@{~~}r@{~~}r@{~~}r}
\toprule
& Q12 & Q13 & Q14 & Q15 & Q16 & Q17 & Q18 & Q19 & Q20 & Q21 & Q22\\
\midrule
\systemname{} \mbox{(Naive/C) -- LLVM}&3037&12794&1289&889&16362&18893&4135&2810&974&11648&1187\\
+Struct Field Removal&2631&11291&812&420&16068&17953&4070&2550&736&10647&970\\
+Domain-Specific Code Motion&2553&9415&786&495&15251&18063&3050&2568&742&10386&985\\
+Data-Structure Specialization&467&2389&291&277&4243&47&2709&62&168&410&300\\
+Date Indices&308&2233&38&40&4737&39&2718&46&168&392&291\\
+String Dictionaries&125&1379&16&52&860&13&2730&20&136&389&299\\
\systemname{}(Opt/C)&120&516&11&46&695&11&133&19&130&388&79\\
\bottomrule
\end{tabular}
\end{small}
\caption{Execution times (in milliseconds) of \tpch queries with individual optimizations applied (as shown in Figure~\ref{fig:optimizations} of this \paper{}). Each listed optimization is applied additionally to the set of optimizations applied in the system specified above it.}
\label{table2}
\end{table}

\begin{table}[h!]
\centering
\begin{small}
\setlength{\tabcolsep}{5pt}
\begin{tabular}[b]{l r@{~~}r@{~~}r@{~~~}r@{~~}r@{~~}r@{~~}r@{~~}r@{~~}r@{~~}r@{~~}r@{~~}r}
\toprule
 & Q1 & Q2 & Q3 & Q4 & Q5 & Q6 & Q7 & Q8 & Q9 & Q10 & Q11\\
\midrule
Memory Consumption& 7.86 &	6.20 &	10.45 &	6.39 &	7.56 &	10.88 &	14.51 &	8.72 &	15.30 &	14.35&	7.53\\
Loading Time (No opt.) & 34 &	7 &	44 &	42 &	43 &	33 &	43 &	46 &	45 & 	44 &	5\\
Loading Time (All opt.) & 38 &	10 &	52 &	47 &	49 &	39 &	55 &	56 &	61 &	52 &	10\\
\compiler{} Optimization & 429 & 633 & 482 & 323 & 663 & 128 & 547 & 918 & 608 & 498 & 317\\
CLang C Compilation & 354 & 509 & 482 & 359 & 418 & 179 & 332 & 346 & 320 & 507 & 378 \\
\bottomrule
\end{tabular}
\end{small}
\end{table}
\begin{table}[h!]
\centering
\begin{small}
\setlength{\tabcolsep}{5pt}
\begin{tabular}[b]{l r@{~~}r@{~~}r@{~~}r@{~~}r@{~~~}r@{~~}r@{~~}r@{~~}r@{~~~}r@{~~}r@{~~}r}
\toprule
& Q12 & Q13 & Q14 & Q15 & Q16 & Q17 & Q18 & Q19 & Q20 & Q21 & Q22\\
\midrule
Memory Consumption& 	9.73 &	8.72 &	11.06 &	11.64 &	1.81 &	9.26 &	10.92 &	7.81 &	11.77 &	7.86 &	5.36\\
Loading Time (No opt.) & 	41 &	9	&36&	34&	7	&34	&42&	35&	38&	41&	9\\
Loading Time (All opt.) & 53 &	135 &	42 &	38 &	10 &	47 &	47 &	52 &	53 &	52 &	13\\
\compiler{} Optimization & 310 & 215 & 295 & 255 & 518 & 248 & 321 & 357 & 420 & 411 & 389\\
CLang C Compilation & 449 & 386 & 454 & 329 & 563 & 461 & 382 &  552 & 566 & 507 & 365\\
\bottomrule
\end{tabular}
\end{small}
\caption{Memory consumption in GB, input data loading time in seconds, and optimization/compilation 
time in milliseconds as shown in Figure~\ref{fig:mem}, Figure~\ref{fig:loading}, and, 
Figure~\ref{compilation-overheads} of this \paper{}, respectively.}
\label{table3}
\end{table}
\newpage
\section{Code Snippet for the Partitioning Transformer}
\label{code:partitioning}
\lstset{style=scala_unnumbered}
Next, we present a portion of the data partitioning transformation, an explanation of which was given
in Section~\ref{subsec:indexingAndPartitioning}. This code corresponds to the join processing for equi-joins (and not the actual partitioning of input data), but similar rules are employed for other join types as well. The aim of this snippet is to demonstrate the ease-of-use of the \compiler compiler. 
\begin{lstlisting}
/* A transformer for partitioning and indexing MultiMap data-structures. As a result, this 
    transformation converts MultiMap operations to native Array operations. */
class HashTablePartitioning extends RuleBasedTransformer {
   val allMaps = mutable.Set[Any]()
   var currentWhileLoop: While = _
   
  /* ---- ANALYSIS PHASE ---- */
  /* Gathers all MultiMap symbols which are holding a record as their value */
  analysis += statement { 
    case sym -> code"new MultiMap[_, $v]" if isRecord(v) => allMaps += sym
  }
  /* Keeps the closest while loop in scope (used in the next analysis rule)*/
  analysis += rule {
    case whileLoop @ code"while($cond) $body" => currentWhileLoop = whileLoop
  }
  /* Maintain necessary information for the left relation */
  analysis += rule {
    case code"($mm: MultiMap[_,_]).addBinding(struct_field($struct, $fieldName),$value)" =>
      mm.attributes("addBindingLoop") = currentWhileLoop
  }
  /* Maintain necessary information for the right relation */
  analysis += rule {
    case code"($mm : MultiMap[_, _]).get(struct_field($struct, $fieldName))" =>
      mm.attributes("partitioningStruct") = struct
      mm.attributes("partitioningFieldName") = fieldName
  }
  
  /* ---- REWRITING PHASE ---- */
  def shouldBePartitioned(mm: Multimap[Any, Any]) =  allMaps.contains(mm)
  	
  /* If the left relation should be partitioned, then remove the `addBinding` and `get` 
      function calls for this multimap, as well as any related loops. Notice that there is 
      no need to remove the multimap itself, as DCE will do so once all of its dependent 
      operations have been removed.*/
  rewrite += remove {
    case code"($mm: MultiMap[Any, Any]).addBinding($elem, $value)" if 	
      shouldBePartitioned(mm) =>
  }
  rewrite += remove {
    case code"($mm: MultiMap[Any, Any]).get($elem)" if shouldBePartitioned(mm) =>
  }
  rewrite += remove {
    case node @ code"while($cond) $body" if allMaps.exists({
      case mm => shouldBePartitioned(mm) && mm.attributes("addBindingLoop") == node
    }) =>
  }
  /* If a MultiMap should be partitioned, instead of the construction of that MultiMap 
      object, use the corresponding partitioned array constructed during data-loading. 
      This can be an 1D or 2D array, depending on the properties and relationships of the 
      primary and foreign keys of that table (described in Section 3.2.1 in more detail). */
  rewrite += statement {
    case sym -> (code"new MultiMap[_, _]") if shouldBePartitioned(sym) => 
      getPartitionedArray(sym)
  }
  
  /* Rewrites the logic for extracting matching elements of the left relation (initially 
      using the HashMap), inside the loop iterating over the right relation. */
  rewrite += rule {
    case code"($mm:MultiMap[_,_]).get($elem).get.foreach($f)" if shouldBePartitioned(mm) =>{
      val leftArray = transformed(mm)
      val hashElem = struct_field(mm.attributes("partitioningStruct"),
					    mm.attributes("partitioningField"))
      val leftBucket = leftArray(hashElem)
      /* In what follows, we iterate over the elements of the bucket, even though the 
          partitioned array may be an 1D-array as discussed in Section 3.1.2. There is 
          another optimization in the pipeline which flattens the for loop of this case. */
      for(e <- leftBucket) {
      	/* Function f corresponds to checking the join condition and creating the join 
            output. This functionality remains the same, thus, we can simply inline the 
            related code here as follows */
        ${f(e)}
      }
    }
  /* For a partitioned relation, there is no need to check for emptiness, due to primary / 
      foreign key relationship. The if (true) is later removed by another optimization. */
  rewrite += rule {
    case code"($mm: MultiMap[Any, Any]).get($elem).nonEmpty" if shouldBePartitioned(mm) => 
      true
  }
}
\end{lstlisting}

{
\bibliographystyle{ACM-Reference-Format-Journals}
\bibliography{references}


 \providecommand\noopsort[1]{}
\begin{thebibliography}{00}


\ifx \showCODEN    \undefined \def \showCODEN     #1{\unskip}     \fi
\ifx \showDOI      \undefined \def \showDOI       #1{{\tt DOI:}\penalty0{#1}\ }
  \fi
\ifx \showISBNx    \undefined \def \showISBNx     #1{\unskip}     \fi
\ifx \showISBNxiii \undefined \def \showISBNxiii  #1{\unskip}     \fi
\ifx \showISSN     \undefined \def \showISSN      #1{\unskip}     \fi
\ifx \showLCCN     \undefined \def \showLCCN      #1{\unskip}     \fi
\ifx \shownote     \undefined \def \shownote      #1{#1}          \fi
\ifx \showarticletitle \undefined \def \showarticletitle #1{#1}   \fi
\ifx \showURL      \undefined \def \showURL       #1{#1}          \fi

\bibitem[\protect\citeauthoryear{Abadi, Madden, and Hachem}{Abadi
  et~al\mbox{.}}{2008}]%
        {rowcolumn1}
{Daniel~J. Abadi}, {Samuel~R. Madden}, {and} {Nabil Hachem}. 2008.
\newblock \showarticletitle{{Column-Stores vs. Row-Stores: How Different Are
  They Really?}}. In {\em the 2008 ACM SIGMOD International Conference on
  Management of Data} {\em (SIGMOD '08)}. ACM, New York, NY, USA, 967--980.
\newblock
\showISBNx{978-1-60558-102-6}
\showDOI{%
\url{http://dx.doi.org/10.1145/1376616.1376712}}


\bibitem[\protect\citeauthoryear{Ackermann, Jovanovic, Rompf, and
  Odersky}{Ackermann et~al\mbox{.}}{2012}]%
        {jet2012}
{Stefan Ackermann}, {Vojin Jovanovic}, {Tiark Rompf}, {and} {Martin Odersky}.
  2012.
\newblock \showarticletitle{Jet: {A}n {E}mbedded {DSL} for {H}igh {P}erformance
  {B}ig {D}ata {P}rocessing}. In {\em International {W}orkshop on {E}nd-to-end
  {M}anagement of {B}ig {D}ata ({B}ig{D}ata 2012)}.
\newblock
\showURL{%
\url{http://infoscience.epfl.ch/record/181673/files/paper.pdf}}


\bibitem[\protect\citeauthoryear{Ahmad and Koch}{Ahmad and Koch}{2009}]%
        {DBLP:journals/pvldb/AhmadK09}
{Yanif Ahmad} {and} {Christoph Koch}. 2009.
\newblock \showarticletitle{DBToaster: A SQL Compiler for High-performance
  Delta Processing in Main-Memory Databases}.
\newblock {\em Proc. VLDB Endow.\/} {2}, 2 (Aug. 2009), 1566--1569.
\newblock
\showISSN{2150-8097}
\showDOI{%
\url{http://dx.doi.org/10.14778/1687553.1687592}}


\bibitem[\protect\citeauthoryear{Ailamaki, DeWitt, Hill, and
  Skounakis}{Ailamaki et~al\mbox{.}}{2001}]%
        {Ailamaki:2001:WRC:645927.672367}
{Anastassia Ailamaki}, {David~J. DeWitt}, {Mark~D. Hill}, {and} {Marios
  Skounakis}. 2001.
\newblock \showarticletitle{Weaving Relations for Cache Performance}. In {\em
  Proceedings of the 27th International Conference on Very Large Data Bases}
  {\em (VLDB\hspace{0.3cm} '01)}. \hspace{0.3cm}Morgan\hspace{0.3cm}
  Kaufmann\hspace{0.3cm} Publishers\hspace{0.3cm} Inc.,
  \hspace{0.3cm}San\hspace{0.3cm} Francisco,\hspace{0.3cm} CA,\hspace{0.3cm}
  USA, 169--180.
\newblock
\showISBNx{1-55860-804-4}
\newblock
\shownote{\\\url{http://research.cs.wisc.edu/multifacet/papers/vldb01_pax.pdf}.}


\bibitem[\protect\citeauthoryear{Armbrust, Xin, Lian, Huai, Liu, Bradley, Meng,
  Kaftan, Franklin, Ghodsi, and Zaharia}{Armbrust et~al\mbox{.}}{2015}]%
        {spark-sql}
{Michael Armbrust}, {Reynold~S. Xin}, {Cheng Lian}, {Yin Huai}, {Davies Liu},
  {Joseph~K. Bradley}, {Xiangrui Meng}, {Tomer Kaftan}, {Michael~J. Franklin},
  {Ali Ghodsi}, {and} {Matei Zaharia}. 2015.
\newblock \showarticletitle{Spark SQL: Relational Data Processing in Spark}
  {\em (SIGMOD '15)}. ACM, New York, NY, USA, 1383--1394.
\newblock
\showISBNx{978-1-4503-2758-9}
\showDOI{%
\url{http://dx.doi.org/10.1145/2723372.2742797}}


\bibitem[\protect\citeauthoryear{Beeri and Kornatzky}{Beeri and
  Kornatzky}{1990}]%
        {alg-opt-oop-ql}
{Catriel Beeri} {and} {Yoram Kornatzky}. 1990.
\newblock \showarticletitle{Algebraic Optimization of Object-Oriented Query
  Languages}.
\newblock In {\em ICDT '90}, {Serge Abiteboul} {and} {Paris~C. Kanellakis}
  (Eds.). Lecture Notes in Computer Science, Vol. 470. Springer Berlin
  Heidelberg, Berlin, Heidelberg, 72--88.
\newblock
\showISBNx{978-3-540-53507-2}
\showDOI{%
\url{http://dx.doi.org/10.1007/3-540-53507-1_71}}


\bibitem[\protect\citeauthoryear{Boncz, Neumann, and Erling}{Boncz
  et~al\mbox{.}}{2014}]%
        {Boncz2014}
{Peter Boncz}, {Thomas Neumann}, {and} {Orri Erling}. 2014.
\newblock {\em TPC-H Analyzed: Hidden Messages and Lessons Learned from an
  Influential Benchmark}.
\newblock Springer International Publishing, Cham, 61--76.
\newblock
\showISBNx{978-3-319-04936-6}
\showDOI{%
\url{http://dx.doi.org/10.1007/978-3-319-04936-6_5}}


\bibitem[\protect\citeauthoryear{Chamberlin, Astrahan, Blasgen, Gray, King,
  Lindsay, Lorie, Mehl, Price, Putzolu, Selinger, Schkolnick, Slutz, Traiger,
  Wade, and Yost}{Chamberlin et~al\mbox{.}}{1981}]%
        {systemR}
{Donald~D. Chamberlin}, {Morton~M. Astrahan}, {Michael~W. Blasgen}, {James~N.
  Gray}, {W.~Frank King}, {Bruce~G. Lindsay}, {Raymond Lorie}, {James~W. Mehl},
  {Thomas~G. Price}, {Franco Putzolu}, {Patricia~Griffiths Selinger}, {Mario
  Schkolnick}, {Donald~R. Slutz}, {Irving~L. Traiger}, {Bradford~W. Wade},
  {and} {Robert~A. Yost}. 1981.
\newblock \showarticletitle{A History and Evaluation of {System R}}.
\newblock {\em Comm. ACM\/} {24}, 10 (1981), 632--646.
\newblock
\showISSN{0001-0782}
\showDOI{%
\url{http://dx.doi.org/10.1145/358769.358784}}


\bibitem[\protect\citeauthoryear{Coutts, Leshchinskiy, and Stewart}{Coutts
  et~al\mbox{.}}{2007}]%
        {Coutts07streamfusion}
{Duncan Coutts}, {Roman Leshchinskiy}, {and} {Don Stewart}. 2007.
\newblock \showarticletitle{Stream Fusion: From Lists to Streams to Nothing at
  All.}. In {\em ICFP} (2007-11-06), {Ralf Hinze} {and} {Norman Ramsey} (Eds.).
  ACM, New York, NY, USA, 315--326.
\newblock
\showISBNx{978-1-59593-815-2}
\showDOI{%
\url{http://dx.doi.org/10.1145/1291151.1291199}}


\bibitem[\protect\citeauthoryear{Crotty, Galakatos, Dursun, Kraska, Binnig,
  Cetintemel, and Zdonik}{Crotty et~al\mbox{.}}{2015}]%
        {crotty-udf-centric}
{Andrew Crotty}, {Alex Galakatos}, {Kayhan Dursun}, {Tim Kraska}, {Carsten
  Binnig}, {Ugur Cetintemel}, {and} {Stan Zdonik}. 2015.
\newblock \showarticletitle{An Architecture for Compiling UDF-centric
  Workflows}.
\newblock {\em Proc. VLDB Endow.\/} {8}, 12 (Aug. 2015), 1466--1477.
\newblock
\showISSN{2150-8097}
\showDOI{%
\url{http://dx.doi.org/10.14778/2824032.2824045}}


\bibitem[\protect\citeauthoryear{Crotty, Galakatos, Dursun, Kraska, Cetintemel,
  and Zdonik}{Crotty et~al\mbox{.}}{2014}]%
        {DBLP:journals/corr/CrottyGDKCZ14}
{Andrew Crotty}, {Alex Galakatos}, {Kayhan Dursun}, {Tim Kraska}, {Ugur
  Cetintemel}, {and} {Stan Zdonik}. 2014.
\newblock \showarticletitle{Tupleware: Redefining Modern Analytics}.
\newblock {\em CoRR\/}  {abs/1406.6667} (2014).
\newblock
\showURL{%
\url{http://arxiv.org/abs/1406.6667}}


\bibitem[\protect\citeauthoryear{Faith, Nyland, and Prins}{Faith
  et~al\mbox{.}}{1997}]%
        {khepera}
{Rickard~E. Faith}, {Lars~S. Nyland}, {and} {Jan~F. Prins}. 1997.
\newblock \showarticletitle{KHEPERA: A System for Rapid Implementation of
  Domain Specific Languages}. In {\em Proceedings of the 1997 Conference on
  Domain-Specific Languages} {\em (DSL' 97)}. USENIX Association, Berkeley, CA,
  USA, 19--19.
\newblock
\showURL{%
\url{https://www.usenix.org/legacy/publications/library/proceedings/dsl97/full_papers/faith/faith.pdf}}


\bibitem[\protect\citeauthoryear{F\"{a}rber, Cha, Primsch, Bornh\"{o}vd, Sigg,
  and Lehner}{F\"{a}rber et~al\mbox{.}}{2012}]%
        {SAPHANA}
{Franz F\"{a}rber}, {Sang~Kyun Cha}, {J\"{u}rgen Primsch}, {Christof
  Bornh\"{o}vd}, {Stefan Sigg}, {and} {Wolfgang Lehner}. 2012.
\newblock \showarticletitle{{SAP HANA} Database -- Data Management for Modern
  Business Applications}.
\newblock {\em SIGMOD Record\/} {40}, 4 (2012), 45--51.
\newblock
\showISSN{0163-5808}
\showDOI{%
\url{http://dx.doi.org/10.1145/2094114.2094126}}


\bibitem[\protect\citeauthoryear{Gill, Launchbury, and Peyton~Jones}{Gill
  et~al\mbox{.}}{1993}]%
        {foldr-fusion-1}
{Andrew Gill}, {John Launchbury}, {and} {Simon~L. Peyton~Jones}. 1993.
\newblock \showarticletitle{A Short Cut to Deforestation}. In {\em Proceedings
  of the Conference on Functional Programming Languages and Computer
  Architecture} {\em (FPCA '93)}. ACM, New York, NY, USA, 223--232.
\newblock
\showISBNx{0-89791-595-X}
\showDOI{%
\url{http://dx.doi.org/10.1145/165180.165214}}


\bibitem[\protect\citeauthoryear{Goel, Pound, Auch, Bumbulis, MacLean,
  F\"{a}rber, Gropengiesser, Mathis, Bodner, and Lehner}{Goel
  et~al\mbox{.}}{2015}]%
        {goel-scalable-analytics}
{Anil~K. Goel}, {Jeffrey Pound}, {Nathan Auch}, {Peter Bumbulis}, {Scott
  MacLean}, {Franz F\"{a}rber}, {Francis Gropengiesser}, {Christian Mathis},
  {Thomas Bodner}, {and} {Wolfgang Lehner}. 2015.
\newblock \showarticletitle{Towards Scalable Real-time Analytics: An
  Architecture for Scale-out of OLxP Workloads}.
\newblock {\em Proc. VLDB Endow.\/} {8}, 12 (Aug. 2015), 1716--1727.
\newblock
\showISSN{2150-8097}
\showDOI{%
\url{http://dx.doi.org/10.14778/2824032.2824069}}


\bibitem[\protect\citeauthoryear{Goldberg and Paige}{Goldberg and
  Paige}{1984}]%
        {Goldberg:1984:SP:800055.802021}
{Allen Goldberg} {and} {Robert Paige}. 1984.
\newblock \showarticletitle{Stream Processing}. In {\em Proceedings of the 1984
  ACM Symposium on LISP and Functional Programming} {\em (LFP '84)}. ACM, New
  York, NY, USA, 53--62.
\newblock
\showISBNx{0-89791-142-3}
\showDOI{%
\url{http://dx.doi.org/10.1145/800055.802021}}


\bibitem[\protect\citeauthoryear{Graefe}{Graefe}{1994}]%
        {Volcano}
{Goetz Graefe}. 1994.
\newblock \showarticletitle{{Volcano -- An Extensible and Parallel Query
  Evaluation System}}.
\newblock {\em IEEE Transactions on Knowledge and Data Engineering\/} {6}, 1
  (Feb 1994), 120--135.
\newblock
\showISSN{1041-4347}
\showDOI{%
\url{http://dx.doi.org/10.1109/69.273032}}


\bibitem[\protect\citeauthoryear{Greer}{Greer}{1999}]%
        {daytona}
{Rick Greer}. 1999.
\newblock \showarticletitle{{Daytona And The Fourth-Generation Language
  Cymbal}}. In {\em the 1999 ACM SIGMOD International Conference on Management
  of Data} {\em (SIGMOD '99)}. ACM, New York, NY, USA, 525--526.
\newblock
\showISBNx{1-58113-084-8}
\showDOI{%
\url{http://dx.doi.org/10.1145/304182.304242}}


\bibitem[\protect\citeauthoryear{Grust, Mayr, Rittinger, and Schreiber}{Grust
  et~al\mbox{.}}{2009}]%
        {ferry-2}
{Torsten Grust}, {Manuel Mayr}, {Jan Rittinger}, {and} {Tom Schreiber}. 2009.
\newblock \showarticletitle{FERRY -- Database-supported Program Execution}. In
  {\em Proceedings of the 2009 ACM SIGMOD International Conference on
  Management of Data} {\em (SIGMOD '09)}. ACM, New York, NY, USA, 1063--1066.
\newblock
\showISBNx{978-1-60558-551-2}
\showDOI{%
\url{http://dx.doi.org/10.1145/1559845.1559982}}


\bibitem[\protect\citeauthoryear{Harizopoulos, Liang, Abadi, and
  Madden}{Harizopoulos et~al\mbox{.}}{2006}]%
        {rowcolumn3}
{Stavros Harizopoulos}, {Velen Liang}, {Daniel~J. Abadi}, {and} {Samuel
  Madden}. 2006.
\newblock \showarticletitle{Performance Tradeoffs in Read-optimized Databases}.
  In {\em Proceedings of the 32nd International Conference on Very Large Data
  Bases} {\em (VLDB '06)}. VLDB Endowment, 487--498.
\newblock
\showURL{%
\url{http://dl.acm.org/citation.cfm?id=1182635.1164170}}


\bibitem[\protect\citeauthoryear{Holk, Pathirage, Chauhan, Lumsdaine, and
  Matsakis}{Holk et~al\mbox{.}}{2013}]%
        {rust-gpu-prog}
{Eric Holk}, {Milinda Pathirage}, {Arun Chauhan}, {Andrew Lumsdaine}, {and}
  {Nicholas~D. Matsakis}. 2013.
\newblock \showarticletitle{GPU Programming in Rust: Implementing High-Level
  Abstractions in a Systems-Level Language}. In {\em Proceedings of the 27th
  IEEE International Symposium on Parallel and Distributed Processing Workshops
  and PhD Forum} {\em (IPDPSW '13)}. IEEE Computer Society, Washington, DC,
  USA, 315--324.
\newblock
\showISBNx{978-0-7695-4979-8}
\showDOI{%
\url{http://dx.doi.org/10.1109/IPDPSW.2013.173}}


\bibitem[\protect\citeauthoryear{Hudak}{Hudak}{1996}]%
        {hudak-dsl}
{Paul Hudak}. 1996.
\newblock \showarticletitle{Building Domain-specific Embedded Languages}.
\newblock {\em ACM Comput. Surv.\/} {28}, 4es (Dec. 1996).
\newblock
\showISSN{0360-0300}
\showDOI{%
\url{http://dx.doi.org/10.1145/242224.242477}}


\bibitem[\protect\citeauthoryear{Humer, Wimmer, Wirth, W\"{o}\ss, and
  W\"{u}rthinger}{Humer et~al\mbox{.}}{2014}]%
        {truffle-dsl}
{Christian Humer}, {Christian Wimmer}, {Christian Wirth}, {Andreas W\"{o}\ss},
  {and} {Thomas W\"{u}rthinger}. 2014.
\newblock \showarticletitle{A Domain-specific Language for Building
  Self-Optimizing AST Interpreters}. In {\em Proceedings of the 2014
  International Conference on Generative Programming: Concepts and Experiences}
  {\em (GPCE 2014)}. ACM, New York, NY, USA, 123--132.
\newblock
\showISBNx{978-1-4503-3161-6}
\showDOI{%
\url{http://dx.doi.org/10.1145/2658761.2658776}}


\bibitem[\protect\citeauthoryear{Hunt and Larus}{Hunt and Larus}{2007}]%
        {singularity}
{Galen~C. Hunt} {and} {James~R. Larus}. 2007.
\newblock \showarticletitle{{Singularity: Rethinking the Software Stack}}.
\newblock {\em SIGOPS Oper. Syst. Rev.\/} {41}, 2 (2007), 37--49.
\newblock
\showISSN{0163-5980}
\showDOI{%
\url{http://dx.doi.org/10.1145/1243418.1243424}}


\bibitem[\protect\citeauthoryear{Jovanovi\'c, Shaikhha, Stucki, Nikolaev, Koch,
  and Odersky}{Jovanovi\'c et~al\mbox{.}}{2014}]%
        {yinyang}
{Vojin Jovanovi\'c}, {Amir Shaikhha}, {Sandro Stucki}, {Vladimir Nikolaev},
  {Christoph Koch}, {and} {Martin Odersky}. 2014.
\newblock \showarticletitle{Yin-Yang: Concealing the Deep Embedding of DSLs}.
  In {\em Proceedings of the 2014 International Conference on Generative
  Programming: Concepts and Experiences} {\em (GPCE 2014)}. ACM, New York, NY,
  USA, 73--82.
\newblock
\showISBNx{978-1-4503-3161-6}
\showDOI{%
\url{http://dx.doi.org/10.1145/2658761.2658771}}


\bibitem[\protect\citeauthoryear{Kallman, Kimura, Natkins, Pavlo, Rasin,
  Zdonik, Jones, Madden, Stonebraker, Zhang, Hugg, and Abadi}{Kallman
  et~al\mbox{.}}{2008}]%
        {hstore-compilation}
{Robert Kallman}, {Hideaki Kimura}, {Jonathan Natkins}, {Andrew Pavlo},
  {Alexander Rasin}, {Stanley Zdonik}, {Evan P.~C. Jones}, {Samuel Madden},
  {Michael Stonebraker}, {Yang Zhang}, {John Hugg}, {and} {Daniel~J. Abadi}.
  2008.
\newblock \showarticletitle{{H-{S}tore: A High-Performance, Distributed Main
  Memory Transaction Processing System}}.
\newblock {\em PVLDB\/} {1}, 2 (2008), 1496--1499.
\newblock
\showISSN{2150-8097}
\showURL{%
\url{http://dl.acm.org/citation.cfm?id=1454159.1454211}}


\bibitem[\protect\citeauthoryear{Kennedy, Broom, Chauhan, Fowler, Garvin,
  Koelbel, McCosh, and Mellor-Crummey}{Kennedy et~al\mbox{.}}{2005}]%
        {telescoping}
{Ken Kennedy}, {Bradley Broom}, {Arun Chauhan}, {Robert~J. Fowler}, {John
  Garvin}, {Charles Koelbel}, {Cheryl McCosh}, {and} {John Mellor-Crummey}.
  2005.
\newblock \showarticletitle{Telescoping Languages: A System for Automatic
  Generation of Domain Languages}.
\newblock {\it Proc. IEEE} {93}, 2 (2005), 387--408.
\newblock
\showDOI{%
\url{http://dx.doi.org/10.1109/JPROC.2004.840447}}


\bibitem[\protect\citeauthoryear{Klonatos, Koch, Rompf, and Chafi}{Klonatos
  et~al\mbox{.}}{2014}]%
        {legobase}
{Yannis Klonatos}, {Christoph Koch}, {Tiark Rompf}, {and} {Hassan Chafi}. 2014.
\newblock \showarticletitle{Building Efficient Query Engines in a High-Level
  Language}.
\newblock {\em {PVLDB}\/} {7}, 10 (2014), 853--864.
\newblock


\bibitem[\protect\citeauthoryear{Klonatos, N\"{o}tzli, Spielmann, Koch, and
  Kuncak}{Klonatos et~al\mbox{.}}{2013}]%
        {ocas}
{Yannis Klonatos}, {Andres N\"{o}tzli}, {Andrej Spielmann}, {Christoph Koch},
  {and} {Victor Kuncak}. 2013.
\newblock \showarticletitle{{Automatic Synthesis of Out-of-core Algorithms}}.
  In {\em the 2013 ACM SIGMOD International Conference on Management of Data}
  {\em (SIGMOD '13)}. ACM, 133--144.
\newblock
\showISBNx{978-1-4503-2037-5}
\showDOI{%
\url{http://dx.doi.org/10.1145/2463676.2465334}}


\bibitem[\protect\citeauthoryear{Koch}{Koch}{2013}]%
        {koch}
{Christoph Koch}. 2013.
\newblock \showarticletitle{{Abstraction without regret in data management
  systems.}}. In {\em CIDR}. www.cidrdb.org.
\newblock
\showURL{%
\url{http://www.cidrdb.org/cidr2013/Papers/CIDR13_Paper149.pdf}}


\bibitem[\protect\citeauthoryear{Koch}{Koch}{2014}]%
        {kochmanifesto}
{Christoph Koch}. 2014.
\newblock \showarticletitle{Abstraction Without Regret in Database Systems
  Building: a Manifesto}.
\newblock {\em IEEE Data Eng. Bull.\/} {37}, 1 (2014), 70--79.
\newblock
\showURL{%
\url{http://sites.computer.org/debull/A14mar/p70.pdf}}


\bibitem[\protect\citeauthoryear{Krikellas, Viglas, and Cintra}{Krikellas
  et~al\mbox{.}}{2010}]%
        {krikellas}
{Konstantinos Krikellas}, {Stratis Viglas}, {and} {Marcelo Cintra}. 2010.
\newblock \showarticletitle{{Generating code for holistic query evaluation}}.
  In {\em Proceedings of the 26th International Conference on Data Engineering}
  {\em (ICDE '10)}. {IEEE Computer Society}, Washington, DC, USA, 613--624.
\newblock
\showISSN{1063-6382}
\showDOI{%
\url{http://dx.doi.org/10.1109/ICDE.2010.5447892}}


\bibitem[\protect\citeauthoryear{Lattner and Adve}{Lattner and Adve}{2004}]%
        {llvm}
{Chris Lattner} {and} {Vikram Adve}. 2004.
\newblock \showarticletitle{LLVM: A Compilation Framework for Lifelong Program
  Analysis \& Transformation}. In {\em Proceedings of the International
  Symposium on Code Generation and Optimization: Feedback-directed and Runtime
  Optimization} {\em (CGO '04)}. IEEE Computer Society, Washington, DC, USA,
  75--86.
\newblock
\showISBNx{0-7695-2102-9}
\showURL{%
\url{http://dl.acm.org/citation.cfm?id=977395.977673}}


\bibitem[\protect\citeauthoryear{Lee, Brown, Sujeeth, Chafi, Rompf, Odersky,
  and Olukotun}{Lee et~al\mbox{.}}{2011}]%
        {delite}
{HyoukJoong Lee}, {Kevin~J. Brown}, {Arvind~K. Sujeeth}, {Hassan Chafi}, {Tiark
  Rompf}, {Martin Odersky}, {and} {Kunle Olukotun}. 2011.
\newblock \showarticletitle{Implementing Domain-Specific Languages for
  Heterogeneous Parallel Computing}.
\newblock {\em IEEE Micro\/} {31}, 5 (Sept. 2011), 42--53.
\newblock
\showISSN{0272-1732}
\showDOI{%
\url{http://dx.doi.org/10.1109/MM.2011.68}}


\bibitem[\protect\citeauthoryear{Manegold, Kersten, and Boncz}{Manegold
  et~al\mbox{.}}{2009}]%
        {monetdb-mammals}
{Stefan Manegold}, {Martin~L. Kersten}, {and} {Peter Boncz}. 2009.
\newblock \showarticletitle{{Database Architecture Evolution: Mammals
  Flourished long before Dinosaurs became Extinct}}.
\newblock {\em PVLDB\/} {2}, 2 (2009), 1648--1653.
\newblock
\showISSN{2150-8097}
\showDOI{%
\url{http://dx.doi.org/10.14778/1687553.1687618}}


\bibitem[\protect\citeauthoryear{Nagel, Bierman, and Viglas}{Nagel
  et~al\mbox{.}}{2014}]%
        {Nagel:2014:CGE:2732977.2732984}
{Fabian Nagel}, {Gavin Bierman}, {and} {Stratis~D. Viglas}. 2014.
\newblock \showarticletitle{Code Generation for Efficient Query Processing in
  Managed Runtimes}.
\newblock {\em Proc. VLDB Endow.\/} {7}, 12 (Aug. 2014), 1095--1106.
\newblock
\showISSN{2150-8097}
\showDOI{%
\url{http://dx.doi.org/10.14778/2732977.2732984}}


\bibitem[\protect\citeauthoryear{Neumann}{Neumann}{2011}]%
        {neumann11}
{Thomas Neumann}. 2011.
\newblock \showarticletitle{Efficiently {C}ompiling {E}fficient {Q}uery {P}lans
  for {M}odern {H}ardware}.
\newblock {\em PVLDB\/} {4}, 9 (2011), 539--550.
\newblock
\showURL{%
\url{http://www.vldb.org/pvldb/vol4/p539-neumann.pdf}}


\bibitem[\protect\citeauthoryear{Odersky and Zenger}{Odersky and
  Zenger}{2005}]%
        {odersky_scalable}
{Martin Odersky} {and} {Matthias Zenger}. 2005.
\newblock \showarticletitle{{Scalable Component Abstractions}}. In {\em the
  20th Annual ACM SIGPLAN Conference on Object-oriented Programming, Systems,
  Languages, and Applications} {\em (OOPSLA '05)}. ACM, New York, NY, USA,
  41--57.
\newblock
\showISBNx{1-59593-031-0}
\showDOI{%
\url{http://dx.doi.org/10.1145/1094811.1094815}}


\bibitem[\protect\citeauthoryear{{Oracle Corporation}}{{Oracle
  Corporation}}{2006}]%
        {TimesTen}
{{Oracle Corporation}}. 2006.
\newblock {\hspace{0.4cm}TimesTen\hspace{0.4cm} In-Memory\hspace{0.4cm}
  Database\hspace{0.4cm} Architectural\hspace{0.4cm} Overview}.
\newblock   (2006).
\newblock
\newblock
\shownote{\url{http://download.oracle.com/otn_hosted_doc/timesten/603/TimesTen-Documentation/arch.pdf}.}


\bibitem[\protect\citeauthoryear{Padmanabhan, Malkemus, Agarwal, and
  Jhingran}{Padmanabhan et~al\mbox{.}}{2001}]%
        {blockprocessing}
{Sriram Padmanabhan}, {Timothy Malkemus}, {Ramesh~C. Agarwal}, {and} {Anant
  Jhingran}. 2001.
\newblock \showarticletitle{{Block oriented processing of Relational Database
  operations in modern Computer Architectures}}. In {\em Proceedings of the
  17th International Conference on Data Engineering} {\em (ICDE '01)}. IEEE
  Computer Society, Washington, DC, USA, 567--574.
\newblock
\showDOI{%
\url{http://dx.doi.org/10.1109/ICDE.2001.914871}}


\bibitem[\protect\citeauthoryear{Raman, Swart, Qiao, Reiss, Dialani, Kossmann,
  Narang, and Sidle}{Raman et~al\mbox{.}}{2008}]%
        {constanttimequeryprocessing}
{Vijayshankar Raman}, {Garret Swart}, {Lin Qiao}, {Frederick Reiss}, {Vijay
  Dialani}, {Donald Kossmann}, {Inderpal Narang}, {and} {Richard Sidle}. 2008.
\newblock \showarticletitle{{Constant-Time Query Processing}}. In {\em
  Proceedings of the 24th International Conference on Data Engineering} {\em
  (ICDE '08)}. IEEE Computer Society, Washington, DC, USA, 60--69.
\newblock
\showISBNx{978-1-4244-1836-7}
\showDOI{%
\url{http://dx.doi.org/10.1109/ICDE.2008.4497414}}


\bibitem[\protect\citeauthoryear{Rao, Pirahesh, Mohan, and Lohman}{Rao
  et~al\mbox{.}}{2006}]%
        {mohan-compilation}
{Jun Rao}, {Hamid Pirahesh}, {C. Mohan}, {and} {Guy Lohman}. 2006.
\newblock \showarticletitle{{Compiled Query Execution Engine using JVM}}. In
  {\em Proceedings of the 22nd International Conference on Data Engineering}
  {\em (ICDE '06)}. IEEE Computer Society, Washington, DC, USA, 23--34.
\newblock
\showISBNx{0-7695-2570-9}
\showDOI{%
\url{http://dx.doi.org/10.1109/ICDE.2006.40}}


\bibitem[\protect\citeauthoryear{Rompf}{Rompf}{2012}]%
        {tiark-phd-thesis}
{Tiark Rompf}. 2012.
\newblock {\em Lightweight {M}odular {S}taging and {E}mbedded {C}ompilers:
  {A}bstraction {W}ithout {R}egret for {H}igh-{L}evel {H}igh-{P}erformance
  {P}rogramming}.
\newblock Ph.D. Dissertation. \'{E}cole {P}olytechnique {F}\'{e}d\'{e}rale de
  {L}ausanne (EPFL).
\newblock
\showDOI{%
\url{http://dx.doi.org/10.5075/epfl-thesis-5456}}


\bibitem[\protect\citeauthoryear{Rompf and Odersky}{Rompf and Odersky}{2010}]%
        {lms}
{Tiark Rompf} {and} {Martin Odersky}. 2010.
\newblock \showarticletitle{{Lightweight Modular Staging: A Pragmatic Approach
  to Runtime Code Generation and Compiled DSLs}}. In {\em the ninth
  international conference on Generative programming and component engineering}
  {\em (GPCE '10)}. ACM, New York, NY, USA, 127--136.
\newblock
\showISBNx{978-1-4503-0154-1}
\showDOI{%
\url{http://dx.doi.org/10.1145/1868294.1868314}}


\bibitem[\protect\citeauthoryear{Rompf, Sujeeth, Amin, Brown, Jovanovic, Lee,
  Jonnalagedda, Olukotun, and Odersky}{Rompf et~al\mbox{.}}{2013}]%
        {rompf13popl}
{Tiark Rompf}, {Arvind~K. Sujeeth}, {Nada Amin}, {Kevin~J. Brown}, {Vojin
  Jovanovic}, {HyoukJoong Lee}, {Manohar Jonnalagedda}, {Kunle Olukotun}, {and}
  {Martin Odersky}. 2013.
\newblock \showarticletitle{Optimizing Data Structures in High-level Programs:
  New Directions for Extensible Compilers based on Staging}. In {\em
  Proceedings of the 40th Annual ACM SIGPLAN-SIGACT Symposium on Principles of
  Programming Languages} {\em (POPL '13)}. ACM, New York, NY, USA, 497--510.
\newblock
\showISBNx{978-1-4503-1832-7}
\showDOI{%
\url{http://dx.doi.org/10.1145/2429069.2429128}}


\bibitem[\protect\citeauthoryear{Sompolski, Zukowski, and Boncz}{Sompolski
  et~al\mbox{.}}{2011}]%
        {vectorizevscompile}
{Juliusz Sompolski}, {Marcin Zukowski}, {and} {Peter Boncz}. 2011.
\newblock \showarticletitle{{Vectorization vs. Compilation in Query
  Execution}}. In {\em the Seventh International Workshop on Data Management on
  New Hardware} {\em (DaMoN '11)}. ACM, New York, NY, USA, 33--40.
\newblock
\showISBNx{978-1-4503-0658-4}
\showDOI{%
\url{http://dx.doi.org/10.1145/1995441.1995446}}


\bibitem[\protect\citeauthoryear{Stonebraker, Abadi, Batkin, Chen, Cherniack,
  Ferreira, Lau, Lin, Madden, O'Neil, O'Neil, Rasin, Tran, and
  Zdonik}{Stonebraker et~al\mbox{.}}{2005}]%
        {rowcolumn2}
{Mike Stonebraker}, {\hspace{0.2cm}Daniel~J. Abadi}, {\hspace{0.2cm}Adam
  Batkin}, {\hspace{0.2cm}Xuedong Chen}, {\hspace{0.2cm}Mitch Cherniack},
  {\hspace{0.2cm}Miguel Ferreira}, {\hspace{0.2cm}Edmond Lau},
  {\hspace{0.2cm}Amerson Lin}, {\hspace{0.2cm}Sam Madden},
  {\hspace{0.2cm}Elizabeth O'Neil}, {\hspace{0.2cm}Pat O'Neil},
  {\hspace{0.2cm}Alex Rasin}, {\hspace{0.2cm}Nga Tran}, {and}
  {\hspace{0.2cm}Stan Zdonik}. 2005.
\newblock \showarticletitle{{C-{S}tore:\hspace{0.2cm} A\hspace{0.2cm}
  Column-oriented\hspace{0.2cm} {DBMS}}}. In {\em
  \hspace{0.2cm}the\hspace{0.2cm} 31st\hspace{0.2cm}
  International\hspace{0.2cm} Conference\hspace{0.2cm} on\hspace{0.2cm}
  Very\hspace{0.2cm} Large\hspace{0.2cm} Data\hspace{0.2cm}
  Bases\hspace{0.2cm}} {\em (VLDB '05)}. VLDB Endowment, 553--564.
\newblock
\showISBNx{1-59593-154-6}
\showURL{%
\url{http://dl.acm.org/citation.cfm?id=1083592.1083658}}


\bibitem[\protect\citeauthoryear{Stonebraker, Madden, Abadi, Harizopoulos,
  Hachem, and Helland}{Stonebraker et~al\mbox{.}}{2007}]%
        {VoltDB}
{Michael Stonebraker}, {Samuel Madden}, {Daniel~J. Abadi}, {Stavros
  Harizopoulos}, {Nabil Hachem}, {and} {Pat Helland}. 2007.
\newblock \showarticletitle{The end of an architectural era: (it's time for a
  complete rewrite)}. In {\em \hspace{0.1cm}the\hspace{0.1cm}
  33rd\hspace{0.1cm} international\hspace{0.1cm} conference\hspace{0.1cm}
  on\hspace{0.1cm} Very\hspace{0.1cm} large\hspace{0.1cm} data\hspace{0.1cm}
  bases} {\em (VLDB '07)}. VLDB Endowment, 1150--1160.
\newblock
\showISBNx{978-1-59593-649-3}
\showURL{%
\url{http://dl.acm.org/citation.cfm?id=1325851.1325981}}


\bibitem[\protect\citeauthoryear{Sujeeth, Gibbons, Brown, Lee, Rompf, Odersky,
  and Olukotun}{Sujeeth et~al\mbox{.}}{2013}]%
        {forge}
{Arvind~K Sujeeth}, {Austin Gibbons}, {Kevin~J Brown}, {HyoukJoong Lee}, {Tiark
  Rompf}, {Martin Odersky}, {and} {Kunle Olukotun}. 2013.
\newblock \showarticletitle{Forge: Generating a High Performance {DSL}
  Implementation from a Declarative Specification}. In {\em Proceedings of the
  12th international conference on Generative programming: concepts \&
  experiences}. ACM, New York, NY, USA, 145--154.
\newblock
\showDOI{%
\url{http://dx.doi.org/10.1145/2517208.2517220}}


\bibitem[\protect\citeauthoryear{Sumii and Kobayashi}{Sumii and
  Kobayashi}{2001}]%
        {anftoast}
{Eijiro Sumii} {and} {Naoki Kobayashi}. 2001.
\newblock \showarticletitle{A Hybrid Approach to Online and Offline Partial
  Evaluation}.
\newblock {\em Higher Order Symbol. Comput.\/} {14}, 2-3 (Sept. 2001),
  101--142.
\newblock
\showISSN{1388-3690}
\showDOI{%
\url{http://dx.doi.org/10.1023/A:1012984529382}}


\bibitem[\protect\citeauthoryear{Svenningsson}{Svenningsson}{2002}]%
        {Svenningsson:2002:SFA:581478.581491}
{Josef Svenningsson}. 2002.
\newblock \showarticletitle{Shortcut Fusion for Accumulating Parameters \&
  Zip-like Functions}. In {\em Proceedings of the Seventh ACM SIGPLAN
  International Conference on Functional Programming} {\em (ICFP '02)}. ACM,
  New York, NY, USA, 124--132.
\newblock
\showISBNx{1-58113-487-8}
\showDOI{%
\url{http://dx.doi.org/10.1145/581478.581491}}


\bibitem[\protect\citeauthoryear{Taha and Sheard}{Taha and Sheard}{2000}]%
        {taha00staging}
{Walid Taha} {and} {Tim Sheard}. 2000.
\newblock \showarticletitle{{MetaML and multi-stage programming with explicit
  annotations}}.
\newblock {\em Theor. Comput. Sci.\/} {248}, 1-2 (2000), 211--242.
\newblock
\showDOI{%
\url{http://dx.doi.org/10.1016/S0304-3975(00)00053-0}}


\bibitem[\protect\citeauthoryear{{The GNOME Project}}{{The GNOME
  Project}}{2013}]%
        {GLib}
{{The GNOME Project}}. 2013.
\newblock {GLib}: Library Package for low-level data structures in {C} -- The
  Reference Manual.
\newblock   (2013).
\newblock
\newblock
\shownote{{\url{https://developer.gnome.org/glib/2.38/}}.}


\bibitem[\protect\citeauthoryear{{Transaction Processing Performance
  Council}}{{Transaction Processing Performance Council}}{1999}]%
        {tpch}
{{Transaction Processing Performance Council}}. 1999.
\newblock {TPC-H, an ad-hoc, decision support benchmark.}
\newblock   (1999).
\newblock
\showURL{%
\url{http://www.tpc.org/tpch}}


\bibitem[\protect\citeauthoryear{van Deursen, Klint, and Visser}{van Deursen
  et~al\mbox{.}}{2000}]%
        {annotated-dsl}
{Arie van Deursen}, {Paul Klint}, {and} {Joost Visser}. 2000.
\newblock \showarticletitle{Domain-specific Languages: An Annotated
  Bibliography}.
\newblock {\em SIGPLAN Not.\/} {35}, 6 (June 2000), 26--36.
\newblock
\showISSN{0362-1340}
\showDOI{%
\url{http://dx.doi.org/10.1145/352029.352035}}


\bibitem[\protect\citeauthoryear{Viglas, Bierman, and Nagel}{Viglas
  et~al\mbox{.}}{2014}]%
        {DBLP:journals/debu/ViglasBN14}
{Stratis Viglas}, {Gavin~M. Bierman}, {and} {Fabian Nagel}. 2014.
\newblock \showarticletitle{Processing\hspace{0.05cm}
  Declarative\hspace{0.05cm} Queries\hspace{0.05cm} Through\hspace{0.05cm}
  Generating\hspace{0.05cm} Imperative\hspace{0.05cm} Code\hspace{0.05cm}
  in\hspace{0.05cm} Managed\hspace{0.05cm} Runtimes}.
\newblock {\em {IEEE} Data Eng. Bull.\/} {37}, 1 (2014), 12--21.
\newblock
\showURL{%
\url{http://sites.computer.org/debull/A14mar/p12.pdf}}


\bibitem[\protect\citeauthoryear{Yu, Isard, Fetterly, Budiu, Erlingsson, Gunda,
  and Currey}{Yu et~al\mbox{.}}{2008}]%
        {dryadlunc}
{Yuan Yu}, {Michael Isard}, {Dennis Fetterly}, {Mihai Budiu}, {\'{U}lfar
  Erlingsson}, {Pradeep~Kumar Gunda}, {and} {Jon Currey}. 2008.
\newblock \showarticletitle{DryadLINQ: A \hspace{0.1cm}System \hspace{0.1cm}for
  \hspace{0.1cm}General-purpose \hspace{0.1cm}Distributed
  \hspace{0.1cm}Data-parallel \hspace{0.1cm}Computing \hspace{0.1cm}Using
  \hspace{0.1cm}a \hspace{0.1cm}High-level Language}. In {\em
  \hspace{0.1cm}Proceedings \hspace{0.1cm}of \hspace{0.1cm}the
  \hspace{0.1cm}8th \hspace{0.1cm}USENIX \hspace{0.1cm}Conference
  \hspace{0.1cm}on \hspace{0.1cm}Operating \hspace{0.1cm}Systems
  \hspace{0.1cm}Design and \hspace{0.1cm}Implementation} {\em (OSDI' 08)}.
  \hspace{0.1cm}USENIX \hspace{0.1cm}Association, \hspace{0.1cm}Berkeley,
  \hspace{0.1cm}CA, \hspace{0.1cm}USA, \hspace{0.1cm}1--14.
\newblock
\showURL{%
\url{http://dl.acm.org/citation.cfm?id=1855741.1855742}}


\bibitem[\protect\citeauthoryear{Zadok, Iyer, Joukov, Sivathanu, and
  Wright}{Zadok et~al\mbox{.}}{2006}]%
        {zadok-incremental-fs}
{Erez Zadok}, {Rakesh Iyer}, {Nikolai Joukov}, {Gopalan Sivathanu}, {and}
  {Charles~P. Wright}. 2006.
\newblock \showarticletitle{On Incremental File System Development}.
\newblock {\em Transactions on Storage\/} {2}, 2 (May 2006), 161--196.
\newblock
\showISSN{1553-3077}
\showDOI{%
\url{http://dx.doi.org/10.1145/1149976.1149979}}


\bibitem[\protect\citeauthoryear{Zaharia, Chowdhury, Franklin, Shenker, and
  Stoica}{Zaharia et~al\mbox{.}}{2010}]%
        {spark-zaharia}
{Matei Zaharia}, {Mosharaf Chowdhury}, {Michael~J. Franklin}, {Scott Shenker},
  {and} {Ion Stoica}. 2010.
\newblock \showarticletitle{Spark: \hspace{0.1cm}Cluster
  \hspace{0.1cm}Computing \hspace{0.1cm}with \hspace{0.1cm}Working
  \hspace{0.1cm}Sets}. In {\em \hspace{0.1cm}Proceedings \hspace{0.1cm}of
  \hspace{0.1cm}the \hspace{0.1cm}2nd \hspace{0.1cm}USENIX
  \hspace{0.1cm}Conference \hspace{0.1cm}on \hspace{0.1cm}Hot
  \hspace{0.1cm}Topics \hspace{0.1cm}in \hspace{0.1cm}Cloud
  \hspace{0.1cm}Computing} {\em (HotCloud' 10)}. \hspace{0.1cm}USENIX
  \hspace{0.1cm}Association, \hspace{0.1cm}Berkeley, \hspace{0.1cm}CA,
  \hspace{0.1cm}USA, \hspace{0.1cm}10--10.
\newblock
\showURL{%
\url{http://dl.acm.org/citation.cfm?id=1863103.1863113}}


\bibitem[\protect\citeauthoryear{Zane, Ballard, Hinshaw, Kirkpatrick, and
  Premanand~Yerabothu}{Zane et~al\mbox{.}}{2008}]%
        {sql-generation-zhao}
{Barry~M. Zane}, {James~P. Ballard}, {Foster~D. Hinshaw}, {Dana~A.
  Kirkpatrick}, {and} {Less Premanand~Yerabothu}. 2008.
\newblock {Optimized {SQL} Code Generation (US Patent 7430549 B2)}.
\newblock WO Patent App. US 10/886,011.   (Sept. 2008).
\newblock
\showURL{%
\url{http://www.google.ch/patents/US7430549}}


\bibitem[\protect\citeauthoryear{Zhang, Debray, and Snodgrass}{Zhang
  et~al\mbox{.}}{2012a}]%
        {microspecialization2}
{Rui Zhang}, {Saumya Debray}, {and} {Richard~T. Snodgrass}. 2012a.
\newblock \showarticletitle{{Micro-Specialization: Dynamic Code Specialization
  of Database Management Systems}}. In {\em the Tenth ACM International
  Symposium on Code Generation and Optimization} {\em (CGO '12)}. ACM, New
  York, NY, USA, 63--73.
\newblock
\showISBNx{978-1-4503-1206-6}
\showDOI{%
\url{http://dx.doi.org/10.1145/2259016.2259025}}


\bibitem[\protect\citeauthoryear{Zhang, Snodgrass, and Debray}{Zhang
  et~al\mbox{.}}{2012b}]%
        {microspecialization1}
{Rui Zhang}, {Richard~T. Snodgrass}, {and} {Saumya Debray}. 2012b.
\newblock \showarticletitle{{Application of Micro-specialization to Query
  Evaluation Operators}}. In {\em Proceedings of the 28th International
  Conference on Data Engineering Workshops} {\em (ICDEW '12)}. IEEE Computer
  Society, Washington, DC, USA, 315--321.
\newblock
\showDOI{%
\url{http://dx.doi.org/10.1109/ICDEW.2012.43}}


\bibitem[\protect\citeauthoryear{Zhang, Snodgrass, and Debray}{Zhang
  et~al\mbox{.}}{2012c}]%
        {microspecialization3}
{Rui Zhang}, {Richard~T. Snodgrass}, {and} {Saumya Debray}. 2012c.
\newblock \showarticletitle{{Micro-Specialization in {DBMS}es}}. In {\em ICDE}.
  IEEE Computer Society, Washington, DC, USA, 690--701.
\newblock
\showISSN{1063-6382}
\showDOI{%
\url{http://dx.doi.org/10.1109/ICDE.2012.110}}


\end{thebibliography}
}

\end{document}